\documentclass[acmtog]{acmart}

\usepackage{multirow}
\usepackage{booktabs}

\usepackage[ruled]{algorithm2e}

\SetAlFnt{\small}
\SetAlCapFnt{\small}
\SetAlCapNameFnt{\small}
\SetAlCapHSkip{0pt}

\usepackage{tikz}
\usepackage{graphicx}
\usepackage[export]{adjustbox}

\acmJournal{TOG}

\begin{document}
	
\title{Appearance-Driven Automatic 3D Model Simplification}

\author{Jon Hasselgren}
\affiliation{\institution{NVIDIA}\country{Sweden}}
\author{Jacob Munkberg}
\affiliation{\institution{NVIDIA}\country{Sweden}}
\author{Jaakko Lehtinen}
\affiliation{\institution{NVIDIA \& Aalto University}\country{Finland}}
\author{Miika Aittala}
\affiliation{\institution{NVIDIA}\country{Finland}}
\author{Samuli Laine}
\affiliation{\institution{NVIDIA}\country{Finland}}


\newcommand{\figSkull}{
\begin{figure*}[t]
	\centering
	\setlength{\tabcolsep}{1pt}
	\begin{tabular}{cccc}    
		\includegraphics[width=0.24\textwidth]{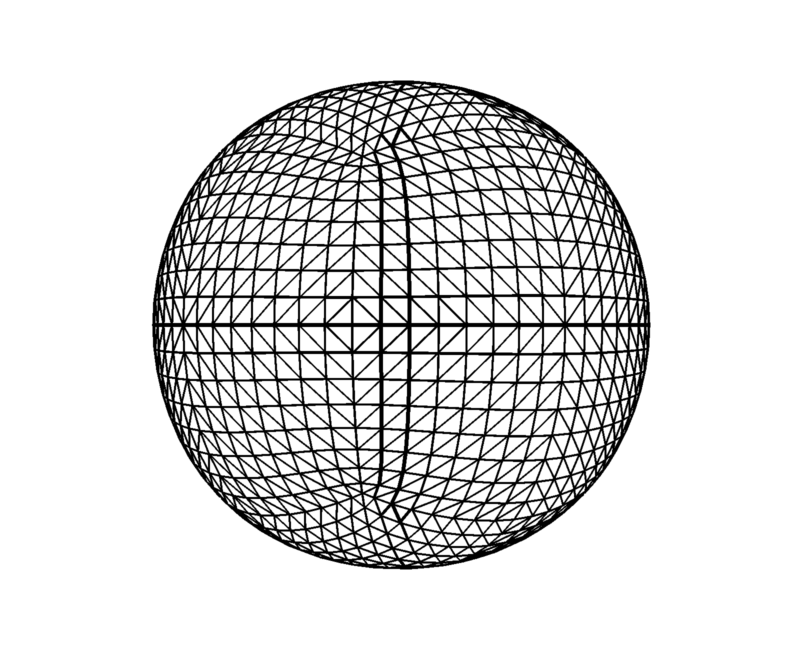} &
		\includegraphics[width=0.24\textwidth]{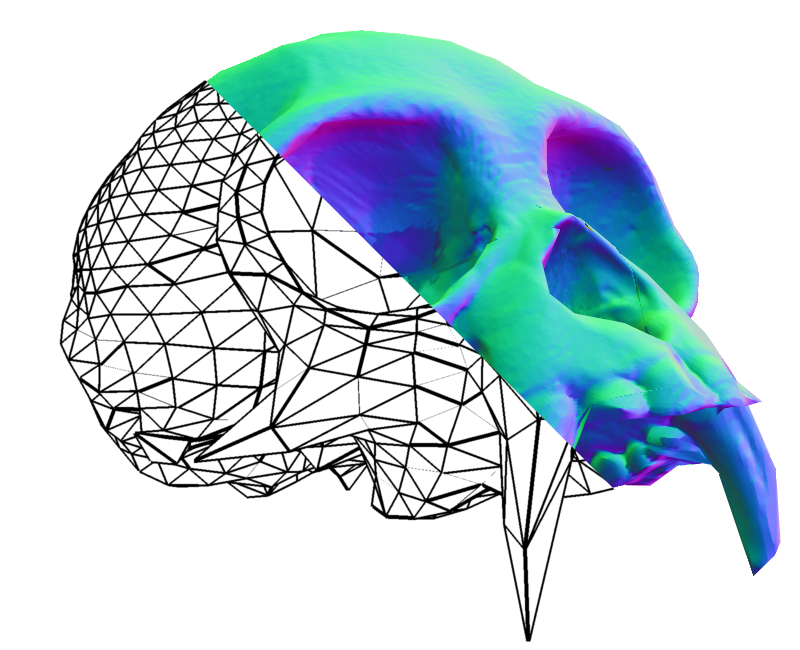} &
		\includegraphics[width=0.24\textwidth]{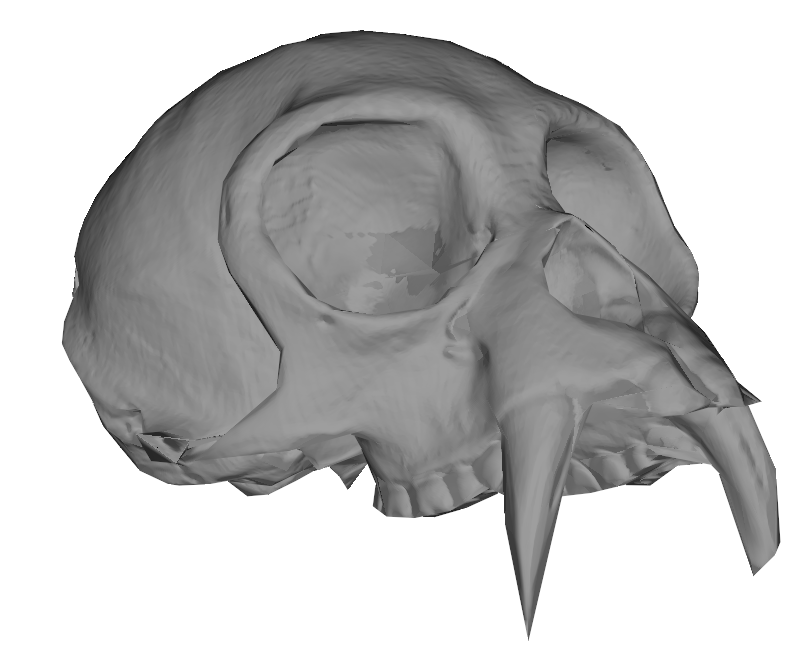} &
		\includegraphics[width=0.24\textwidth]{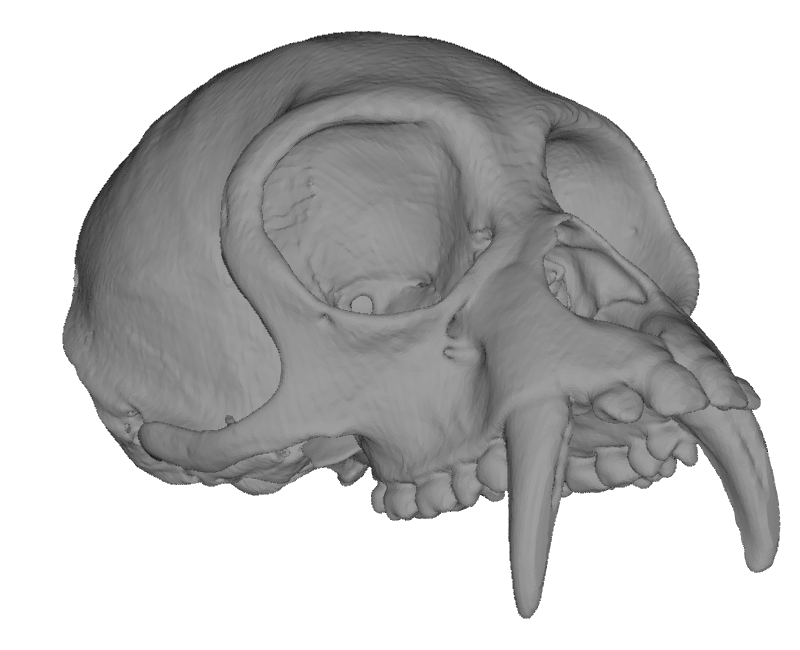} \\
		\small{Initial guess (3k tris)} & \small{Optimized positions \& normals} 
		& \small{Our (3k tris), PSNR: 24.96~dB} & \small{Ref (735k tris)} \\   
		\includegraphics[width=0.24\textwidth]{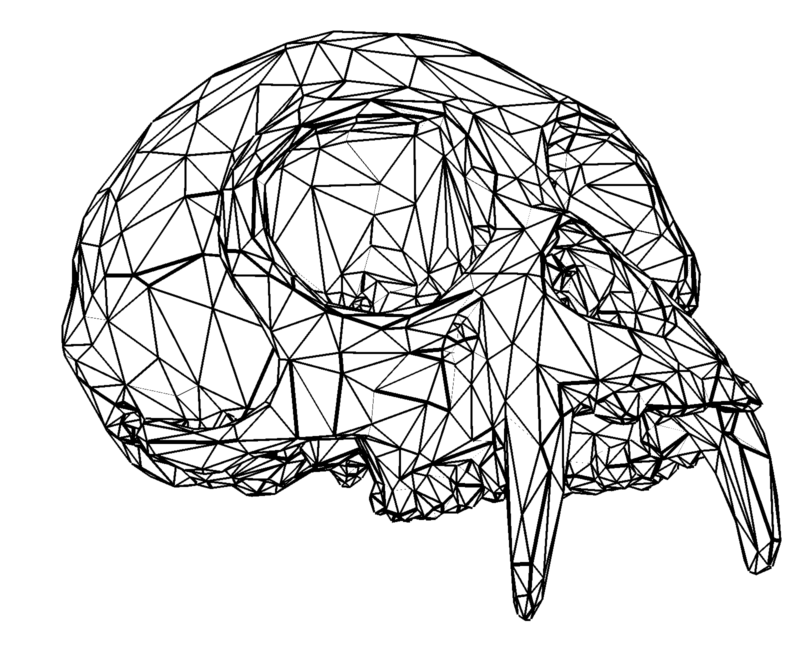} &
		\includegraphics[width=0.24\textwidth]{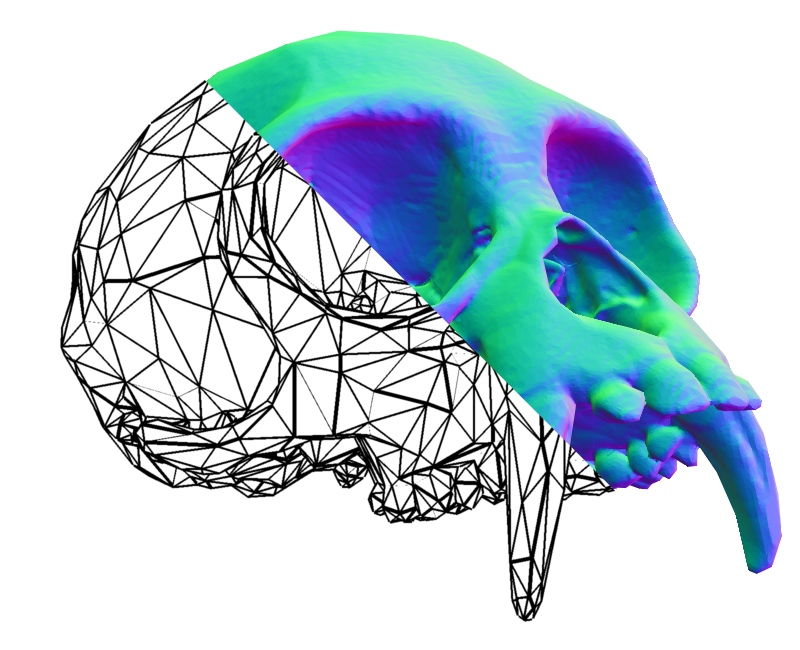} &
		\includegraphics[width=0.24\textwidth]{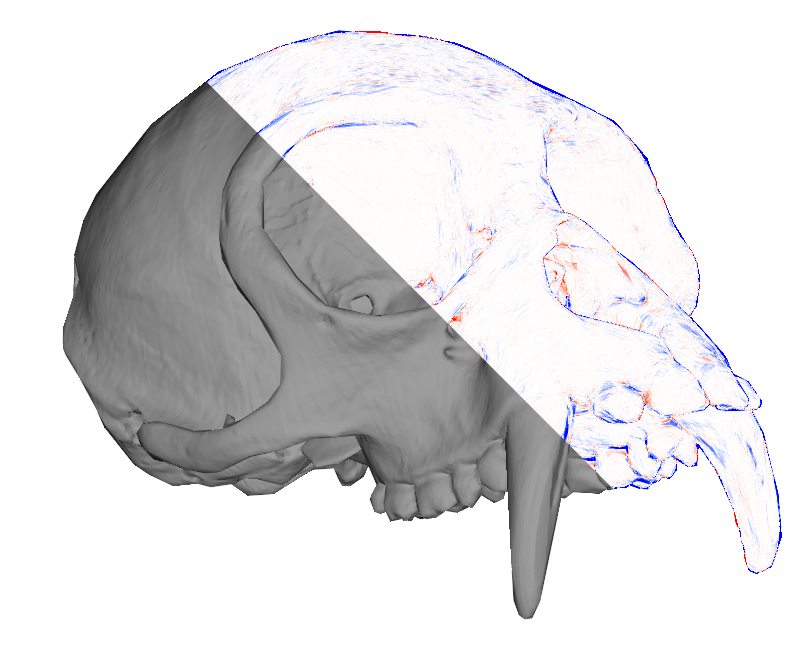} &
		\includegraphics[width=0.24\textwidth]{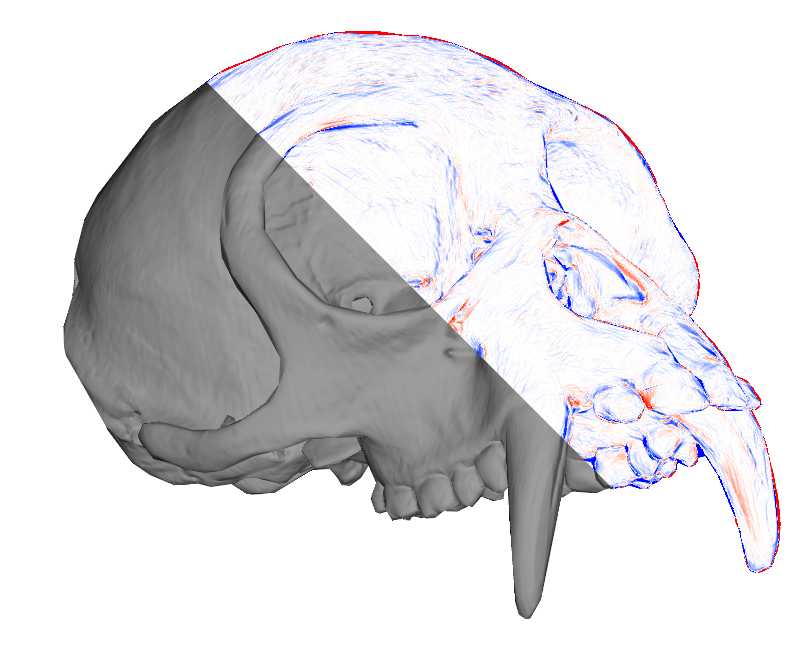} \\
		\small{Initial guess (9k tris)} & \small{Optimized positions \& normals} &
		\small{Our (9k tris), PSNR: 28.30~dB} & \small{Simplygon (9k tris), PSNR: 27.07~dB}
	\end{tabular}
	\vspace*{-2mm}
	\caption{\textbf{Top:} Starting from a sphere, we jointly optimize vertex positions and tangent space normal maps based 
		on image observations of a reference mesh. 
		\textbf{Bottom:} Starting from a reduced mesh with 9k triangles. We compare against the normal map baker in Simplygon and 
		show split images with difference images (red/blue = too bright/dim compared to reference).
		The mesh is courtesy of the Smithsonian 3D Digitization project~\shortcite{Smithsonian2020}.}
	\label{fig:skull}
\end{figure*}
}


\newcommand{\dispbox}[5]
{{\begin{tikzpicture}
		\node[anchor=south west,inner sep=0] (image) at (0,0) {\includegraphics[width=0.245\columnwidth]{#1}};
		\begin{scope}[x={(image.south east)},y={(image.north west)}]
		\draw[orange,thick] (#2,#3) rectangle (#4,#5);
		\end{scope}
		\end{tikzpicture}}}

\newcommand{\figDisplacementMapping}{
\begin{figure}
	\centering
	\setlength{\tabcolsep}{1pt}
	\begin{tabular}{cccc}
		\includegraphics[width=0.245\columnwidth]{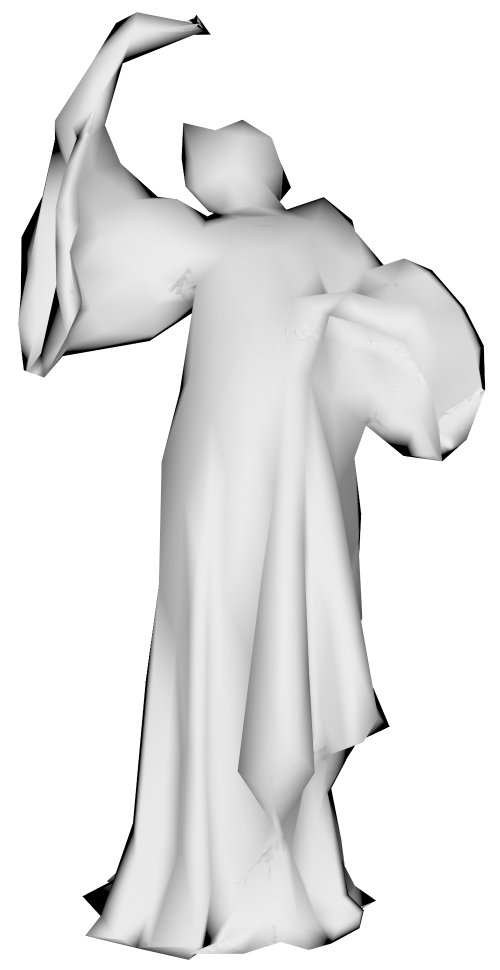} &
		\includegraphics[width=0.245\columnwidth]{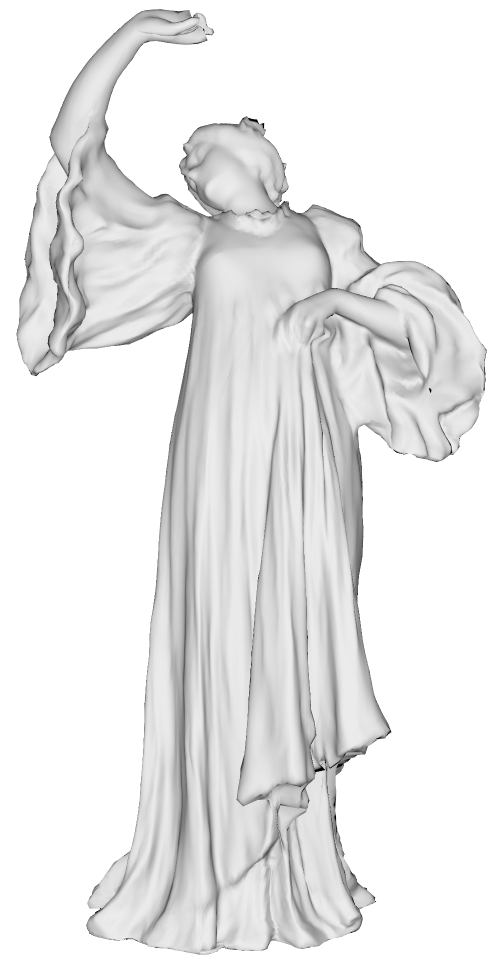} &
		\includegraphics[width=0.245\columnwidth]{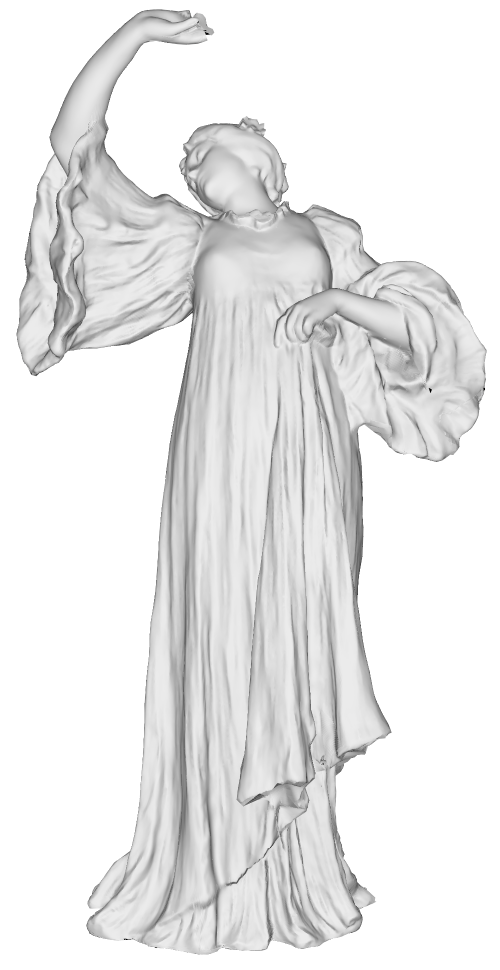} &
		\dispbox{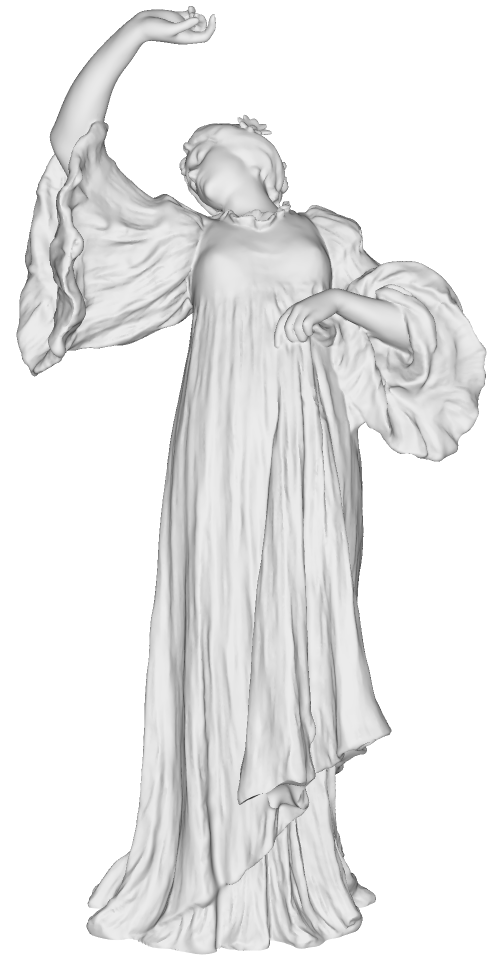}{0.53}{0.61}{0.77}{0.73} \\
		\includegraphics[width=0.245\columnwidth]{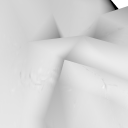} &
		\includegraphics[width=0.245\columnwidth]{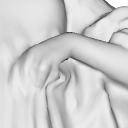} &
		\includegraphics[width=0.245\columnwidth]{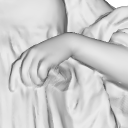} &
		\includegraphics[width=0.245\columnwidth]{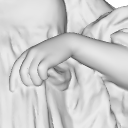} \\
		\small{Base mesh (1k tris)} & \multicolumn{2}{c}{\small{Displaced (64k tris)}} & \small{Ref (370k tris)} \\
		& \small{w/o normal map} & \small{w/ normal map} & \\
	\end{tabular}
	\vspace*{-2mm}
	\caption{
		We jointly optimize a base mesh, displacement map, and normal map to match the appearance of the 
		dancer, courtesy of the Smithsonian 3D Digitization project~\shortcite{Smithsonian2020}. 
		This is a complex optimization problem, with displacement constrained to the normal direction, 
		and a coarsely tessellated base mesh. Still, the appearance after optimization matches the reference closely. 
		Notably, some small details in the insets
		are baked into the normal map, even though they could be easily represented by displacement. We speculate 
		that view parallax is minimal from the range of camera distances used during optimization, causing displacement 
		and normal mapping to be equally viable. 
		The initial decimated mesh was generated using the mesh simplifier in Autodesk Maya 2019.
	}
	\label{fig:displacement}
\end{figure}
}


\newcommand{\figEwer}{
\begin{figure}
	\centering
	\setlength{\tabcolsep}{1pt}
	\begin{tabular}{cc}
		\includegraphics[width=0.49\columnwidth]{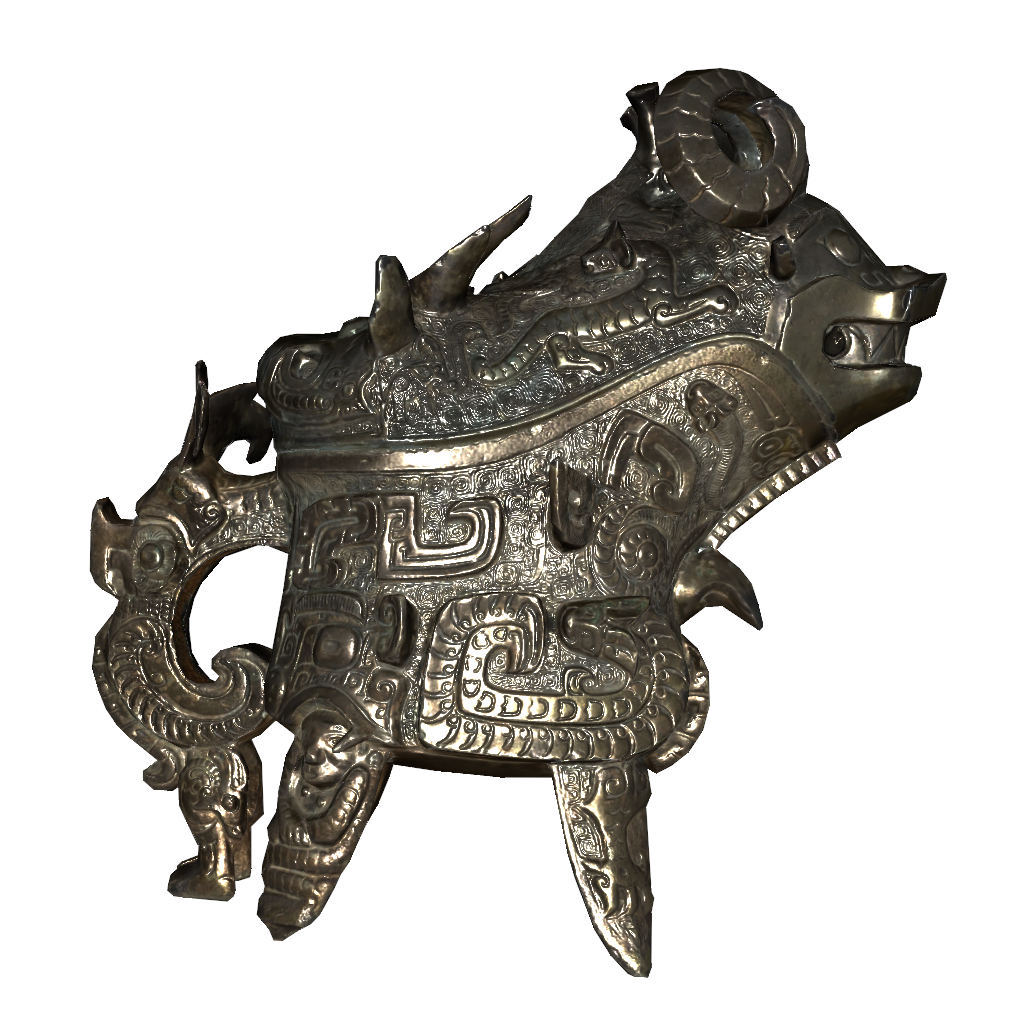} &
		\includegraphics[width=0.49\columnwidth]{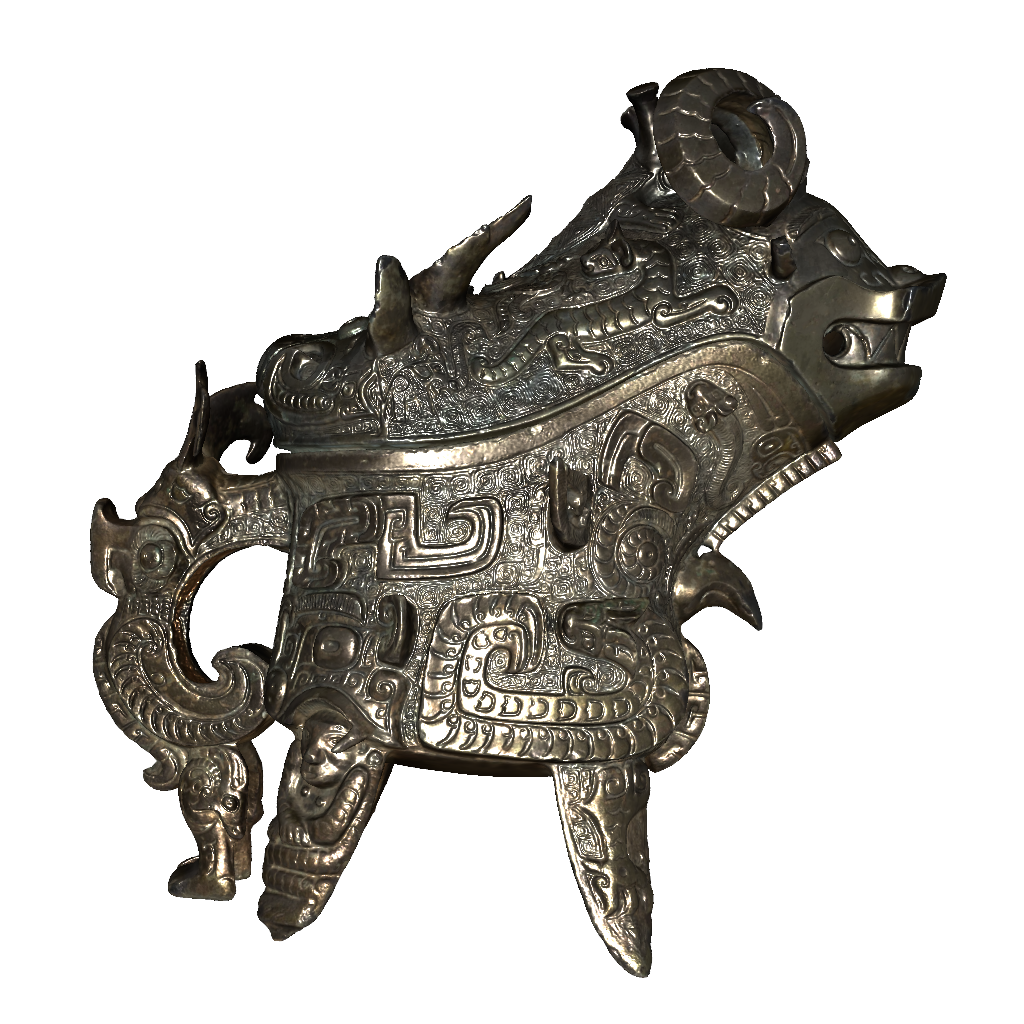} \\
		\small{Our (7k tris)} & \small{Reference (300k tris)}
	\end{tabular}
	\vspace*{-2mm}
	\caption{The Ewer bronze sculpture courtesy of the Smithsonian 3D Digitization project~\shortcite{Smithsonian2020}. 
		The reference consists of 300k triangles, normal maps, textured base color and a bronze metal material.
		We obtain a high quality result, approximating the reference with only 2.3\% of the triangles.
		The initial guess for the geometry was generated using Simplygon Free~8.3. See Figure~\ref{fig:ewerbreakdown} for a more detailed view.}
	\label{fig:ewer}
\end{figure}
}

\newcommand{\ewerfig}[1]{\includegraphics[width=0.188\linewidth]{#1}}
\newcommand{\ewerrow}[2]{\rotatebox{90}{\makebox[#1\linewidth]{\centering\small #2}}}
\newcommand{\ewercrop}[1]{\adjincludegraphics[width=0.188\linewidth,trim={0 {.05\height} 0 {.05\height}},clip]{#1}}


\newcommand{\figEwerBreakdown}{
\begin{figure*}
	\centering
	\setlength{\tabcolsep}{1pt}
	\begin{tabular}{@{}c@{\hspace*{1mm}}ccc@{\hspace*{3.5mm}}cc@{}}
		\ewerrow{0.188}{Geometry} &
		\ewerfig{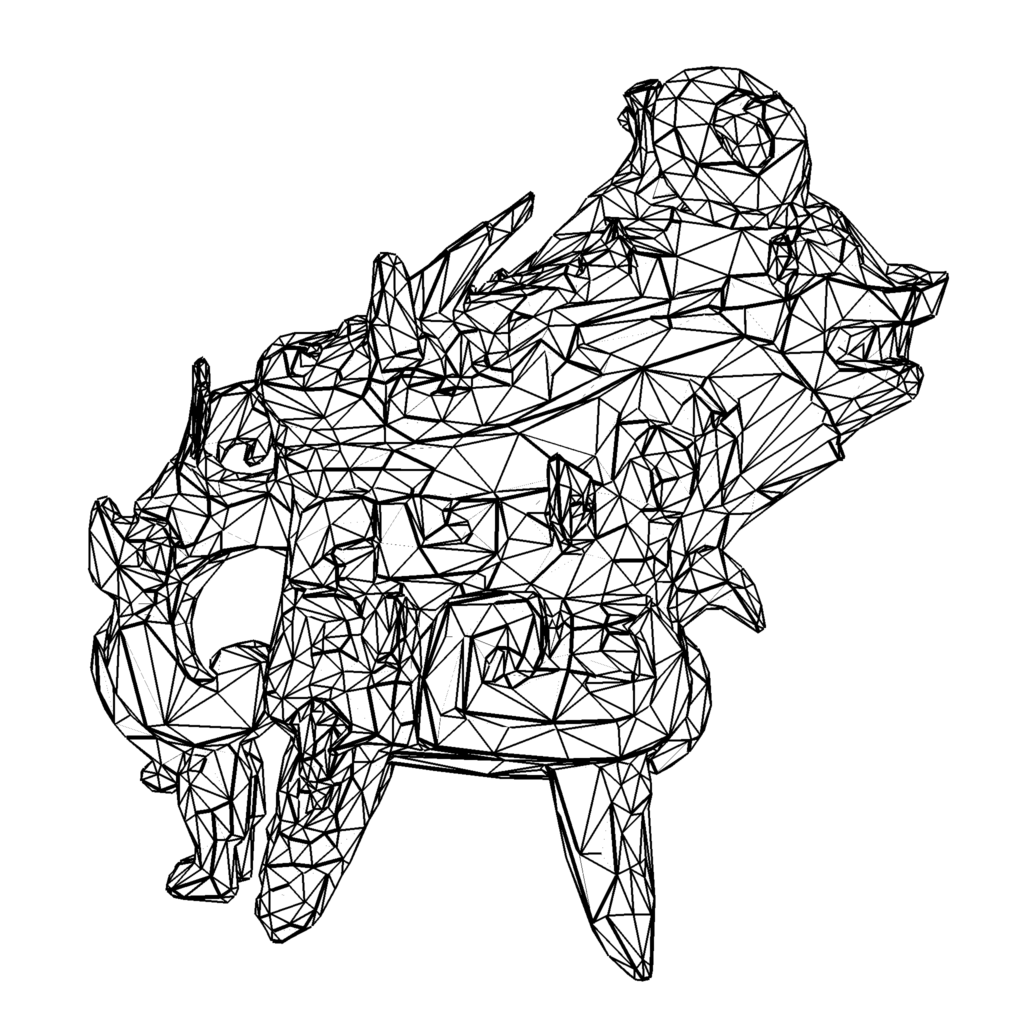} &
		\ewerfig{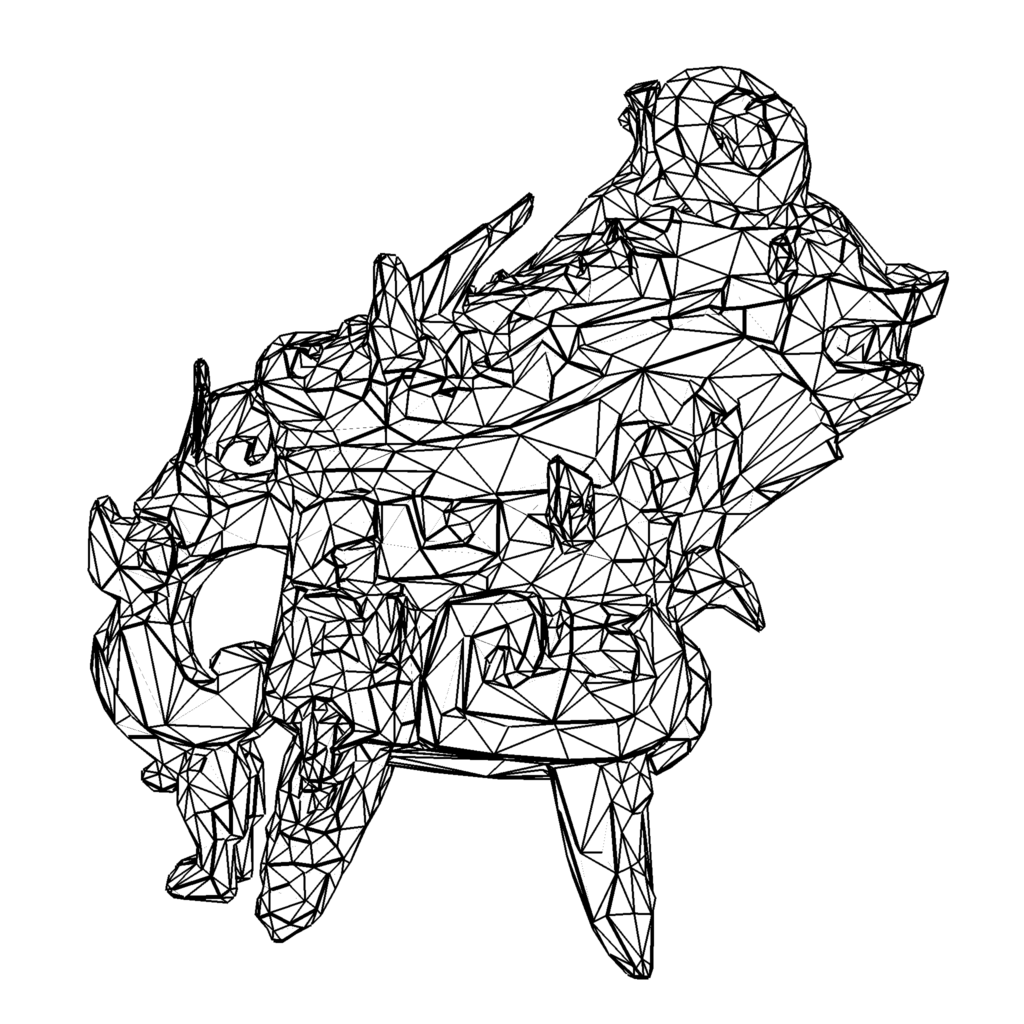} &
		\ewerfig{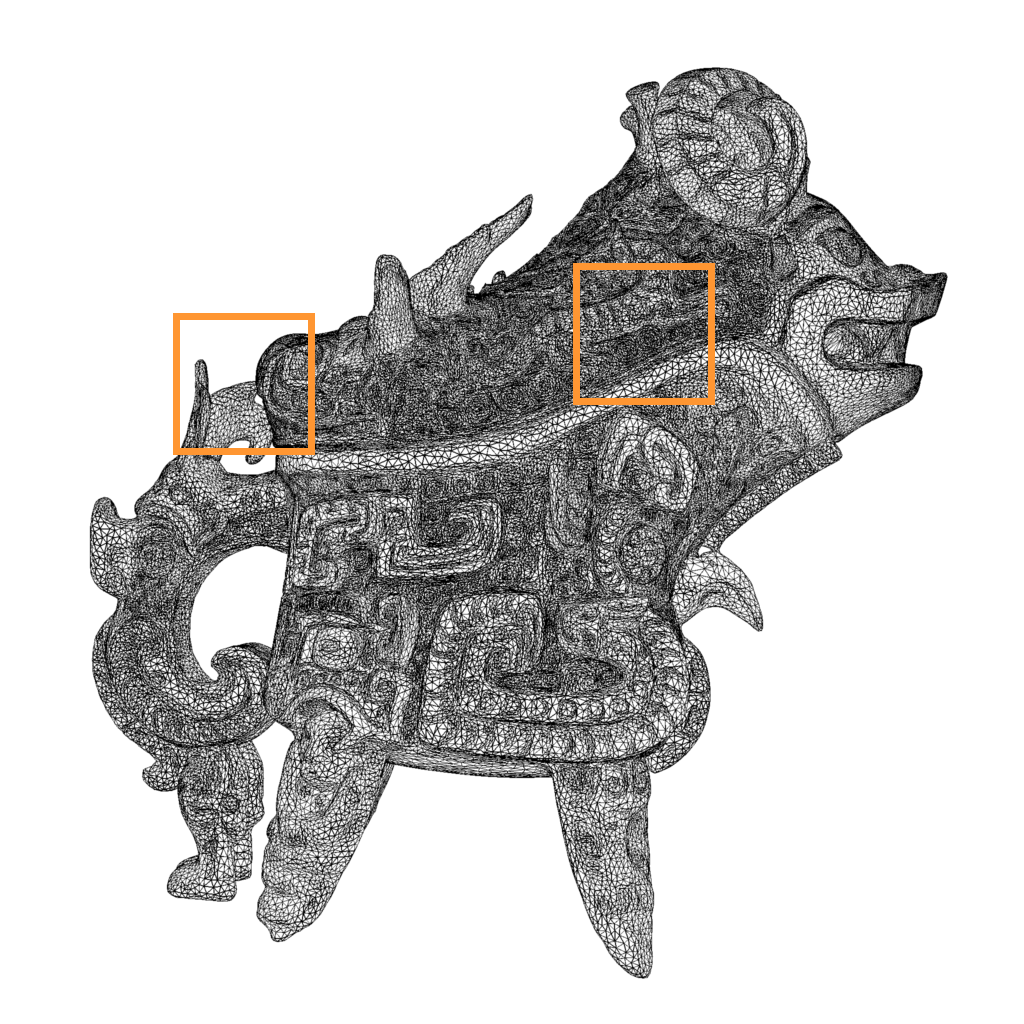} &
		\ewerfig{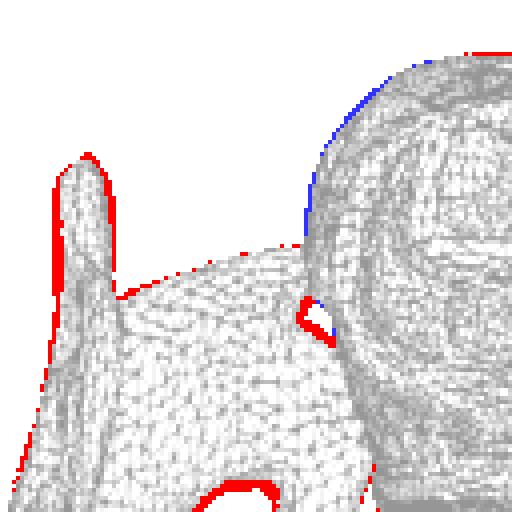} &
		\ewerfig{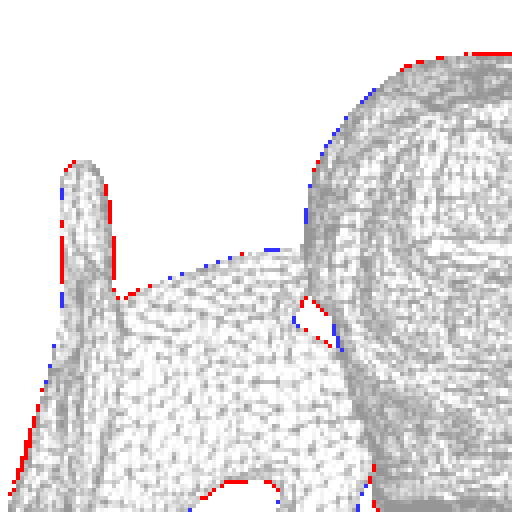} \\
		\ewerrow{0.169}{Normals} &
		\ewercrop{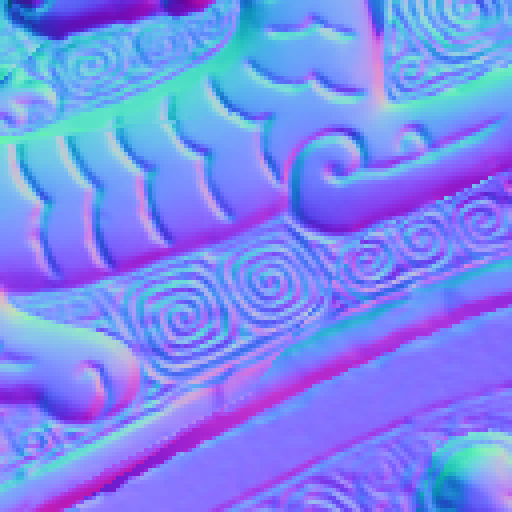} &
		\ewercrop{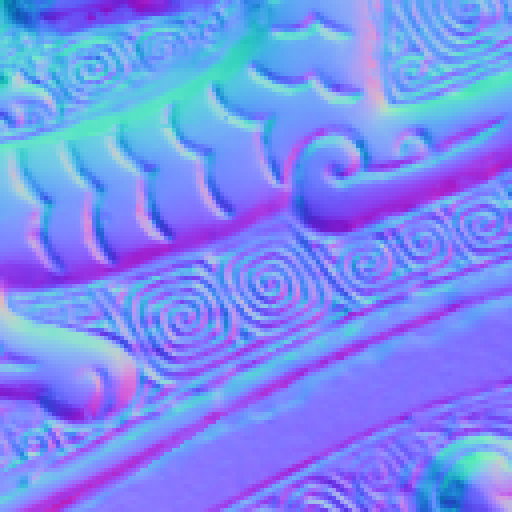} &
		\ewercrop{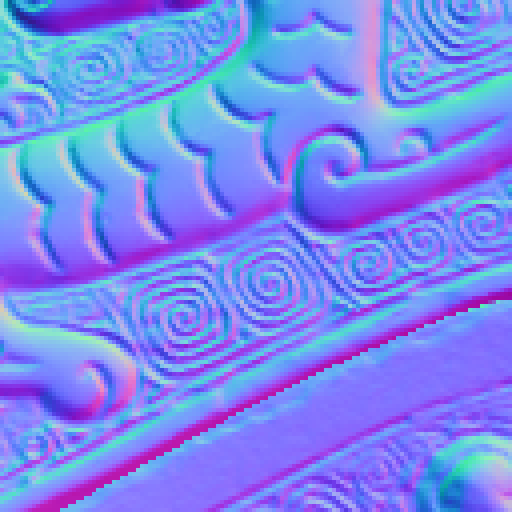} &
		\ewercrop{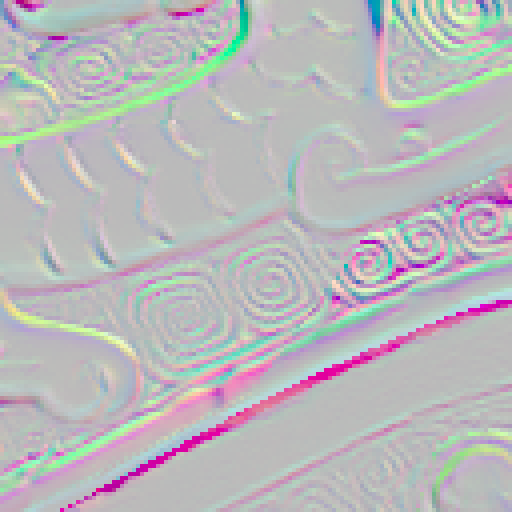} &
		\ewercrop{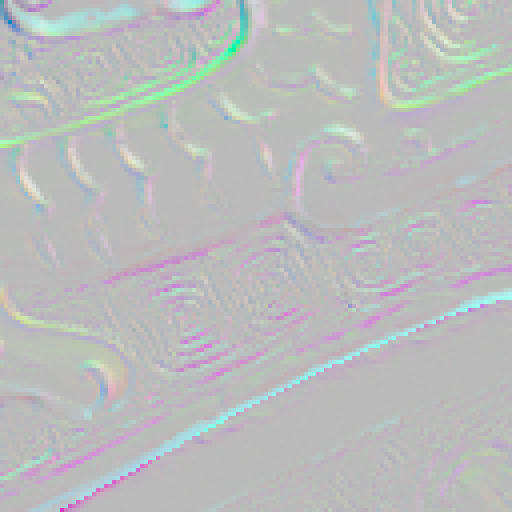} \\
		\ewerrow{0.169}{Full shading} &
		\ewercrop{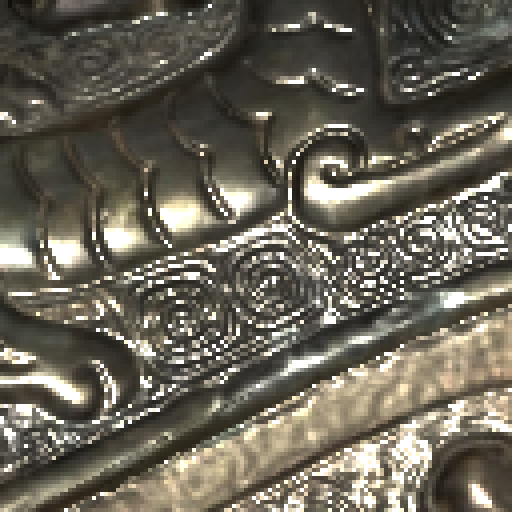} &
		\ewercrop{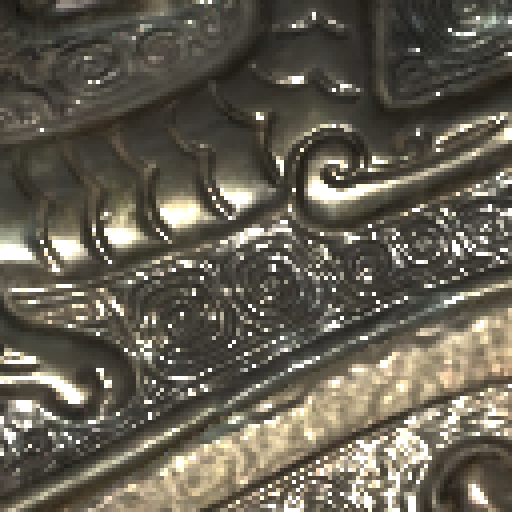} &
		\ewercrop{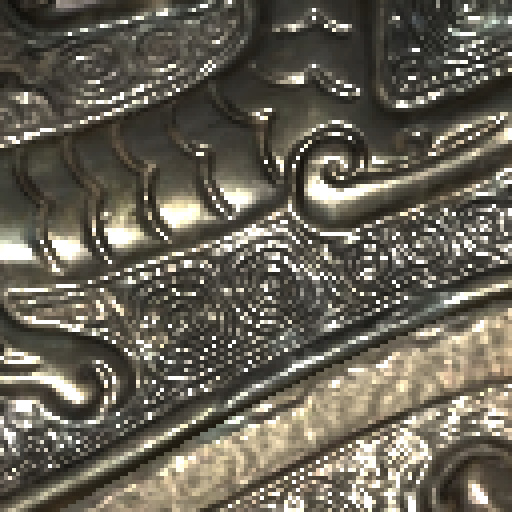} &
		\ewercrop{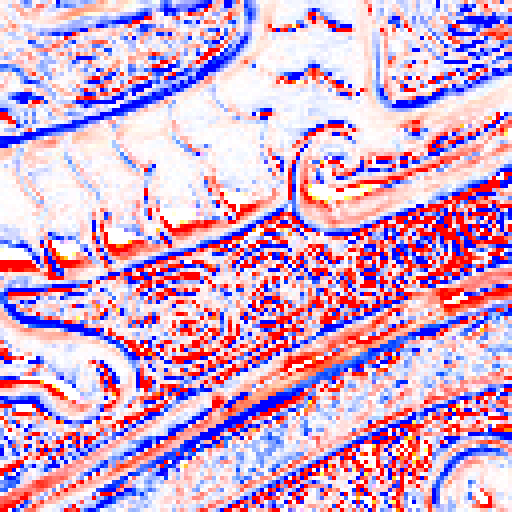} &
		\ewercrop{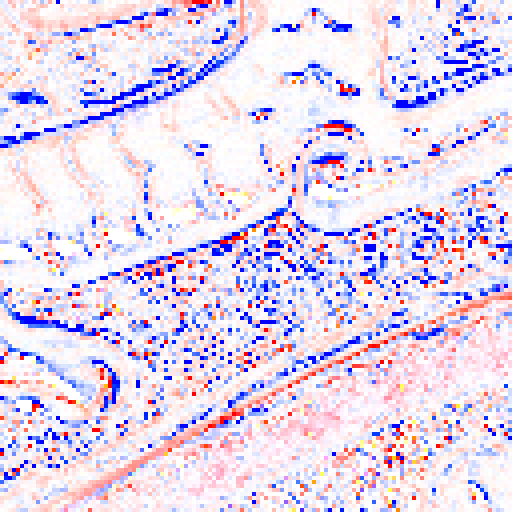}\vspace*{-.5mm}\\
		&
		\small{Simplygon (7k tris)} & 
		\small{Our (7k tris)} & 
		\small{Reference (300k tris)} &
		\small{Simplygon vs reference} & 
		\small{Our vs reference}
	\end{tabular}
	\vspace*{-2mm}
	\caption{\label{fig:ewerbreakdown}
		Analysis of the Ewer sculpture mesh.
		\textbf{Top row:} 
		Initial simplified geometry by Simplygon, our optimized geometry, and original geometry.
		Difference images on the right show the coverage errors of the simplified meshes overlaid on the reference mesh.
		Red indicates pixels covered by the simplified mesh but not the reference mesh, and blue indicates the opposite situation.
		As can be seen, optimizing the geometry improves the silhouettes considerably, although its effects are otherwise subtle.
		\textbf{Middle row:}
		Closeups of normals generated by Simplygon, our optimized normals, and reference normals, followed by difference images.
		Note that our normals are not optimized directly against reference normals but discovered via optimizing the full shading result against the reference.
		\textbf{Bottom row:}
		Full shading results rendered at 1~spp with difference images (red/blue = too bright/dim compared to reference).
		Our result exhibits fewer overly bright pixels because the rendering is optimized to match the reference on average, thereby reducing aliasing-induced sparkling.
		Some of the improvement over Simplygon's output may be attributed to the use of more versatile shading model, but that does not explain the improvement in, e.g., normals or silhouettes.
	}
	\vspace*{-2mm}
\end{figure*}
}


\newcommand{\gruxbox}[5]
{{\begin{tikzpicture}
		\node[anchor=south west,inner sep=0] (image) at (0,0) {\includegraphics[width=0.32\columnwidth]{#1}};
		\begin{scope}[x={(image.south east)},y={(image.north west)}]
		\draw[orange,thick] (#2,#3) rectangle (#4,#5);
		\end{scope}
		\end{tikzpicture}}}

\newcommand{\figGrux}{
\begin{figure}
	\centering
	\setlength{\tabcolsep}{1pt}
	\begin{tabular}{ccc}    
		\includegraphics[width=0.33\columnwidth]{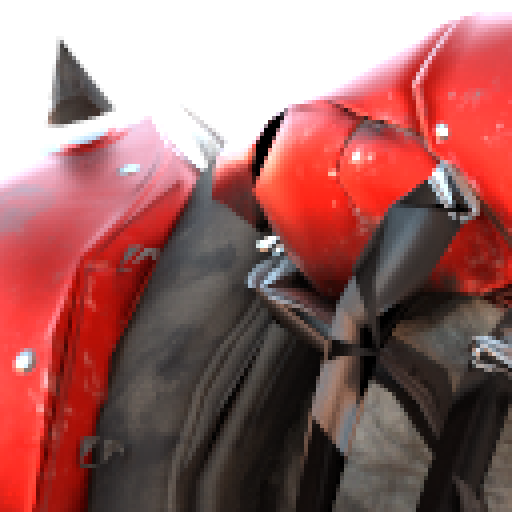} &
		\includegraphics[width=0.33\columnwidth]{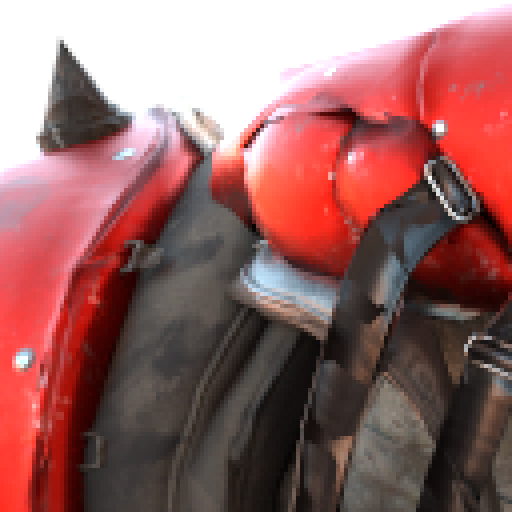} &
		\includegraphics[width=0.33\columnwidth]{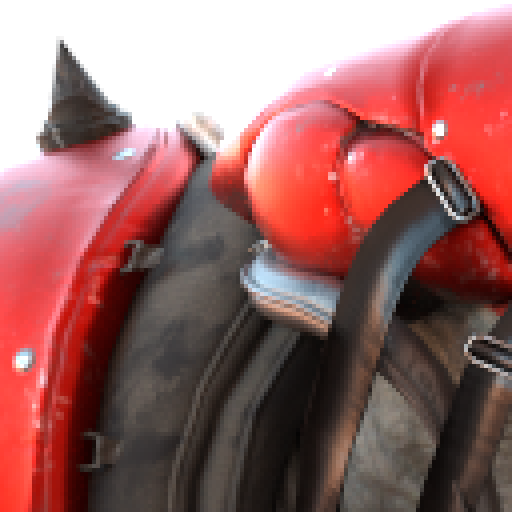} \\
		\includegraphics[width=0.33\columnwidth]{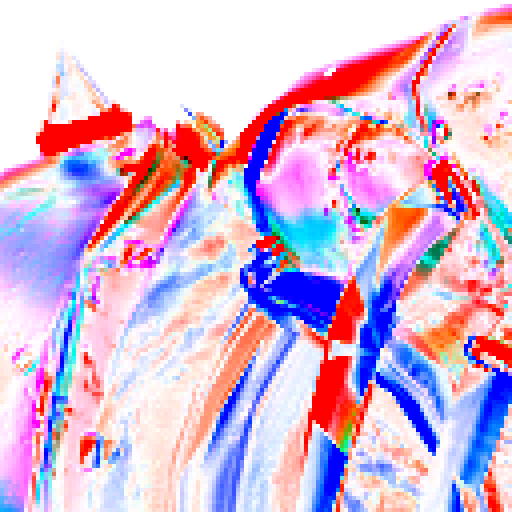} &
		\includegraphics[width=0.33\columnwidth]{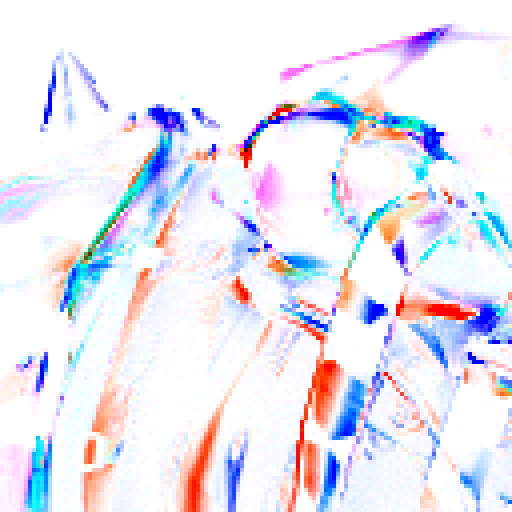} &
		\gruxbox{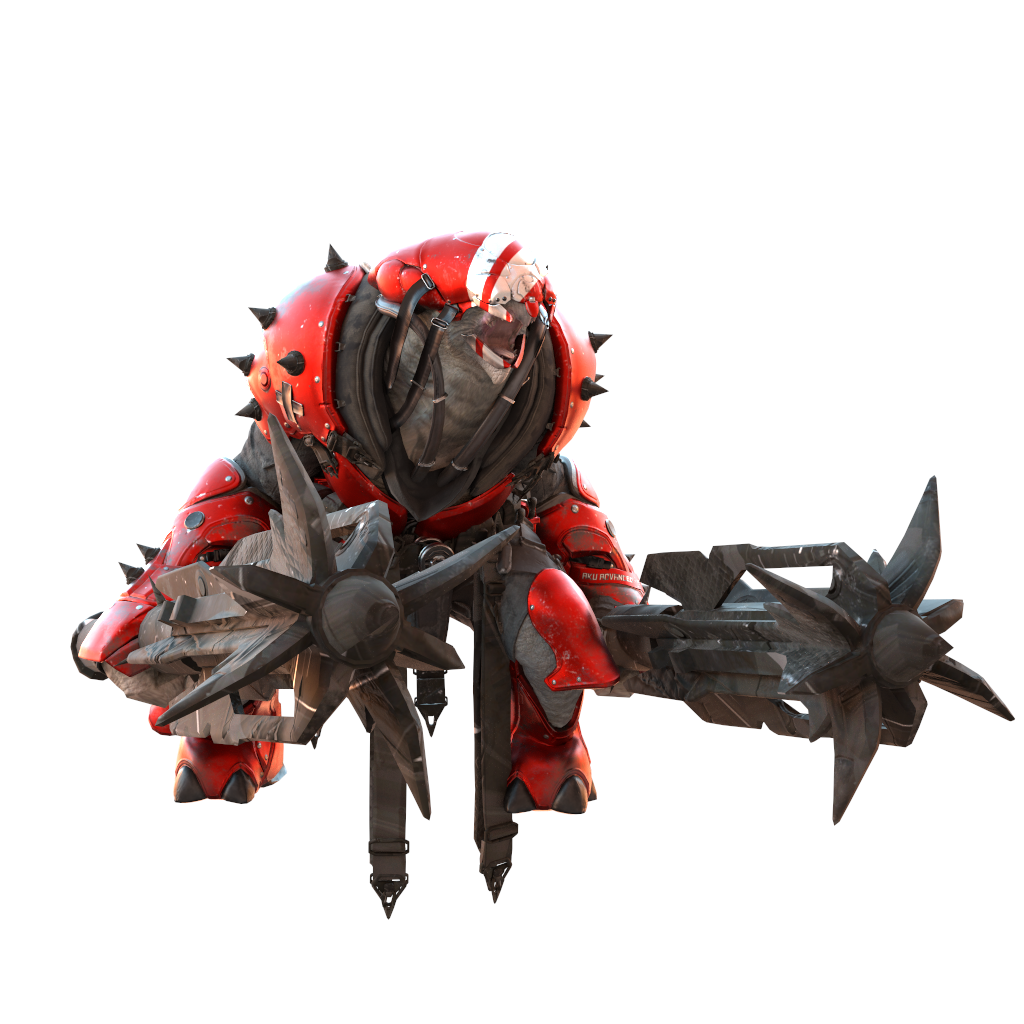}{0.31}{0.65}{0.44}{0.78} \\
		\small{Reduced (15k tris)} & \small{Our (15k tris)} & \small{Ref (81k tris)}
	\end{tabular}
	\vspace*{-2mm}
	\caption{
		A character from the Unreal Engine Paragon asset~\shortcite{Paragon2018}. We clear up some of the artifacts
		introduced by the automatically reduced mesh (generated in Autodesk Maya 2019), reattach geometry, and repair 
		texture. The bottom row shows difference images (red/blue = too bright/dim compared to reference). Please refer 
		to the supplemental material for full images.}
	\label{fig:grux}
	\vspace*{-2mm}
\end{figure}
}


\newcommand{\figPrefilter}{
\begin{figure*}[h]
	\setlength{\tabcolsep}{1pt}
	\begin{tabular}{ccccccccc}
		\rotatebox[origin=c]{90}{\small{$64\times64$ pixels}} &
		\raisebox{-0.5\height}{\includegraphics[width=0.12\textwidth]{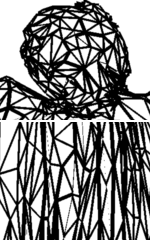}} &
		\raisebox{-0.5\height}{\includegraphics[width=0.12\textwidth]{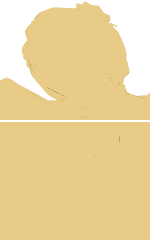}} &
		\raisebox{-0.5\height}{\includegraphics[width=0.12\textwidth]{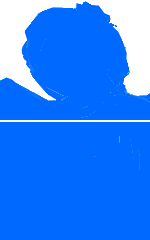}} &
		\raisebox{-0.5\height}{\includegraphics[width=0.12\textwidth]{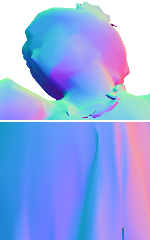}} &
		\raisebox{-0.5\height}{\includegraphics[width=0.12\textwidth]{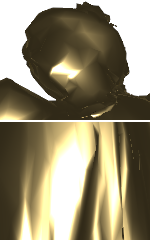}} &
		\raisebox{-0.5\height}{\includegraphics[width=0.12\textwidth]{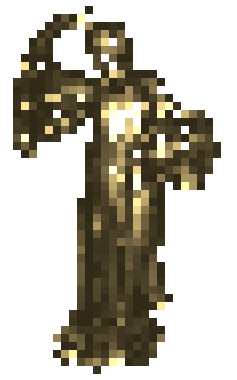}} &
		\raisebox{-0.5\height}{\includegraphics[width=0.12\textwidth]{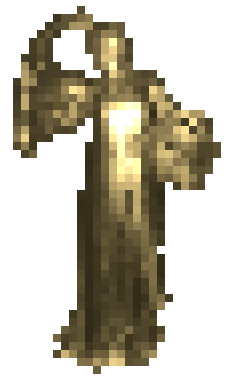}} &
		\raisebox{-0.5\height}{\includegraphics[width=0.12\textwidth]{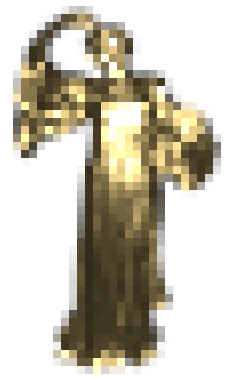}} \\
		\rotatebox[origin=c]{90}{\small{$512\times512$ pixels}} &
		\raisebox{-0.5\height}{\includegraphics[width=0.12\textwidth]{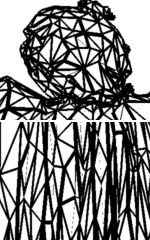}} &
		\raisebox{-0.5\height}{\includegraphics[width=0.12\textwidth]{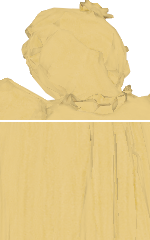}} &
		\raisebox{-0.5\height}{\includegraphics[width=0.12\textwidth]{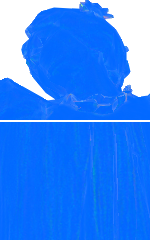}} &
		\raisebox{-0.5\height}{\includegraphics[width=0.12\textwidth]{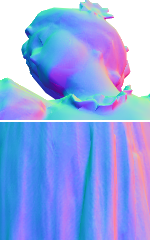}} &
		\raisebox{-0.5\height}{\includegraphics[width=0.12\textwidth]{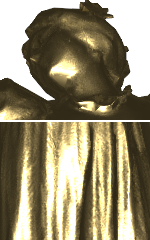}} &
		\raisebox{-0.5\height}{\includegraphics[width=0.12\textwidth]{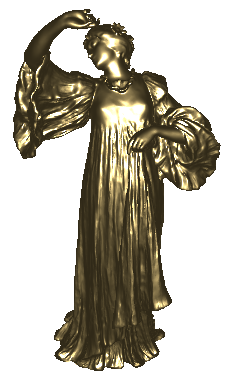}} &
		\raisebox{-0.5\height}{\includegraphics[width=0.12\textwidth]{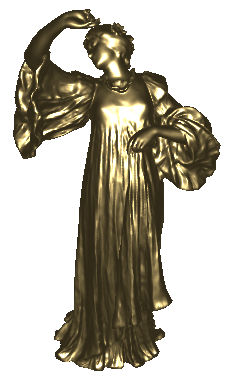}} &
		\raisebox{-0.5\height}{\includegraphics[width=0.12\textwidth]{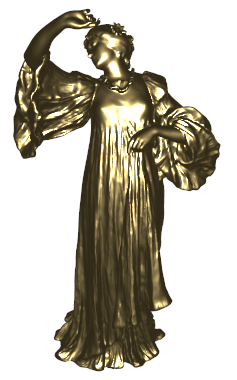}} \\
		& \small{Mesh} & \small{Diffuse map} & \small{Specular Map} & \small{Normals} & \small{Shaded} & \small{Ref: 1~spp} & \small{Our: 1~spp} & \small{Ref: 256~spp} \\
		& & & & & & \small{(370k tris)} & \small{(8k tris)} & \small{(370k tris)} \\
	\end{tabular}
	\vspace*{-2mm}
	\caption{We can jointly prefilter shape and appearance for a certain rendering resolution. The dancer in the \textbf{top row} is optimized for a resolution of $64\times64$ pixels,
		and for  $512\times512$ pixels in the \textbf{bottom row}. Note in the top row how the normals are smoothed to band-limit shading. The three rightmost images show the 
		models rendered at the intended resolution. We match the appearance of the super-sampled reference well with a single sample per pixel.}
	\label{fig:pre_filter}
\end{figure*}
}


\newcommand{\tyronebox}[5]
{{\begin{tikzpicture}
		\node[anchor=south west,inner sep=0] (image) at (0,0) {\includegraphics[width=0.32\columnwidth]{#1}};
		\begin{scope}[x={(image.south east)},y={(image.north west)}]
		\draw[orange,thick] (#2,#3) rectangle (#4,#5);
		\end{scope}
		\end{tikzpicture}}}

\newcommand{\figAnimation}{
\begin{figure}
	\centering
	\setlength{\tabcolsep}{1pt}
	\begin{tabular}{cccc}    
		\rotatebox[origin=c]{90}{\small{Our (5k tris)}} & 
		\raisebox{-0.5\height}{\tyronebox{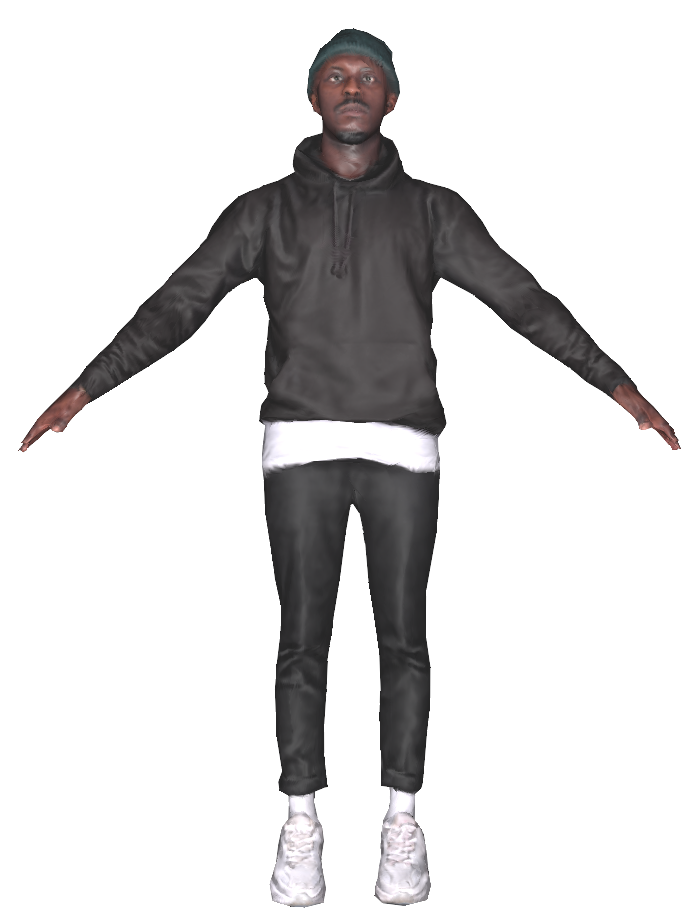}{0.15}{0.71}{0.32}{0.59}} &
		\raisebox{-0.5\height}{\tyronebox{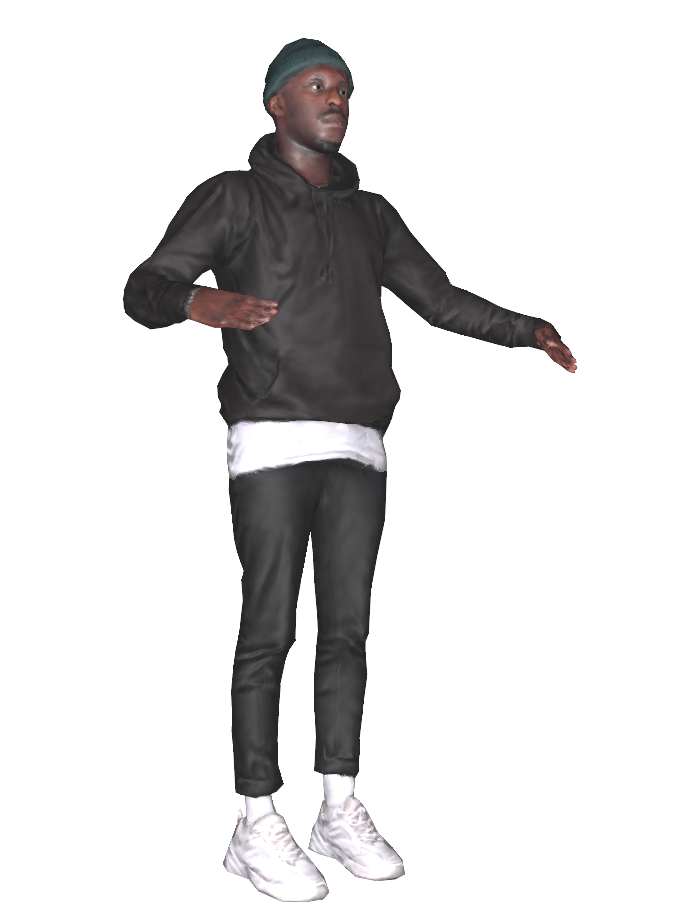}{0.15}{0.75}{0.33}{0.63}} &
		\raisebox{-0.5\height}{\tyronebox{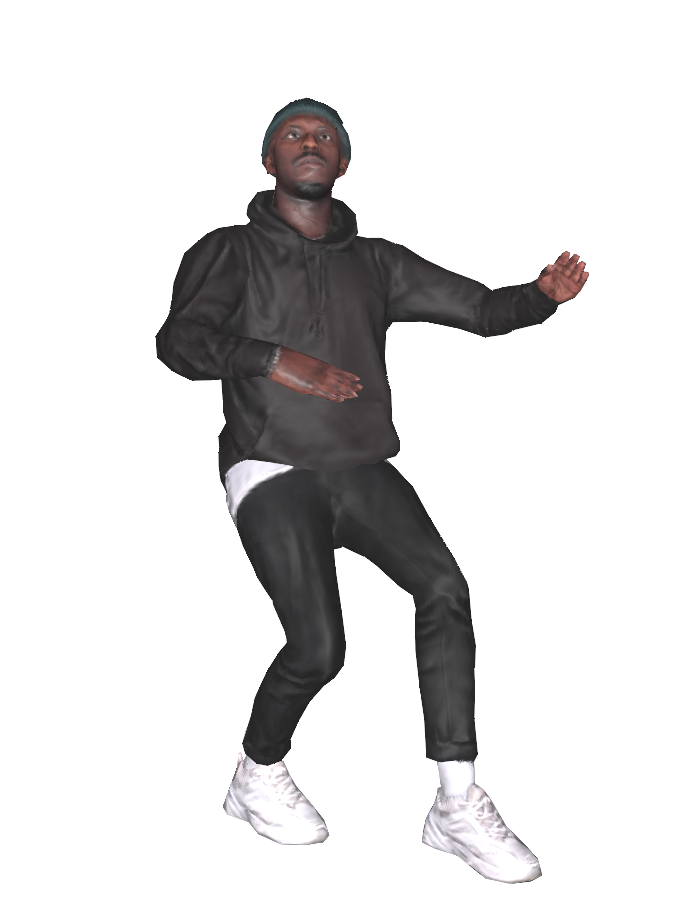}{0.20}{0.7}{0.37}{0.57}} \\
		& \small{T-Pose} & \small{Frame 0} & \small{Frame 39}
	\end{tabular}
	\begin{tabular}{ccccccc}
		\rotatebox[origin=c]{90}{\small{Our (5k tris)}} &     
		\raisebox{-0.5\height}{\includegraphics[width=0.16\columnwidth]{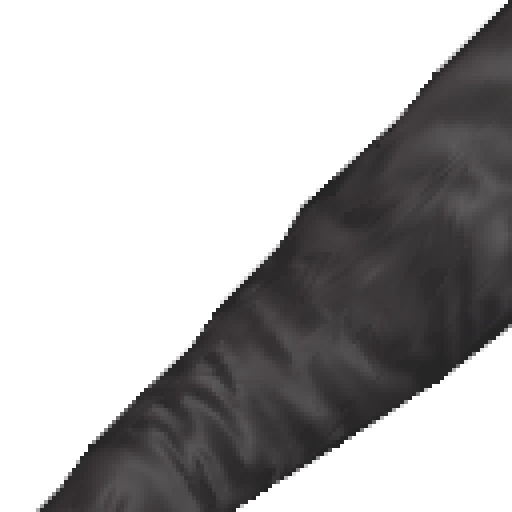}} &
		\raisebox{-0.5\height}{\includegraphics[width=0.16\columnwidth]{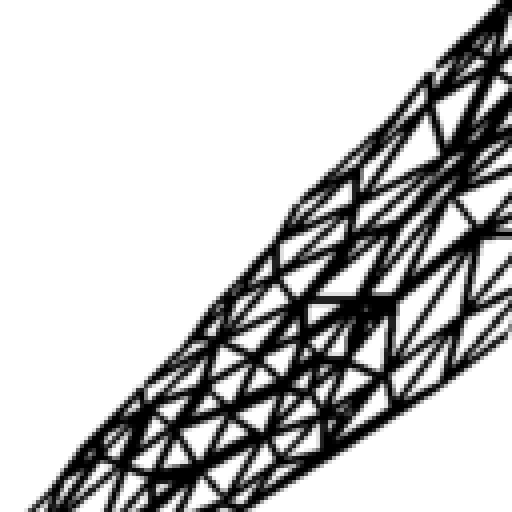}} &
		\raisebox{-0.5\height}{\includegraphics[width=0.16\columnwidth]{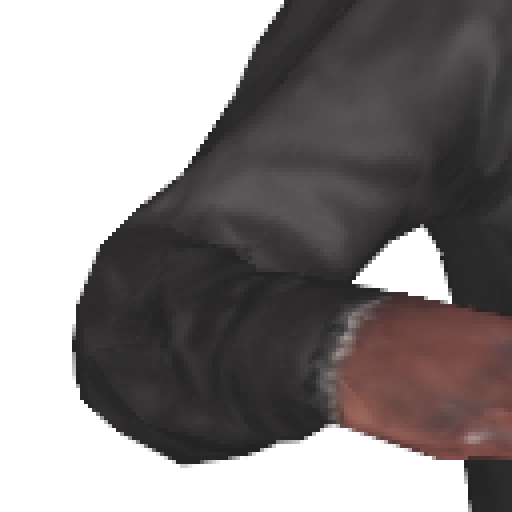}} &
		\raisebox{-0.5\height}{\includegraphics[width=0.16\columnwidth]{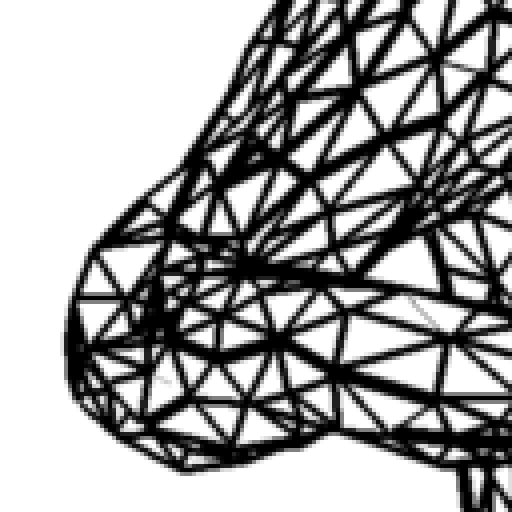}} &
		\raisebox{-0.5\height}{\includegraphics[width=0.16\columnwidth]{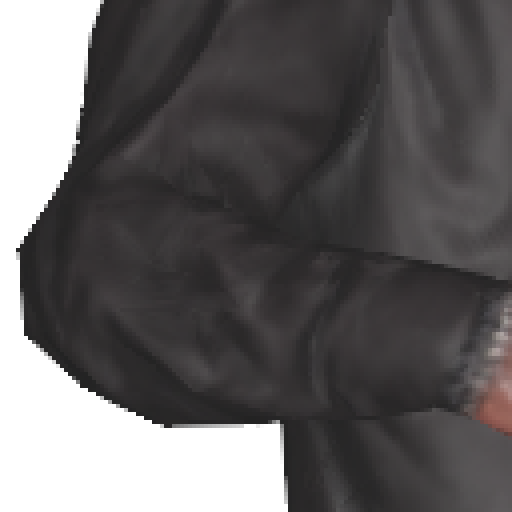}} &
		\raisebox{-0.5\height}{\includegraphics[width=0.16\columnwidth]{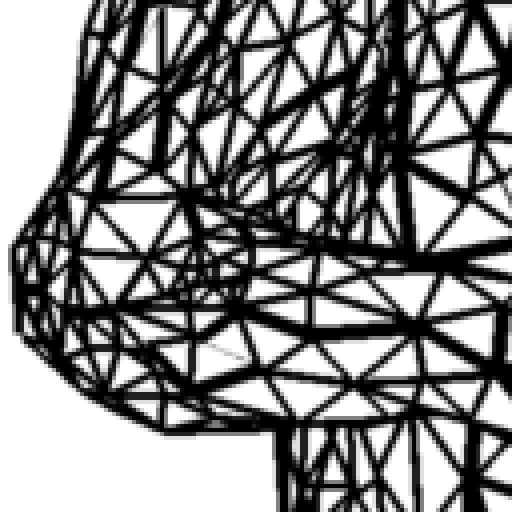}} \\
		\rotatebox[origin=c]{90}{\small{Ref (19k tris)}} &     
		\raisebox{-0.5\height}{\includegraphics[width=0.16\columnwidth]{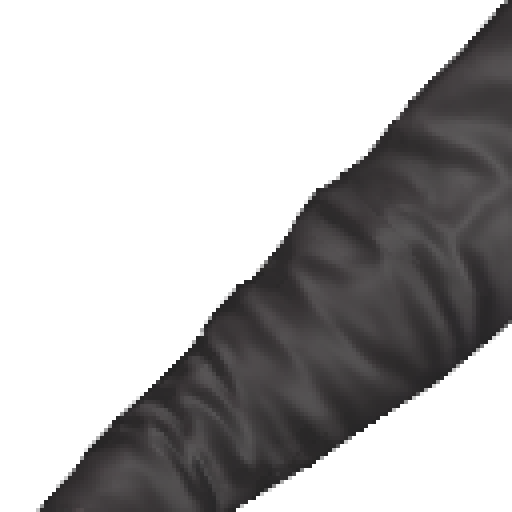}} &
		\raisebox{-0.5\height}{\includegraphics[width=0.16\columnwidth]{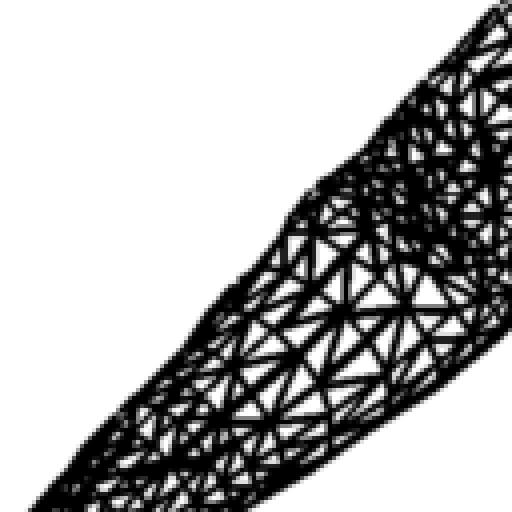}} &
		\raisebox{-0.5\height}{\includegraphics[width=0.16\columnwidth]{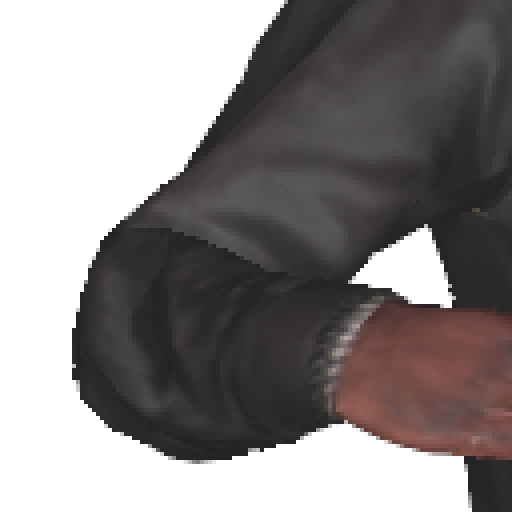}} &
		\raisebox{-0.5\height}{\includegraphics[width=0.16\columnwidth]{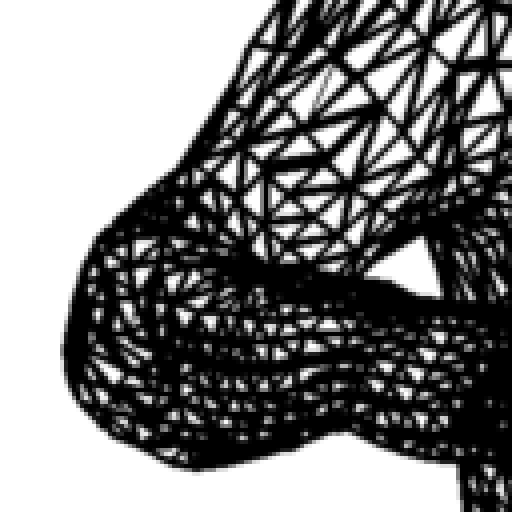}} &
		\raisebox{-0.5\height}{\includegraphics[width=0.16\columnwidth]{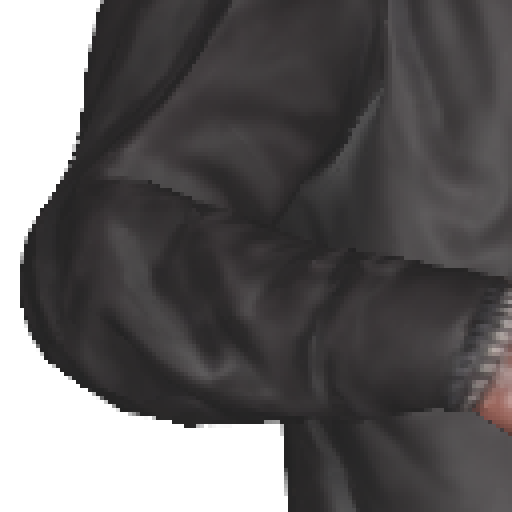}} &
		\raisebox{-0.5\height}{\includegraphics[width=0.16\columnwidth]{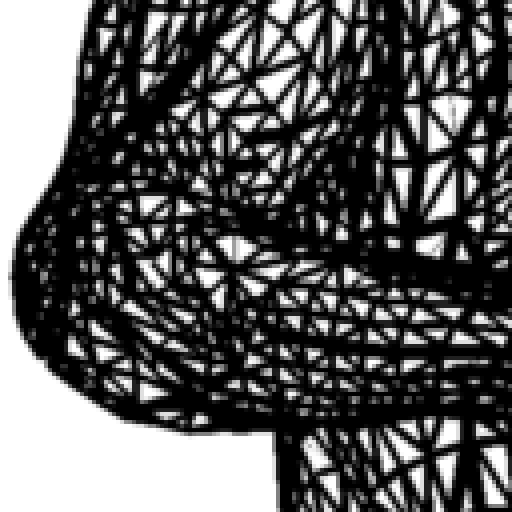}}
	\end{tabular}
	\vspace*{-2mm}
	\caption{Mesh decimation applied to an animated, rigged character from RenderPeople~\shortcite{RenderPeople2020}. 
		We reduce the mesh and optimize vertex positions, skinning weights, normal maps, and material parameters over the animation.
	}
	\label{fig:tyrone}
\end{figure}
}


\newcommand{\figAggregate}{
\begin{figure*}
	\centering
	\setlength{\tabcolsep}{1pt}
	\begin{tabular}{ccccc}
		\includegraphics[width=0.19\textwidth]{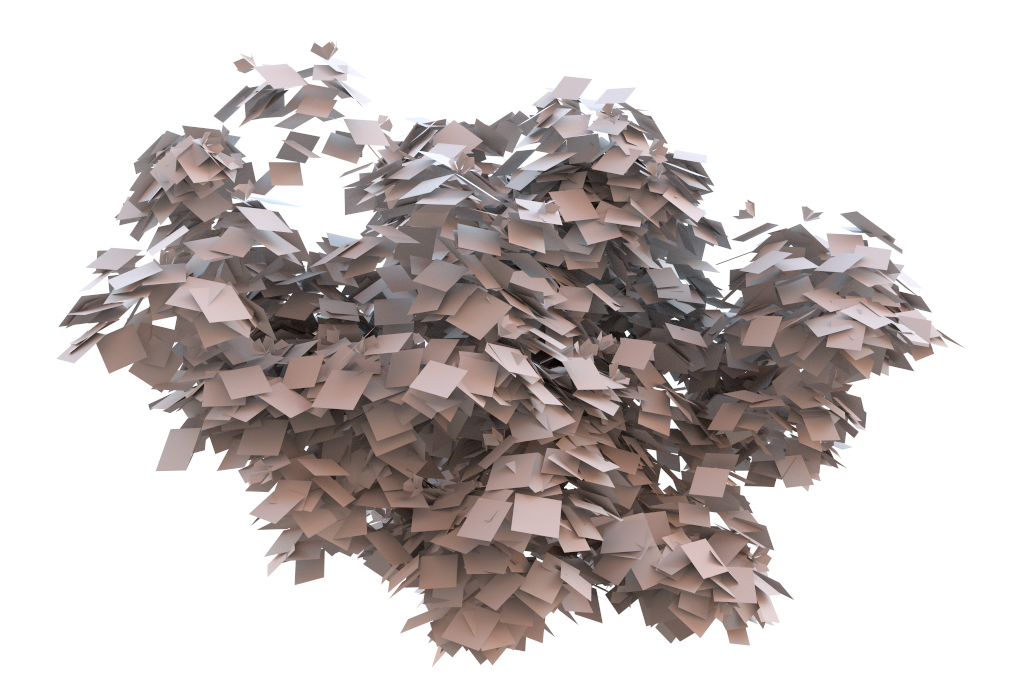} &
		\includegraphics[width=0.19\textwidth]{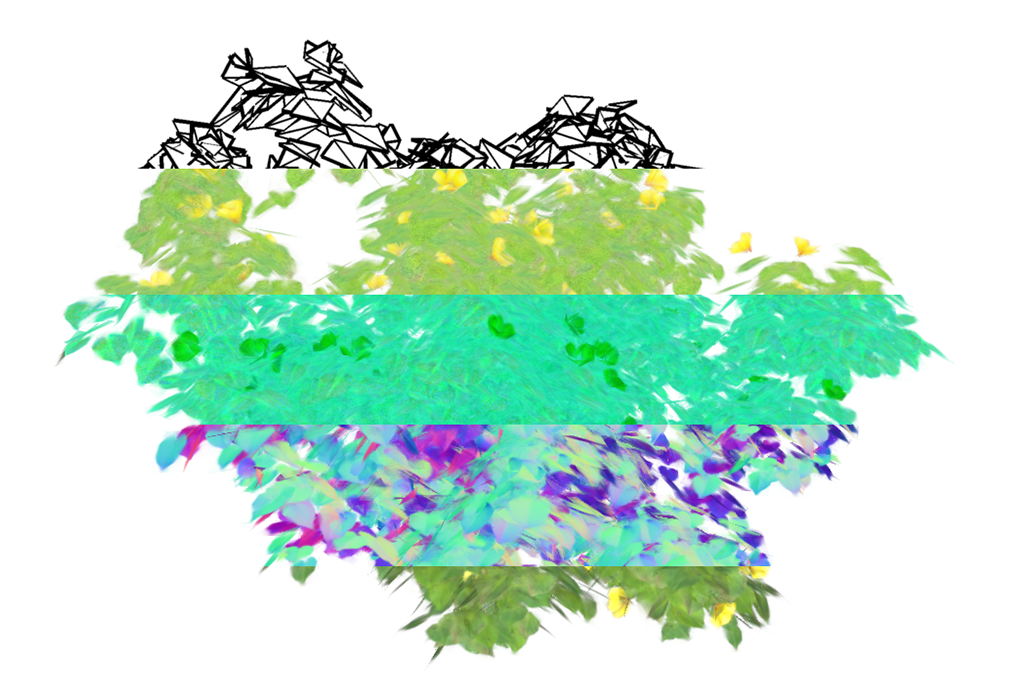} &
		\includegraphics[width=0.19\textwidth]{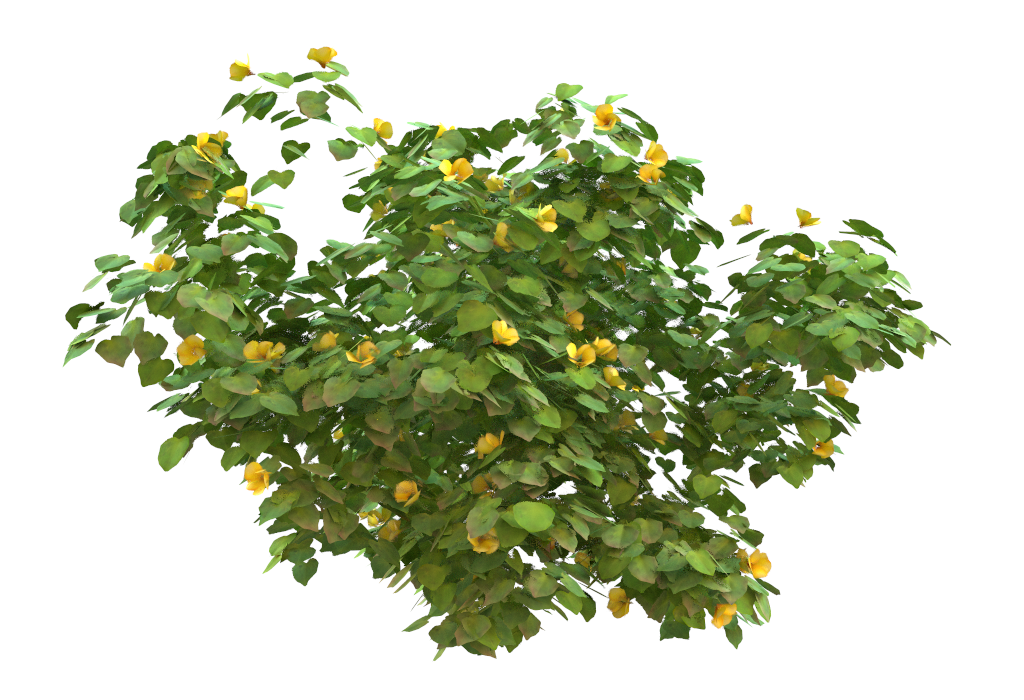} &
		\includegraphics[width=0.19\textwidth]{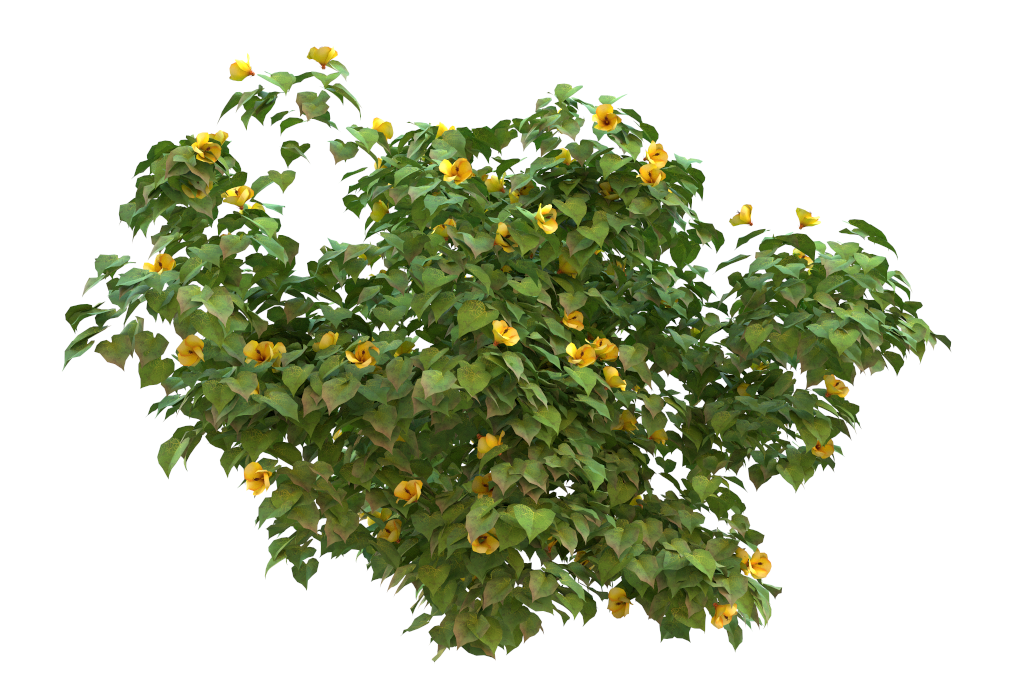} &
		\includegraphics[width=0.19\textwidth]{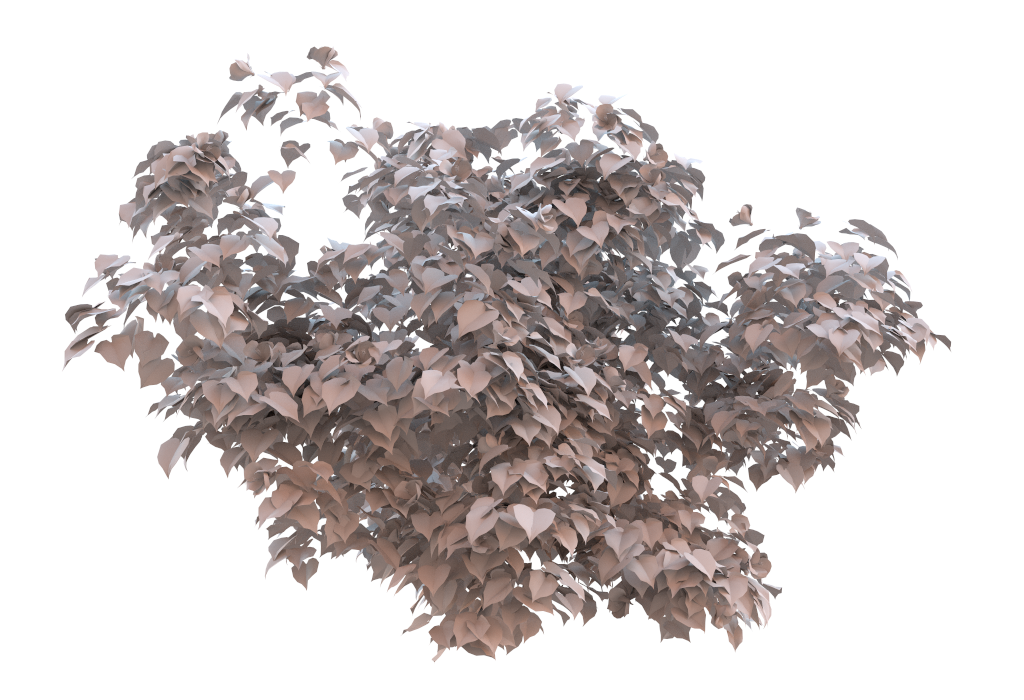} \\
		Initial guess (11k tris) & Optimized parameters & Our (11k tris) & Reference (1.2M tris) & Reference (1.2M tris) \\ 
		\includegraphics[width=0.19\textwidth]{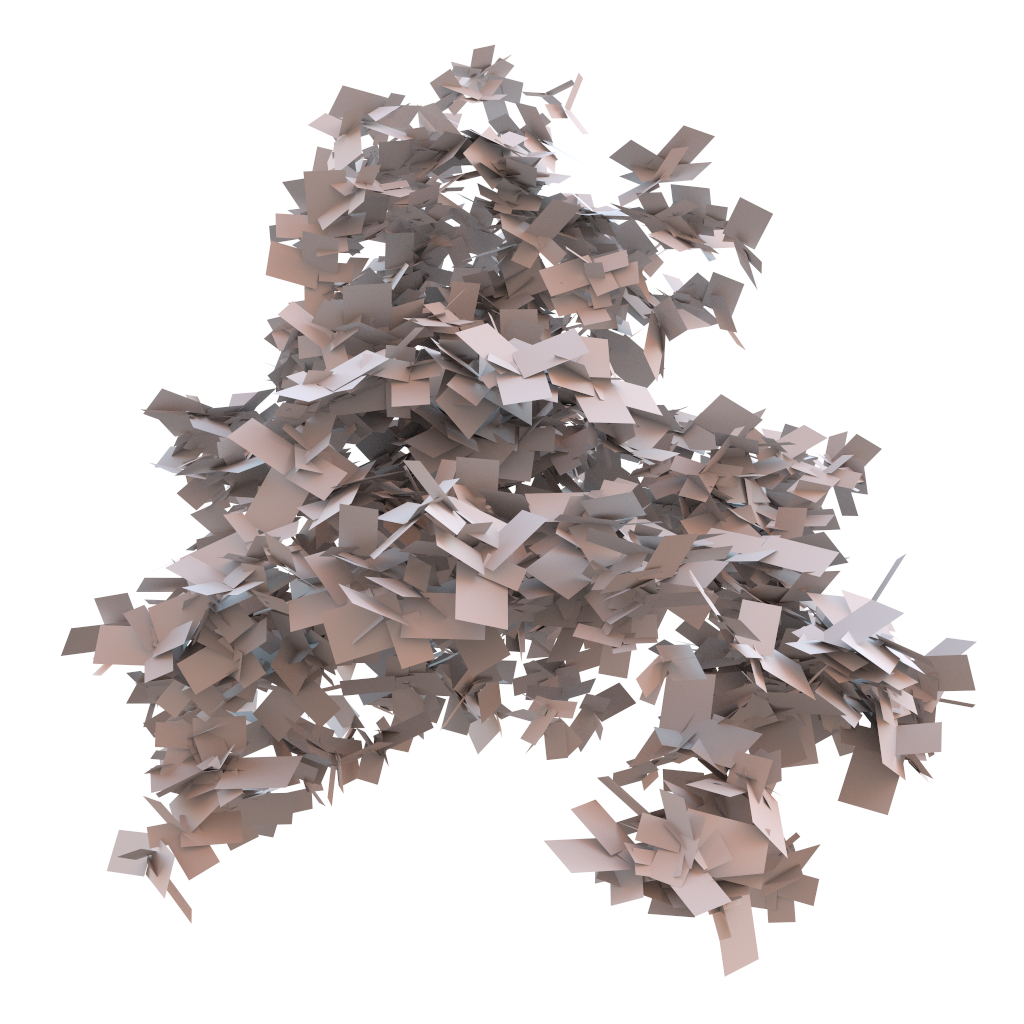} &
		\includegraphics[width=0.19\textwidth]{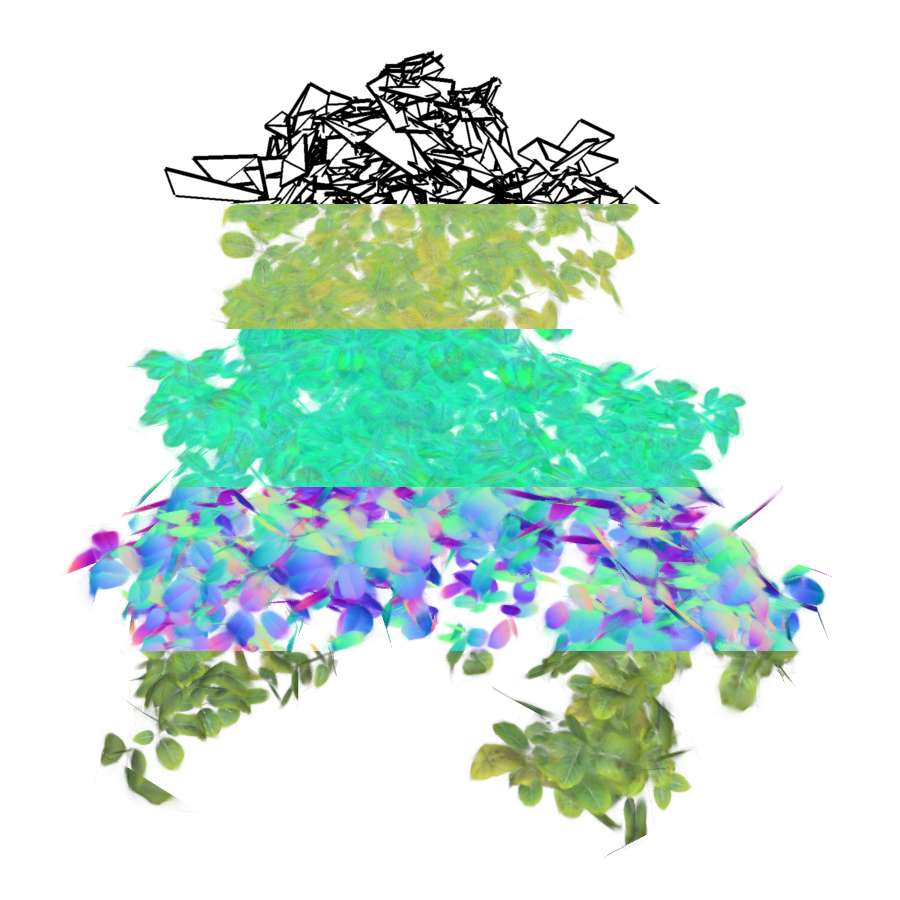} &  
		\includegraphics[width=0.19\textwidth]{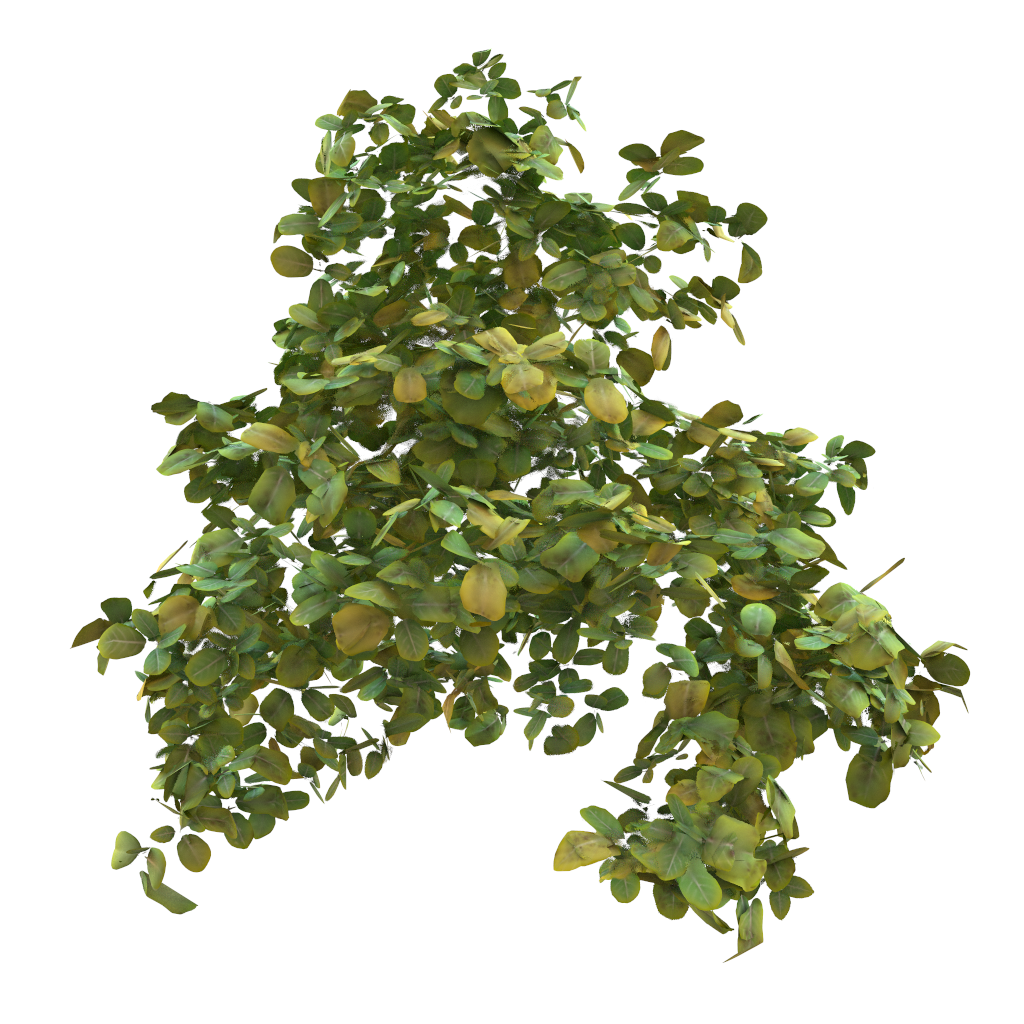} &
		\includegraphics[width=0.19\textwidth]{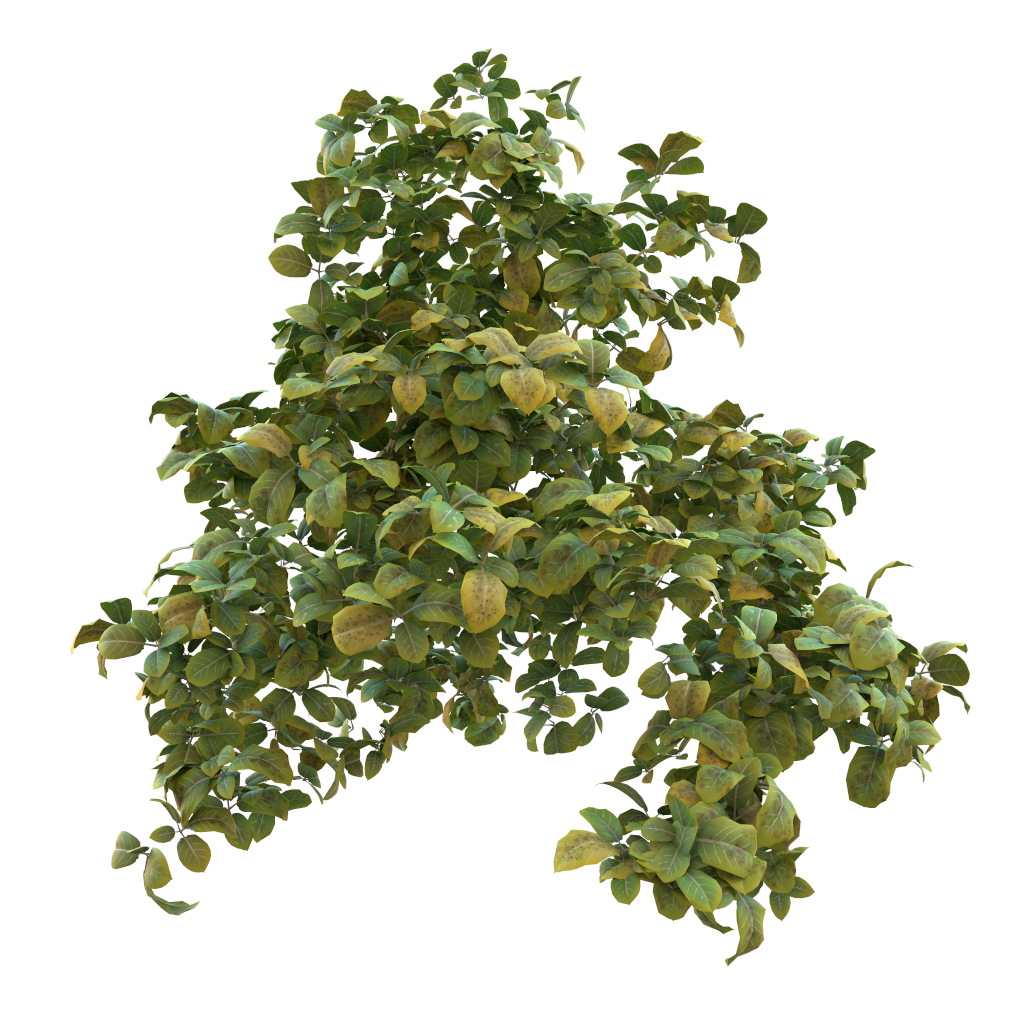} &
		\includegraphics[width=0.19\textwidth]{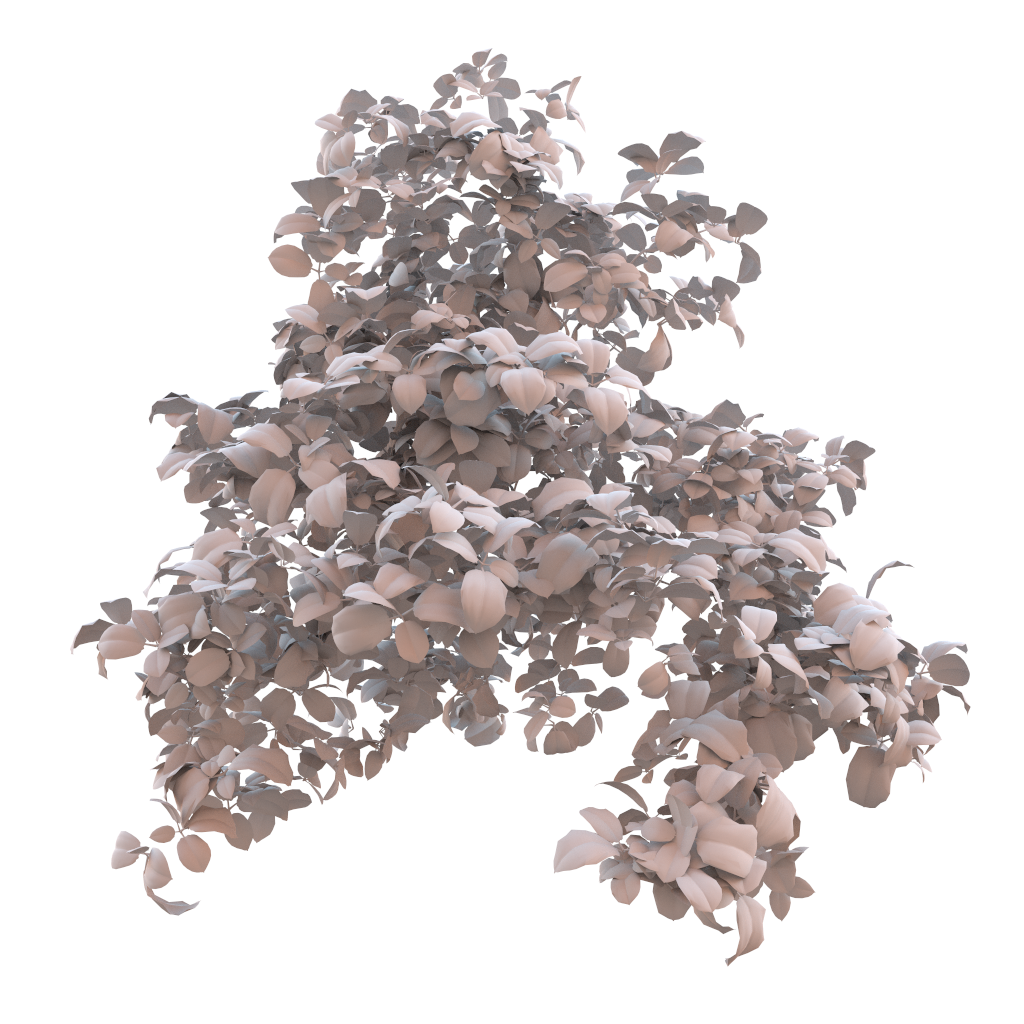} \\
		Initial guess (6.5k tris) & Optimized parameters & Our (6.5k tris) & Reference (1.7M tris) & Reference (1.7M tris) 
	\end{tabular}
	\vspace*{-2mm}
	\caption{Approximating aggregate geometry.
		We start from a low-polygon mesh and jointly optimize shape, material parameters, and transparency. The shaded
		results are rendered in a path tracer to illustrate that our results generalize across renderers.
		\textbf{Top row:} The leaves and flowers of the ``isHibiscus`` asset (1.2M triangles), approximated by 11k tris.
		\textbf{Bottom row:} The leaves from the ``isGardenia`` asset (1.7M triangles), approximated by 6.5k triangles. 
		The models are taken from the Moana Island Scene~\shortcite{Moana2018}, a publicly available data set courtesy 
		of Walt Disney Animation Studios.}
	\label{fig:aggregate}
\end{figure*}
}


\newcommand{\figSDF}{
\begin{figure}
	\centering
	\setlength{\tabcolsep}{1pt}
	\begin{tabular}{ccc}
		\includegraphics[width=0.32\columnwidth]{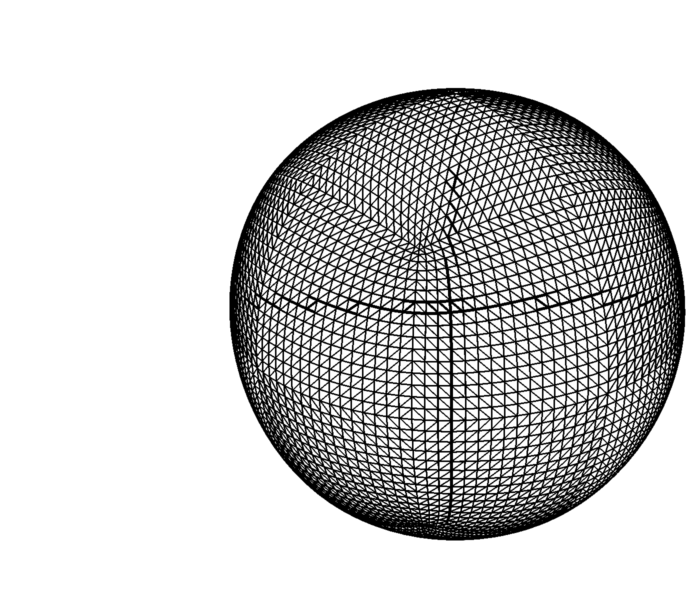} &
		\includegraphics[width=0.32\columnwidth]{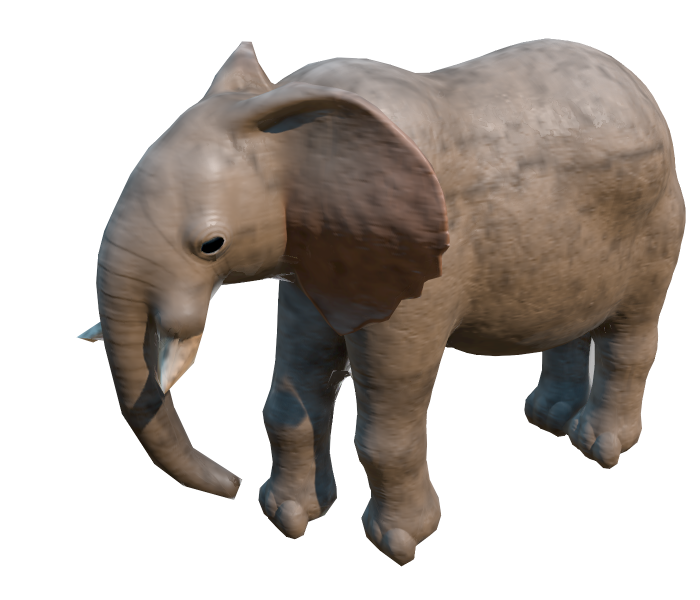} &
		\includegraphics[width=0.32\columnwidth]{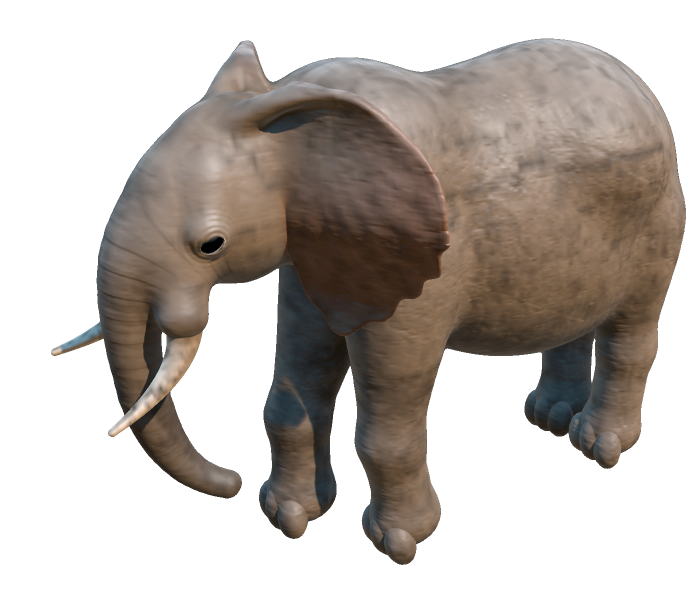} \\
		{\small Initial guess} &{\small Our rasterized model} & {\small Ray marched} \\
		{\small 12k triangles}  &{\small PSNR: 26.51~dB} & {\small implicit surface}\\
		\includegraphics[width=0.32\columnwidth]{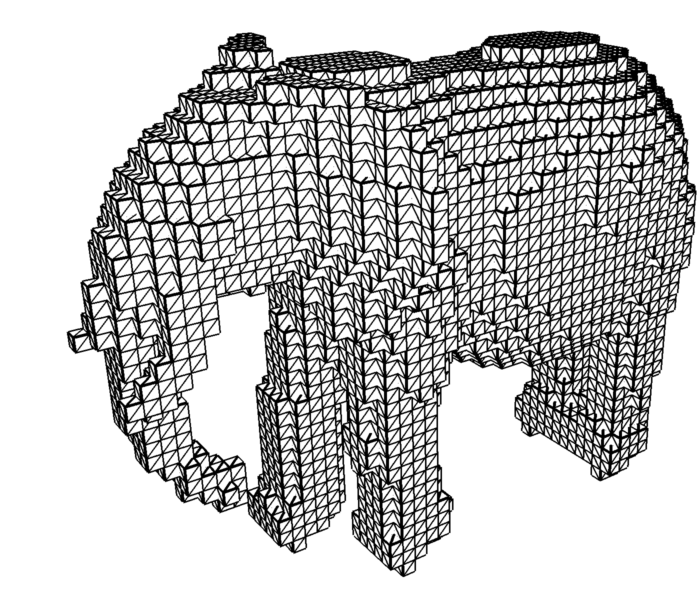} &
		\includegraphics[width=0.32\columnwidth]{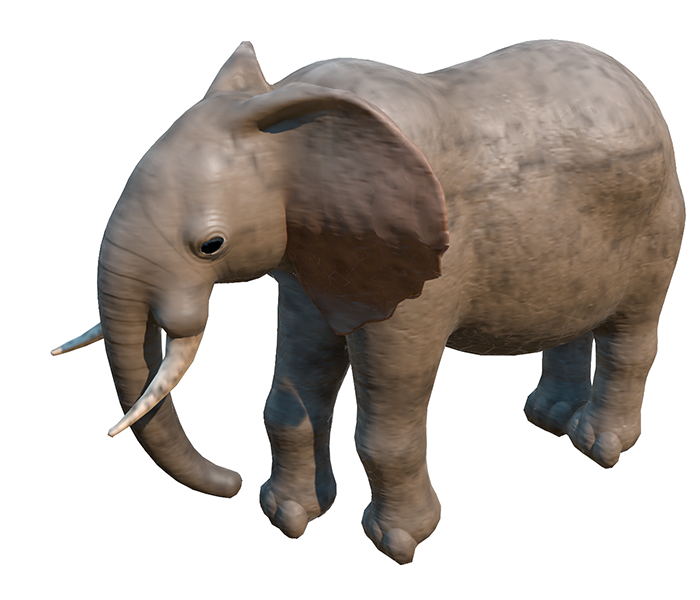} &
		\includegraphics[width=0.32\columnwidth]{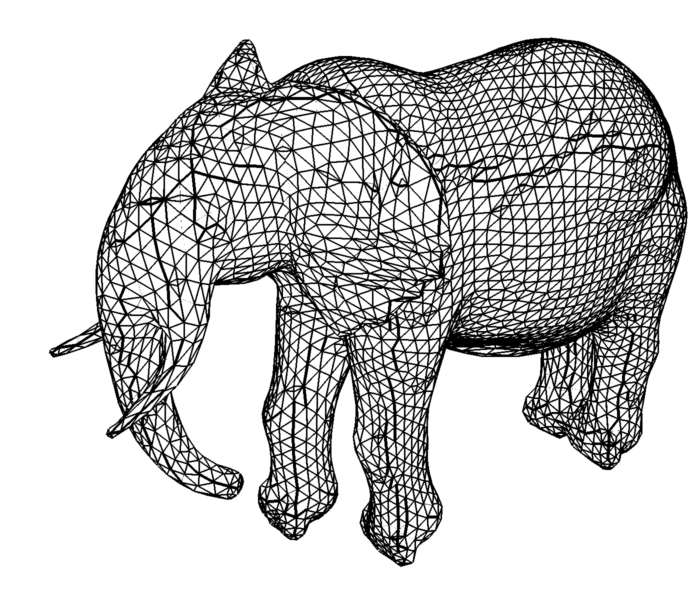} \\
		{\small Initial guess} & {\small Our rasterized model} & {\small Our optimized mesh} \\
		{\small 12k triangles} & {\small PSNR: 29.99~dB} & \\

	\end{tabular}
	\vspace*{-2mm}
	\caption{We convert a ray marched implicit model, an adapted version of the ShaderToy ``Elephant'' \copyright 
		Inigo Quilez, to a mesh with materials by optimizing for visual loss
		in a differentiable rasterizer. We use a tessellated sphere (12k triangles) as initial guess 
		and jointly optimize shape and appearance. We also show the corresponding results using a better 
		initial guess, produced by marching cubes. Note the improvements on sharp details, 
		e.g., the tusks.}
	\label{fig:implicit_mike}
\end{figure}
}


\newcommand{\figVISII}{
\begin{figure}
	\centering
	\setlength{\tabcolsep}{1pt}
	\begin{tabular}{cc}
		\includegraphics[width=0.495\columnwidth]{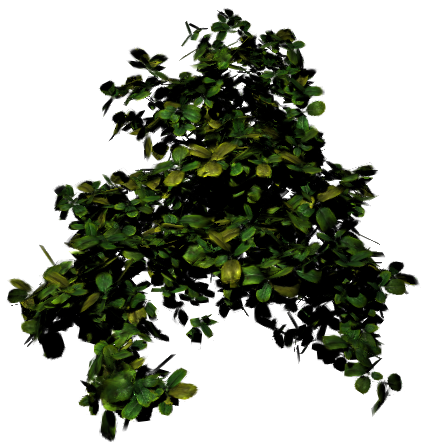} &
		\includegraphics[width=0.495\columnwidth]{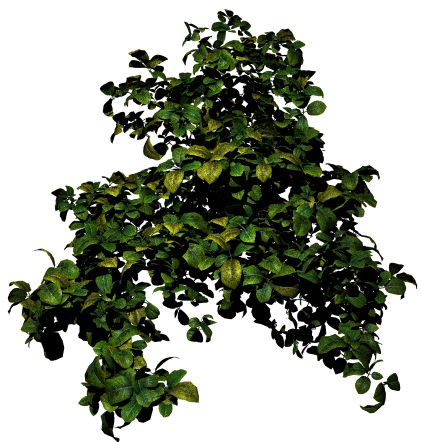} \\
		{\small Our rasterized model (6.5k tris)} & {\small Path traced reference (1.7M tris)} \\
	\end{tabular}
	\vspace*{-2mm}
	\caption{
		We optimize a mesh and materials to have the appearance of a path traced reference, when rendered in a rasterizer. 
		This can be used to convert between different material models, or as a simple way of baking shading 
		into material terms.
		Our optimized model is rasterized with 1~spp + post-processing antialiasing. 
		The path traced reference is rendered with 256~spp.
	}
	\label{fig:visii}
\end{figure}
}

\begin{abstract}
We present a suite of techniques for jointly optimizing triangle meshes and shading models to match the appearance of reference scenes. This capability has a number of uses, including appearance-preserving simplification of extremely complex assets, conversion between rendering systems, and even conversion between geometric scene representations. 

We follow and extend the classic analysis-by-synthesis family of techniques: enabled by a highly efficient differentiable renderer and modern nonlinear optimization algorithms, our results are driven to minimize the image-space difference to the target scene when rendered in similar viewing and lighting conditions. As the only signals driving the optimization are differences in rendered images, the approach is highly general and versatile: it easily supports many different forward rendering models such as normal mapping, spatially-varying BRDFs, displacement mapping, etc. Supervision through images only is also key to the ability to easily convert between rendering systems and scene representations.

We output triangle meshes with textured materials to ensure that the models render efficiently on modern graphics hardware and benefit from, e.g., hardware-accelerated rasterization, ray tracing, and filtered texture lookups. Our system is integrated in a small Python code base, and can be applied at high resolutions and on large models. We describe several use cases, including mesh decimation, level of detail generation, seamless mesh filtering
and approximations of aggregate geometry.
\end{abstract}



\begin{teaserfigure}
  \centering
  \setlength{\tabcolsep}{1pt}
  \includegraphics[width=\linewidth]{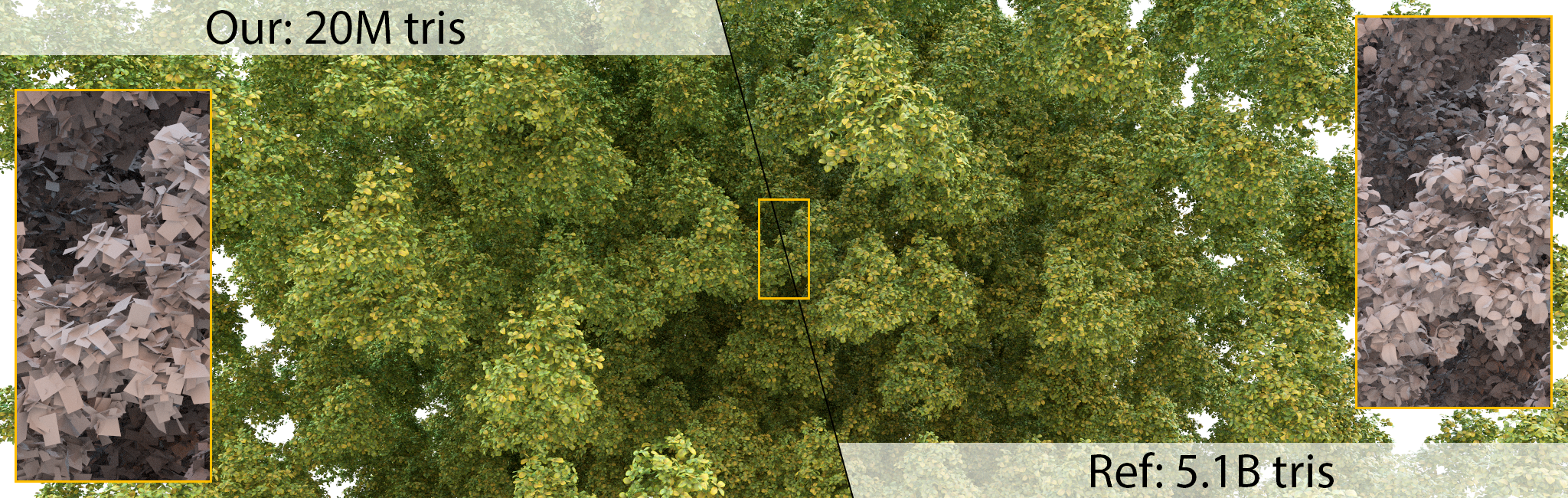}
   \caption{We automatically approximate shape and appearance of detailed 3D scenes. Here, we approximate foliage 
   	with low-poly proxy geometry and textures, instance it 3000 times, and render the scene in a standalone path tracer. Compared 
   	to the reference, we accurately capture the appearance of the scene at just a fraction (0.4\%) of the triangle 
   	count. The geometry is illustrated in the insets.
   The assets are from the Moana Island Scene, a publicly available data set courtesy of Walt Disney
Animation Studios.}
\Description{Our approximation of aggregate geometry.}
\label{fig:teaser}
\end{teaserfigure}

\maketitle

\section{Introduction}
\label{sec:intro}

Synthesizing images of objects with complex shapes and appearances is a central goal in computer graphics. 
The problem can be broken down into choosing suitable representations for shape and appearance,
modeling the scene according to the chosen representations, and finally, rendering it efficiently.

Creating a shape and appearance model for a particular 3D scene is inherently an inverse problem: we seek a representation that will, once fed through the renderer, result in an image that looks the way we want. 
Yet, most modeling tools turn the problem around: instead of providing the user with means to specify the image they want, they provide tools for editing the scene representation, leaving the modeler to iteratively proceed toward their goal. 

The goal in this work is to automatically find shape and appearance representations that match, when rendered, a reference scene provided by the user. This approach is often called inverse rendering or analysis-by-synthesis. 
In contrast to algorithms like multi-view stereo that must make do with a small number of reference images, we focus on applications where it is possible to programmatically synthesize reference views of the target scene under 
arbitrary, controllable viewing and lighting conditions. Within this scope, we present inverse rendering techniques for
\begin{itemize}
\item \textbf{Geometric simplification (LOD)\@.} Optimizing for the shape of a lower-resolution mesh to combat geometric aliasing and increase rendering efficiency. 
\item \textbf{Joint shape-appearance simplification.} Optimizing the shape and surface appearance model (mesh geometry, displacement maps, normal maps, spatially-varying BRDFs) to mimic the appearance of a more complex asset. 
\item \textbf{Simplification of aggregate geometry.} Dramatically simplifying complex foliage assets with little impact in visual quality. See Figure~\ref{fig:teaser} for an example.
\item \textbf{Animation.} Joint optimization of shape and skinning weights on reduced geometry to match a target animation.
\item \textbf{Conversion between rendering systems.} Optimizing the scene representation to match images rendered by an entirely different system. 
\item \textbf{Conversion between shape representations.} Finding a mesh geometry and associated appearance model that captures the appearance of objects given by other shape representations, such as signed distance fields (SDF)\@.
\end{itemize}

These kind of goals have been previously pursued with specialized algorithms (Section~\ref{sec:prev_work}) that typically focus on a single task, and consider only specific parts of the object representation. 
Since our approach is based on inverse rendering and nonlinear optimization it easily generalizes over all the different regimes while allowing joint optimization of all aspects of the representation that affect the final appearance.

In all our applications, the search for the shape and appearance is driven by image-space error. This, combined with performing shape and appearance optimization together, has the significant benefit that the mechanisms of the forward 
rendering model can each specialize for the effects they capture best, ``negotiating'' how to achieve the desired outcome together. As an example, this leads to a natural division of labor between the geometry (mesh) and a normal map: 
geometric detail is allowed to move between the representations by, e.g., locally smoothing a mesh and baking geometric detail into the normal map or other parameters of a physically based shading model~\cite{Karis2013}.

Our source code will be publicly available at~\url{https://github.com/NVlabs/nvdiffmodeling}.

\section{Previous Work}
\label{sec:prev_work}

\begin{figure*}[t]
	\centering
	\includegraphics[width=0.99\textwidth]{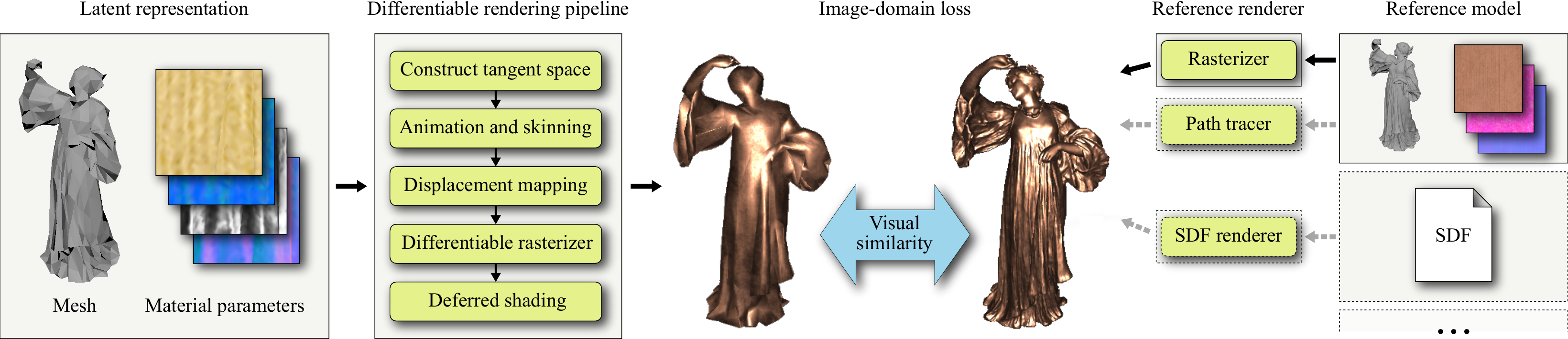}
	\caption{
		An overview of our method. We render the latent representation (mesh and material parameters) in a differentiable 
		rendering pipeline: a sequence of mesh operations followed by rasterization and deferred shading.
		Similarly, a target image is generated by the reference renderer under identical viewing and lighting conditions, and an 
		image-domain loss is computed on the two images. During optimization, we iterate over a large 
		number of image pairs with random viewing and lighting conditions.
		Using back-propagation and stochastic gradient descent, the latent representation is gradually morphed 
		to produce images close to the reference. 
	}
	\label{fig:framework}
\end{figure*}

\paragraph*{Mesh decimation} For a detailed overview of this mature research topic, we refer the reader to 
the book by Luebke~et~al.~\shortcite{Luebke2002}. Commonly used algorithms include
\emph{vertex decimation}~\cite{Schroeder1997}, 
\emph{vertex clustering}~\cite{Low1997} and \emph{edge contraction}~\cite{Garland97}. 
The error metric is typically geometry based. 
A notable exception is view-dependent simplification~\cite{Luebke1997} which optimizes for silhouette quality. 

Lindstrom and Turk present a framework for image-driven simplification~\shortcite{Lindstrom2000}.
They use rendered images to decide which portions of a mesh to simplify. 
Please refer to Corsini et~al.~\shortcite{Corsini2013} for a survey of perceptual metrics for triangle meshes.
Similarly, our objective function is based on visual differences,
but we leverage gradient-based optimization through differentiable rendering. 

Cohen et~al.~\shortcite{Cohen1996} introduce \emph{simplification envelopes} for generating a 
hierarchy of LODs for polygonal meshes. They build envelopes around the mesh to avoid self-intersections 
and can guarantee a distance tolerance between the original and simplified meshes. 
For our continuous level of detail application, we similarly use a sequence of meshes to represent the 
LODs. In contrast, we optimize for visual error in image space. 

Cook et~al.~\shortcite{Cook2007} introduce a decimation technique that stochastically removes a subset
of the geometric elements and alter the remaining elements by, e.g., scale and contrast adjustments, in order
to preserve the overall appearance. This technique is particularly suited for unstructured objects such as foliage. 
We similarly exploit the scene graph when approximating aggregate geometry, but instead of using heuristics for preserving visual appearance, we optimize for it directly.

\paragraph*{Appearance prefiltering} For a summary of surface appearance prefiltering techniques, please refer to 
Bruneton~et~al.~\shortcite{Bruneton2012}. 
Linear texture prefiltering methods are incorrect when applied to
spatially varying BRDF (\emph{bidirectional reflectance distribution function}) and surface normal maps.
Common approaches instead filter the \emph{normal distribution function} (NDF)~\cite{Fournier1992, Toksvig2005,Han2007,Olano2010}, 
where the challenge is how to compactly represent the filtered NDF\@.
Common representations include Gaussians~\cite{Toksvig2005} with one or more lobes, 
moment statistics~\cite{Olano2010}, and spherical harmonics~\cite{Han2007}.

Further work has considered, e.g., displacement mapping, and masking and shadowing effects~\cite{Dupuy2013, Wu2019}.
Loubet and Neyret~\shortcite{Loubet2017} prefilter meshes and materials by absorbing fine details
into a volumetric microflake representation. 
Zhao et al.~\shortcite{Zhao2016} downsample volumetric materials by optimizing for rendered similarity to the original.

Our work can be applied to prefiltering. In our experiments, we restrict the shading model to one  
diffuse lobe and one specular GGX lobe~\cite{Walter2007}, and let optimization adjust the mesh shape and the material parameters
so that the rendered result matches a highly supersampled reference. This model is indeed less expressive than many of those
in prefiltering literature, but has the benefit that there is no change to 
a typical game engine or any runtime overhead.  
The approach is flexible, as it treats any target surface and material representation in a unified manner:
only the final visual appearance is observed, and it does not matter
what combination of, e.g., mesh shape, displacement, normal, and material parameters produced it.

\paragraph*{Appearance and geometry capture}

Appearance capture can be framed as seeking a digital asset (e.g., an SVBRDF map and a mesh)
whose renderings visually match some real-world object.
This is conceptually similar to our setup, with the exception that our 
targets are other digital assets.

Much of the difficulty in appearance capture originates from the desire to limit the acquisition effort for the user. 
Exhaustively measuring real-world appearance under all lighting and view directions is
prohibitively expensive for most purposes. When only a sparse sampling is available,
typical approaches employ multistage processing involving, e.g.,
clustering~\cite{Lensch2003} or
multi-view stereo~\cite{Nam2018} to find a solution that reproduces the measurements and generalizes well to unseen conditions. 
Many approaches use special viewing configurations and lighting 
patterns~\cite{Gardner2003, Ghosh2009} to recover more varied reflectance information in each measurement. 
For recent surveys, see Dong~\shortcite{Dong2019}, Guarnera et al.~\shortcite{Guarnera2016}, and 
Weinmann and Klein~\shortcite{Weinmann2015}.

We are free to render our targets in as many viewing and lighting conditions as needed.
This eliminates much of
the complexity, and allows us to directly end-to-end optimize over the material parameters and 
vertex positions using a visual similarity loss.

Many appearance capture methods can be viewed as using 
simple differentiable renderers to match their predictions to the observations.
For example, Gao et al.~\shortcite{Gao2019} and Guo et al.~\shortcite{Guo2020} optimize for
SVBRDF maps that reproduce a handful of target photographs upon rendering. They assume planar geometry, and
use neural network based regularization to encourage plausible generalization to unseen conditions.

Sztrajman et al.~\shortcite{Sztrajman2017} convert materials between 
different analytic BRDF representations
by optimizing for their rendered visual similarity. Similar to us,
their target images do not originate from the real world, but from renderings of the
target asset. However, they do not consider geometry as part of the optimization.

\paragraph*{Scene acquisition with neural networks}

NeRF~\cite{Mildenhall2020} represents the scene as a neural radiance field. 
The quality of the reconstructed views are impressive, but not yet suitable for interactive
applications as the radiance field needs to be densely sampled at inference time, 
and the generated radiance field is static. 
Similar to our approach, they use many observations of the scene to train the model. 
In contrast, we generate triangle meshes and materials, which render
in real-time in a standard 3D engine.

Thies et al.~\shortcite{Thies2019} achieve a similar goal by learning a latent texture 
map and an associated image-space decoder CNN that allows high-quality view 
interpolation on meshes reconstructed by multi-view stereo in fixed lighting environments.

BSP-Net~\cite{Chen2020} generates compact meshes via binary space partitioning but does not consider materials or fine detail such as normal maps.

Pixel2Mesh~\cite{Wang2018b} generates a triangular mesh from a single color image by 
progressively deforming an ellipsoid. They use a graph convolutional neural network
to represent the mesh, and a set of losses to ensure that the mesh is well-formed, including 
a Laplacian regularization term.
Our shape optimization is also based on deforming an existing triangular mesh, but we rely on multiple image observations to allow higher-quality reconstruction that includes materials.
In addition, our approach is different in that we seek a representation directly, instead of training a neural network that would perform the conversion.

\paragraph*{Differentiable rendering} 
For an overview of current approaches, please refer to Laine~et~al.~\shortcite{Laine2020}.
They also introduce a flexible set of differentiable rasterization primitives which can be used
together with PyTorch or TensorFlow to build custom differentiable rendering pipelines. Several differentiable rendering  packages~\cite{Murthy2019, Valentin2019, Ravi2020} provide more built-in functionality but are less 
customizable. 

Chen et~al.~\shortcite{Chen2019,Zhang2020image} optimize over vertex positions, colors, normals, light directions and 
texture coordinates through a variety of lighting models, purely from 2D observations.
We extend this approach to handle more complex shading models (GGX), more complex geometry, normal maps,
displacement maps, and transparency, for higher visual quality. Additionally, we optimize shape and 
appearance jointly instead of training in stages, and show additional use cases including
mesh filtering and animation. Chen et~al.~focus primarily on predicting 3D objects from single images, while 
we use multiple image observations to capture the shape and appearance, which leads to higher-quality reconstructions.

Li~et~al.~\shortcite{Li2018} present a differentiable Monte Carlo ray tracer, 
and Mitsuba 2~\cite{Nimier2019} implements a full differentiable path tracer. 
In this paper, we use and extend the rasterization primitives from Laine~et~al.~\shortcite{Laine2020} 
for quicker iteration times and ease of customization. Combining our approach
with a differential path tracer would enable additional 
applications, but at a significantly higher computational cost.
Given recent progress in differentiable rendering performance~\cite{Nimier2020}, we hope
to leverage a differentible path tracer in future work.

\section{Our Method}
\label{sec:system}

Our goal is to jointly optimize shape and material parameters to match the visual appearance
of images from a reference renderer. 
We follow the common practice of representing the shape as a triangle mesh and using a spatially varying BRDF for materials.
This ensures that
our optimized representation renders directly in modern game engines and can readily exploit
hardware-accelerated rasterization, ray tracing, and filtered texture-lookups.

Figure~\ref{fig:framework} outlines our method. The \emph{latent representation} consists of a triangle mesh and a set of textures
describing spatially varying material parameters from a physically based shading model\footnote{We use the word ``latent'' with care; though our representation and its decoder (the renderer) are fixed-function and completely interpretable, the representation \emph{is} unobserved, i.e., latent, and we seek to infer it from the input images. This choice of word encompasses potential future extensions with learnable latent representations and decoders in a natural way.}, its exact form varies slightly between applications.
During optimization, we render the latent representation using a differentiable rendering pipeline: a sequence of mesh operations, 
a rasterizer, and a deferred shading stage. 
An image-space loss is then computed between the resulting image and a target image produced by a reference renderer under identical 
viewing and lighting conditions.
Because the rendering pipeline is fully differentiable, we can compute the gradient of the loss with respect to parameters of the
latent representation, i.e., vertex positions and texture contents, and consequently optimize these to improve the visual similarity.

We iterate over a large number of image pairs with randomized camera and a single randomized point light, similar
to a virtual photo-goniometer. Using stochastic gradient descent, 
the latent representation is gradually morphed to match the appearance of the reference model.

More formally, let $\theta$ denote the parameters of our latent representation (e.g., mesh vertex positions and spatially varying material parameters).
The rendered image $I_{\theta}(c,l)$ is a function of $\theta$, camera, $c$, and light, $l$.
The reference render is another function $I_{\mathrm{ref}}(c,l)$, parameterized by the camera and light.
Given an image space loss function $L$, we minimize the empirical risk
\begin{equation}
\underset{\theta}{\mathrm{argmin}}\ \mathbb{E}_{c,l}\big[L\big(I_{\theta}(c, l), I_{\mathrm{ref}}(c, l)\big)\big]
\end{equation}
using stochastic gradient descent based on gradients w.r.t.~the latent parameters,
$\partial L/\partial\theta$, which are obtained through differentiable rendering.

Apart from the ability to place the camera and light, we consider the reference renderer as a black box. The only information 
communicated between the reference renderer and our latent representation are the target images that are used in the image-domain loss. Note, in particular, that this means that the reference renderer does not need to be differentiable, or even implemented in the same framework.
This allows us to convert models across 
renderers, and even between different geometrical representations---e.g., from signed distance fields (SDF) to triangle meshes.

Note that the choice of reference rendering algorithm depends on the intended use of the final model.
Our pipeline does not render shadows or other global effects, so if the reference renderings feature no such effects either---as can usually be 
arranged---the optimization will converge on materials that are as close as possible to the reference model.
This is because both sides of the process in Figure~\ref{fig:framework} agree upon which effects in the final image are due to the model and which are due to the rendering algorithm.
How the model is ultimately used in production is an orthogonal question, and at that point, shadows or path tracing can be enabled if desired.
Alternatively, if the references are rendered with, say, ambient occlusion or path tracing enabled, the optimization process will bake these into the material parameters so that rendering \emph{without} these effects produces a reasonable approximation.
We demonstrate both approaches later in Sections~\ref{sec:aggregate} and~\ref{sec:generalizations}.

As often in nonlinear optimization, good initial guesses may have a dramatic effect on the speed of convergence and eventual quality of the result. When a high-resolution mesh of the target object is available, we use off-the-shelf mesh decimation tools to produce the initial guess for the latent triangle mesh.  In some cases, e.g., when baking foliage as billboard clouds, we draw on prior domain knowledge and explicitly specify a suitable initial mesh. However, we find that starting from a tessellated sphere often yields surprisingly good results.
Similarly, if available, we may use texture maps from the reference scene as an initial guess for 
material parameters. If unavailable, we start from randomly initialized texture maps. We currently do not optimize 
the topology or texture coordinates of the latent representation. Hence, we require the initial mesh to have a reasonable 
level of tessellation and non-overlapping texture parameterization.

Our rendering pipeline combines the differentiable rasterization primitives by Laine~et~al.~\shortcite{Laine2020} 
with mesh operations and deferred shading in PyTorch~\shortcite{Paszke2017}, a modern autodifferentiation framework that greatly simplifies
the implementation. 
To encourage well-formed meshes, we include a Laplacian regularizer~\cite{Sorkine2005}
in our objective function.
Each component is further detailed in Section~\ref{sec:components}. 
This setup is flexible and allows for differentiable rendering at high resolutions and high polygon counts at interactive rates. 

\figSkull
\figDisplacementMapping
\figEwer
\figEwerBreakdown
\figGrux
\figPrefilter

\section{Applications}

In this section, we present several use cases for our method:
joint simplification of shape and appearance, prefiltering shape and appearance to reduce aliasing, geometric simplification of skinned character animation,
approximation of aggregate geometry, and 3D mesh extraction from implicit surfaces. Unless otherwise noted,
we optimize for 10k iterations at a resolution of $2048\times 2048$ pixels,
where each iteration uses a random camera and light position. We defer discussion of implementation details until Section~\ref{sec:components}. 
Please refer to our interactive image viewer for detailed image comparisons.

\subsection{Joint Shape-Appearance Simplification}
\label{sec:joint}

Simplifying complex assets with minimal loss in visual fidelity is our most straightforward application. We present three variants that demonstrate joint optimization over different combinations of shape and appearance: normal map baking, joint simplification that also accounts for surface reflectance, and approximating complex meshes with displacement maps applied on a coarse base domain.

\paragraph{Normal map baking.} As an initial example, we start from a sphere with 3k triangles
and optimize shape and a tangent-space normal map to approximate a highly 
detailed reference mesh with 735k triangles. 
Besides the normals, the material is otherwise fixed as diffuse uniform gray.
Our result is shown in the top row of Figure~\ref{fig:skull}.
While some high-frequency detail is missing, the result is nonetheless encouraging, considering 
that the optimization process is entirely automatic and uses no direct information about the reference model.
In the bottom row of the figure, we repeat the experiment starting from a reduced version of the reference mesh with 9k triangles, as produced by Simplygon Free~\mbox{8.3}.
As expected, this improves the results considerably, as it is now sufficient to fine-tune the geometry instead of discovering it from scratch, and we slightly outperform the Simplygon normal map baker.
Please refer to the supplemental material for a study of how quality is impacted by the triangle count of the initial mesh.

\paragraph{Displacement map baking.}
In addition to normal maps, displacement mapping is an increasingly popular technique for representing complex shapes in real-time settings~\cite{Nanite2020}. 
It achieves a compact representation by tessellating a coarse \emph{base mesh} on the fly, and displacing the resulting vertices in the direction of the interpolated surface normal by amounts read from the displacement map texture. 

Our approach enables using displacement maps for approximating geometry by simply implementing the tessellation and displacement steps in our forward rendering pipeline.
Figure~\ref{fig:displacement} shows a displacement mapped version of the dancer mesh, rendered with diffuse 
shading to make the geometrical impact more apparent. Our result is obtained by jointly optimizing the pre-tessellation shape of the base mesh, the normal map, and the displacement map. As shown in the insets, our process yields a natural ``division of labor'' between the representations: the base mesh models the overall shape, the displacement map models mid-scale detail, and the finest detail that is not representable by the displaced surface is captured by the normal map.

While the above example is optimized for a single, fixed tessellation level, dynamic tessellation is often used for level of detail selection. We can additionally optimize a displacement mapped mesh to look good under multiple levels of tessellation. Please refer to the supplemental material for further results.

\paragraph{Simplification with complex materials.}
Next, we upgrade to a physically based shading model with one diffuse lobe and one isotropic GGX specular lobe~\cite{Walter2007} commonly used in modern game engines. 
Figure~\ref{fig:ewer} shows a bronze sculpture of high geometric complexity, complex texture mapped materials, and normal maps. 
Here, we optimize jointly for shape and appearance under random views and point light directions.
The specular term introduces higher frequencies and higher dynamic range, both of 
which make the optimization process more challenging. Hence, we use an image-domain loss robust to large floating-point values. 
Please refer to Section~\ref{sec:loss} for details.
Figure~\ref{fig:ewerbreakdown} shows further comparisons to the reference and to the initial mesh reduced using Simplygon.
Our method, optimizing based on visual differences, closely matches the true silhouette, finds accurate normals, and 
captures the shaded appearance of the reference. The specular highlights in our results are somewhat blurred, but 
it could be argued that this is preferable to the aliasing that the reference mesh shows when rendered at 1~spp.

\paragraph{Automatic cleanup.}
As a final example, Figure~\ref{fig:grux} demonstrates cleanup of the result of an unsuccessful mesh 
decimation operation performed in another software package. In this test, we reduced a game character from the Unreal Engine Paragon asset~\shortcite{Paragon2018} 
from 81k to 15k triangles in Autodesk Maya 2019. Note that the automatic reduction slightly decreases the volume of the mesh, 
detaches geometric elements, suffers from incorrect texturing, and produces some self-intersecting geometry. 
After optimization, we regain the volume and automatically clean up most of the geometry and texturing issues. 
We do not support topology changes, so our optimized mesh is still not watertight and retains some of the artifacts, but
the rendered appearance is nonetheless clearly improved. Please refer to the interactive viewer for full resolution images.

\subsection{Shape and Appearance Prefiltering}
\label{sec:prefilter}

In the previous section, our goal was to create faithful representations of complex assets with reduced triangle counts. A closely related variant of this problem is to find efficiently renderable approximations to original assets that are so complex that they require a substantial amount of supersampling to produce alias-free images. We call this problem joint prefiltering of shape and appearance. The resulting optimized models have the property that they reproduce, when rendered at only one sample per pixel, the appearance of assets that require potentially hundreds of samples per pixel for alias-free reproduction. The only practical differences in the optimization are specifying a typically smaller target image resolution, and rendering the reference images with enough supersampling to ensure lack of aliasing; all previously demonstrated freedoms in choosing the shape and appearance models still apply. We address some technical challenges associated with high triangle counts in Section~\ref{sec:diffrast}.

Figure~\ref{fig:pre_filter} shows joint shape and appearance prefiltering on the golden statue, targeting 
rendering resolutions of $64\times64$ and $512\times512$ pixels. 
Note that the result prefiltered for the smaller resolution of $64\times64$ pixels has, as one would expect, considerably smoother 
normals that account, together with the specular map, for the effect of averaging present in supersampling. Also note the geometric smoothing, e.g., the flower on the statue's head. 
When rendered at the intended resolution, the low-resolution mesh matches the appearance 
of the reference well, with no apparent aliasing. 

To obtain an appropriate amount of prefiltering at different rendering resolutions (i.e.,~rendering distances), it is not sufficient to optimize for one resolution only.
For materials, this is easy to achieve by treating each mipmap level as an independent latent variable.
This yields resolution-specific material representations that a trilinear texture lookup will automatically interpolate between.
Even though linear interpolation between material parameters is not generally correct~\cite{Bruneton2012}, the optimization process will find a representation that, assuming trilinear texture fetches will be used, matches the target images as well as possible.

For geometry, we can store multiple sets of vertex positions and choose between these based on distance to mesh, average projected edge length, or a similar heuristic.
As with mipmapping, linear interpolation between levels can be used to eliminate popping artifacts.
To simplify the experiments, we have opted to keep the topology fixed, so no reduction in triangle count is obtained in our tests---for geometric simplification, a mesh LOD scheme would need to be incorporated into the optimization.
Please see the accompanying video for an example animation of continuous, distance-dependent prefiltering.

\subsection{Animation and Skinning}

\figAnimation
\figAggregate

Having so far focused on static scenes, we now study appearance-driven simplification of animated articulated characters over entire animation sequences. More precisely, given a high-resolution reference mesh animated by skeletal subspace deformation (SSD), we optimize over the bind-pose vertex positions, normals, SVBRDF, and skinning weights (bone-vertex attachments) of a simplified model in an attempt to replicate the appearance of the reference animation. In contrast to simplifying the character in the bind pose (T-pose) only, this holds promise for being able to strike compromises to distribute the error evenly among the frames by adjusting the geometry, skinning weights, and materials appropriately.

Implementation is straightforward: the only addition required is blending transformed vertex positions using the skinning weights, a simple linear operation.
An example is shown in Figure~\ref{fig:tyrone}, where we decimate a rigged animated mesh in a completely 
automated process. We include examples on rigged meshes from RenderPeople~\shortcite{RenderPeople2020} 
using skeletal animations from the CMU motion capture database~\shortcite{CMU2020} in the accompanying video.

We assume the time-varying bone transformations are known, and treat them as constants during optimization. 
Joint optimization of both bone transformations and skinning weights~\cite{James2005} is a possible direction for future work. Furthermore, we assume that normal maps and SVBRDFs are constant in time. Modeling dynamic behavior like wrinkles opens another interesting future direction.

Above we use a reduced mesh as initial guess, but to thoroughly 
battle-test our ability to optimize skinning weights we instead start from a sphere and 
morph it into an animated figure with known skeletal animation, jointly optimizing shape, materials and 
skinning weights. Please refer to the video for results.

\subsection{Approximating Aggregate Geometry}
\label{sec:aggregate}

Stochastic aggregate geometry, such as foliage, are particularly difficult to simplify: as the overall appearance emerges from the combined effect of many small, disjoint components, techniques such as mesh decimation are ineffective.

Cook et~al.~\shortcite{Cook2007} introduce a stochastic decimation technique that exploits a known 
scene graph, and randomly remove a subset of the geometric elements and alter the remaining elements, 
e.g., by scaling and contrast adjustments, to preserve the overall appearance 
of a scene.
We approach the same problem from another angle, drawing inspiration from the billboard clouds of D\'{e}coret et~al.~\shortcite{Decoret2003}: instead of stochastically pruning the procedural scene graph, 
we replace the complex leaf geometries with textured quads. With the quads providing an initial guess, we then 
jointly optimize material parameters, shape, and \emph{transparency} based on visual loss of rendered images.
For this, we extended the differentiable rasterization primitives of Laine~et.~al~\shortcite{Laine2020} to 
support order-independent transparency through depth peeling~\cite{Everitt2001}. Please refer to Section~\ref{sec:diffrast} for details.

In Figure~\ref{fig:aggregate}, we show two examples of simplification of aggregate geometry from the 
Disney Moana asset~\shortcite{Moana2018}.
In both cases, we create plausible approximations from only 0.8\% and 0.4\% of the triangles of the reference mesh,
respectively. For this application, we used squared $L_2$ as objective function. 

In Figure~\ref{fig:teaser}, we instance our 
approximation 3000$\times$ and render in a path tracer. The result closely resembles the reference with a 
massive reduction in geometric complexity.
Our supplemental material includes a comparison with stochastic simplification~\cite{Cook2007}. Please also refer 
to our video for animated results.

\subsection{Generalizations}
\label{sec:generalizations}

\figSDF

All examples so far have used the same geometric representation for both the latent representation and the reference model. Moreover, the reference images have been produced by the same renderer as used for optimization.
Enabled by supervising the optimization strictly in image space, we now demonstrate generalization across surface representations and rendering systems.
While the full scope of potential applications is vast, we illustrate this by two examples: converting a 
ray marched implicit surface to a mesh, and transferring a path traced rendered model to a rasterizer.

\paragraph{Textured meshes from implicit surfaces.}
In the first example, shown in Figure~\ref{fig:implicit_mike}, we adapt an implicit surface pixel shader 
from ShaderToy~\cite{Jeremias2014} to isolate the main object and match the lighting and camera model of our renderer. 
We automatically convert this ray marched implicit surface to a triangle mesh with materials. We use a tessellated sphere as the initial guess for the geometry. For this example, we also add an ambient material term to our latent representation to better
match the lighting of the implicit surface renderer, which uses custom ambient lighting.
We also use a static light position, so shadowing is captured and baked into the material
parameters in the optimization process. View-dependent shading effects, e.g., specular highlight, are
still captured. Please refer to the supplemental material for additional results.

\paragraph{Baking path traced lighting.}
The second example is shown in Figure~\ref{fig:visii}. Here, we use the ViSII path tracer~\cite{Morrical2020} 
to generate reference images for the aggregate geometry decimation example of Section~\ref{sec:aggregate}. 
We use a static light position during optimization: materials are effectively
converted to our material model and shadows are baked into the textures, creating a plausible rasterized approximation
of the path traced reference. Note that we control the viewing conditions in the reference images, and use matching configurations in the optimization.

\figVISII

\section{Implementation}
\label{sec:components}

This section covers the implementation details of our method. We first describe the differentiable 
mesh operations that are applied to the mesh before rendering. The second part details our differentiable 
renderer, and the final part provides details of the optimization process.

\subsection{Mesh Operations}

Referring to Figure~\ref{fig:framework}, our pipeline includes differentiable mesh operations for 
tangent space computation, animation \& skinning, and displacement mapping. 

\paragraph*{Tangent space} To optimize tangent space normal maps on deforming geometry, the tangent frame must be 
differentiable and dynamically updated to reflect any change in vertex position. We compute smooth vertex normals and 
derive tangent and bi-tangent vectors from the vertex positions and texture coordinates~\cite{Mikkelsen08}.
Using mesh-derived smooth normals is a not a limitation because creases or other sharp features can be handled by the
normal map.

\newcommand{\vi}{\boldsymbol{v}_i}

\paragraph*{Animation \& skinning} We support skinning for Universal Scene Description (USD)~\shortcite{USD2016} meshes
and rely on the  USD API to evaluate skeleton animation. We optimize the skinning weights of animated meshes, and 
therefore implement a differentiable skinning operator according to:
\begin{equation}
\vi^s = \sum_{b\in \mathcal{B}} w_{ib} M_b \vi,
\end{equation}
where $\mathcal{B}$ is the set of bones, $M_b$ is the bone transform matrix for the current frame, and 
$w_{ib}$ is the skinning weight of bone $b$ influencing vertex $\vi$. Weights are typically 
stored using a sparse indexed representation, but we implement the full dense skinning operator to support 
any vertex-bone association during optimization. 

\paragraph*{Displacement mapping} Here, the latent representation consists of a coarse base mesh
and a scalar displacement map. The mesh is subdivided, and
the displacement map is used to displace the tessellated vertices along the interpolated normal direction. 
The tessellator uses edge-midpoint subdivision~\cite{Wang2018b}, 
where each triangle is split into four new triangles by 
inserting a new vertex along each edge. This operation changes topology, which is not a differentiable operation in
our current renderer. This is simple to work around by using a tessellation criteria that does not depend on any parameters requiring gradients.
For the example in Section~\ref{sec:joint}, we select a constant tessellation factor and precompute the topology of the tessellated mesh before optimization. 
The position of each vertex created by the tessellation is recomputed every iteration to ensure that gradients are propagated correctly. 

\newcommand{\ti}{\boldsymbol{t}_{\hspace*{-0.15mm}i}}
\renewcommand{\ni}{\boldsymbol{n}_i}

For displacement lookups we deploy the differentiable texture primitive of Laine~et~al.~\shortcite{Laine2020}, 
and displace each vertex according to:
\begin{equation}
\vi^d = \vi + \textrm{tex2d}(\ti) \cdot \ni, 
\end{equation}
where $\vi$ is the original tessellated vertex position, $\ni$ is the interpolated normal, and $\ti$ is the texture coordinate.

\subsection{Differentiable Renderer}
\label{sec:diffrast}
Our renderer is based on the differentiable rasterization primitives of Laine~et~al.~\shortcite{Laine2020}. 
We employ deferred shading based on a G-buffer that stores 3D position, normal, tangent, bi-tangent, and texture coordinates for each pixel. 

\newcommand{\kd}{\boldsymbol{k}_{\hspace*{-0.1mm}d}}
\newcommand{\ks}{\boldsymbol{k}_{\hspace*{-0.1mm}s}}
For materials, we use a variant of the physically based model  
from Disney~\cite{Burley12} that is common in modern game engines~\cite{Karis2013,Lagarde2014}.
This allows us to easily import game assets and lets us render our optimized meshes 
directly in existing engines without modifications. Material models for real-time rendering commonly
combine a diffuse term with an isotropic, specular GGX lobe~\cite{Walter2007}. 
The parameters for the diffuse lobe $\kd$ are provided in a four-component texture, 
where the optional fourth channel $\alpha$ represents transparency. 
The specular lobe is described by a roughness value $r$ for the GGX normal 
distribution function and a metalness factor $m$ which interpolates between plastic and metallic
appearance by computing a specular highlight color according to 
$\ks = (1-m) \cdot 0.04 + m \cdot \kd$~\cite{Karis2013}.

For appearance filtering, we want additional flexibility in suppressing the specular lobe.
To that end, we add a parameter $\gamma$
that scales the specular lobe according to: $\ks' = (1-\gamma) \ks$.
Our specular parameters are thus represented by a three-component texture $(\gamma, r, m)$
where each component may vary spatially. This representation is purposely chosen to resemble 
the commonly used ORM (occlusion, roughness, metalness) textures, where we have replaced the occlusion
channel with $\gamma$. 

Note that we have chosen to reparameterize a single, fixed BSDF model for appearance filtering
to make results easily adoptable in game engines and to show that true prefiltering can be made a part of the asset 
pipeline. However, it would be straightforward to replace the BSDF with multi-lobe models~\cite{Bruneton2012}, 
spherical harmonics representations~\cite{Han2007}, or a variant of the neural representation proposed 
by M\"{u}ller~et~al.~\shortcite{Muller2019} for more powerful representations.

\paragraph*{Antialiasing} 
For the prefiltering applications in Section~\ref{sec:prefilter}, we want to 
match the appearance of a highly supersampled target image to a 1~spp rendering of our latent representation.
In practice, we experience instabilities when optimizing for low rendering resolutions due to approximations in silhouette gradient 
computations~\cite{Laine2020}. We work around this by rendering our
latent representation using multisampled antialiasing (MSAA), i.e., we sample visibility at a higher rate
but shade only once per pixel. This setup improves visibility gradients at silhouettes, but still enforces that our latent representation
matches the shading of the supersampled reference with just a single shading sample. 

This solution has the limitation that MSAA correctly integrates the visibility term, 
which is not expected for 1~spp rendering. This creates a small mismatch between our optimization setup and 
final rendering of the optimized model at 1~spp, but we have not found this to be an issue in practice.
MSAA is only used during training, and all our result images and animations indicating 1~spp are rendered without it, i.e., 
they represent true 1~spp rendering without any antialiasing unless otherwise mentioned.

\paragraph*{Order-independent transparency} Transparency is required for the aggregate geometry 
application in Section~\ref{sec:aggregate}. We implement order-independ\-ent transparency through \emph{depth peeling}~\cite{Everitt2001}, 
i.e., by rendering multiple passes where each pass peels off the front-most depth layer. We 
extend the rasterization primitive of Laine~et~al.~\shortcite{Laine2020} by adding a two-sided depth test to 
their fragment shader and passing the depth output of previous rasterization pass as a parameter to the next. While 
we rasterize the depth layers front-to-back, we perform blending in back-to-front order (starting with the background) 
as this works best with their antialiasing primitive~\cite{Laine2020}.
Eight passes of depth peeling were used in our experiments.

\subsection{Optimization}

\paragraph{Objective function}  
\label{sec:loss}
Our renderer uses physically based shading and produces images with high dynamic range.
Therefore, the objective function must be robust to the full range of floating-point values.
Following recent work in HDR image 
denoising~\cite{Munkberg2020}, our image space loss, $L_\mathrm{image}$, computes
the $L_1$ norm on tone mapped colors. As tone map operator, 
we transform linear radiance values, $x$, according to $x' = \Gamma(\log(x + 1))$,
where $\Gamma(x)$ is the sRGB transfer function~\cite{srgb96}:
\begin{eqnarray}
\Gamma(x) &=&  
\begin{cases}
12.92x & x \leq 0.0031308 \\
(1+a)x^{1/2.4} -a & x > 0.0031308
\end{cases} \\ 
a &=& 0.055.  \nonumber
\end{eqnarray}

\newcommand{\di}{\boldsymbol{\delta}_i}
\newcommand{\dip}{{\di}{\!\!'\hspace*{0.2mm}}}
\newcommand{\vj}{\boldsymbol{v}_{\hspace*{-0.2mm}j}}

In addition, we use a Laplacian regularizer~\cite{Sorkine2005} on the triangle mesh in our latent representation. 
This is important in the beginning of optimization to keep the mesh surface intact when gradients are large. 
The uniformly-weighted differential $\di$ of vertex $\vi$ is given by 
$\di = \vi - \frac{1}{\left|N_i\right|}\sum_{j \in N_i} \vj$, 
where $N_i$ is the one-ring neighborhood of vertex $\vi$.
We follow Laine et~al.~\shortcite{Laine2020} and use a Laplacian regularizer term given by
\begin{equation}
L_{\boldsymbol{\delta}} = \frac{1}{n}\sum_{i=1}^n\left\lVert\di - \dip\right\rVert^2,
\end{equation}
where $\dip$ is the uniformly-weighted differential of the input mesh (i.e., our initial guess). 
When the input mesh is a poor approximation, e.g., a sphere, we use an \emph{absolute} regularizer and set $\dip = \boldsymbol{0}$. 

\newcommand{\lt}{\lambda_t}

Our combined objective function is:
\begin{equation}
L_\mathrm{opt} = L_\mathrm{image} + \lt L_{\boldsymbol{\delta}},
\end{equation}
where $\lt$ is the regularization weight that depends on the current optimization iteration $t$.
We gradually reduce $\lt$ during optimization according to $\lt = (\lambda_{t-1} - \lambda_{min})\cdot 10^{-k t} + \lambda_{min}$. 
Here, $k=10^{-6}$, and $\lambda_{min}$ is chosen as
2\% of the initial weight, $\lambda_0$. 
The uniform Laplacian regularizer depends on tessellation, 
whereas image-domain loss does not.
Hence, the image loss must be balanced against the Laplacian loss as our applications include meshes with greatly varying triangle counts. 
The initial weight, $\lambda_0$, can either be specifed by the user
or by a simple heuristic: We evaluate the Laplacian error at the start of optimization, 
and set $\lambda_0 = 0.25 \ L_{\mathrm{image}} / L_{\boldsymbol{\delta}}$, which 
has worked well for most of our examples. 
Please refer to the supplemental material for  an example of how the regularizer improves mesh quality.

\paragraph*{Learning rate and runtime}
We optimize the latent representation using Adam~\cite{Kingma2014} with default parameters. 
Learning rate is scheduled as $\mathrm{lr}_i = \mathrm{lr}_0 \cdot 10^{-k t}$, where $t$ is the iteration,
$\mathrm{lr}_0$ is the initial learning rate, and $k=0.0002$.
When starting from a coarse initial guess, e.g., 
a sphere or billboard cloud, we typically use a high initial learning rate, e.g., $\mathrm{lr}_0=0.01$,
to allow for large mesh deformations. When starting from an auto-decimated mesh, we want to fine-tune the result
without jumping out of the already good local minima, which leads us to using lower values for $\mathrm{lr}_0 \in [0.001, 0.003]$.
We note that mini-batching is highly beneficial to reduce gradient noise, particularly during the early phases of optimization.
Gradient noise is problematic for vertex positions as it can cause the mesh to 
fold or self-intersect, especially in highly tessellated regions.
In practice, we use batch size between one and eight.
For the final results, we run optimization at a resolution of 2048$\times$2048 pixels for 10k steps,
which typically takes a few hours on a single NVIDIA V100 GPU\@.

\section{Limitations}
\label{sec:limitations}
Apart from technical limitations, e.g., large memory consumption, in particular when using many layers for depth peeling 
or when rendering at high resolutions, we note three main limitations in our approach. 

\paragraph*{Hyperparameter tuning}
For each use case, we need to tune the constants for the Laplacian regularizer and learning rate. 
This is a common problem in shape and appearance optimization, but it is pronounced here as we work with a wide range of 
tessellation rates and initial meshes of varying quality. We use a simple heuristic to set the Laplacian regularizer weight and learning rate parameters, 
but robustly configuring these hyperparameters for all use cases remains an open challenge. 

\begin{figure}
	\centering
	\includegraphics[width=0.95\columnwidth]{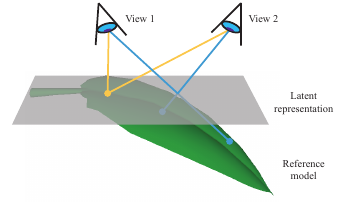}
	\vspace*{-2mm}
	\caption{View parallax can sometimes be an issue. This is most noticeable in our aggregate geometry 
		application where a tessellated leaf is represented as a single quad. There is no problem where shape 
		approximation is good, as illustrated by the yellow rays. However, as the blue rays show, we introduce 
		parallax when the latent representation and reference model deviate. This leads to blurry normal maps and textures. 
		The effect is exaggerated here for illustration purposes---in this situation, the optimization of vertex positions 
		would tend to move the latent geometry closer to the reference model, partially because this also reduces the
		parallax-related discrepancies.
		}
	\label{fig:view_dependent}
\end{figure}

\paragraph*{View parallax}
The second limitation is inherent in our problem formulation. We optimize for image-domain loss, which sometimes
suffers from view parallax between the latent representation and reference model, as illustrated by Figure~\ref{fig:view_dependent}. 
With $L_1$/$L_2$ image losses, discrepancies due to view parallax will converge to the median/mean value during 
optimization, which tend to blur the material parameter textures somewhat in cases where the geometry differences are large. 
This is particularly prominent for aggregate geometry, where transparency helps amplify the effect, and where we do the most
extreme polygon reduction. 

As an attempt to mitigate this problem, we experimented with perceptual VGG 
loss~\cite{Ledig2016} and adversarial losses~\cite{Radford2015} that are known to preserve high-frequency detail in 
image reconstruction tasks and should be less sensitive to small translations.
Unfortunately we noted no or limited benefit, with both strategies adding details resembling film grain or noise patterns instead of improving the visual quality of results.
These losses were also less robust in an HDR setting and did not properly preserve the high dynamic range of the results.

\paragraph*{Rasterization-based rendering model}
As our renderer is based on rasterization, it cannot handle effects such as refractive materials, specular interreflections, or subsurface scattering.
If the reference renderer outputs images with these effects, our system will end up baking them into the available material parameters as well as possible,
but highly view-dependent non-trivial effects such as translucency or subsurface scattering can be captured only on average.

However, differentiable rendering is a highly active field of research, and combining our approach with a differentiable path tracer~\cite{Nimier2020}
could allow handling these kind of effects in the future.

\section{Conclusions and Future Work}

We have demonstrated that a wide spectrum of modeling tasks, including simplification, conversion, and prefiltering, can be achieved in a common inverse rendering framework that supports various shape and appearance models. While we show improvements in individual examples such as mesh decimation, we believe the main strength of our approach lies in the ability to jointly optimize over shape, appearance, and 
animation parameters. 
Additionally, coupling automatic differentiation and optimization makes extensions to 
new applications and latent representations easy. 

There are wide opportunities for future work.
To keep a clear focus for this paper, we have only evaluated applications within the limitations of triangle meshes and a commonly used material model, 
so that the optimized representation can be used unmodified in existing renderers.
However, an obvious line of extensions is to use our framework to augment traditional fixed-function graphics models with learned representations, effectively creating hybrids between traditional graphics and the currently popular fully learned renderers~\cite{Thies2019}.

Another direction is to further automate and improve our optimization process. In particular, highly tessellated
meshes are difficult to optimize, as a large neighborhood of vertices must often be moved in tandem to achieve a large-scale deformation.
It might be much easier to guide optimization using a hierarchical approach where more triangles are progessively generated as the result converges.
This progression could further provide a natural LOD representation for the final mesh.
For applications where no initial guess for the mesh can be obtained, it may be possible to improve results by using alternate geometrical representations, 
e.g., by using a volumetric grid and tessellating it using marching cubes.

We also envision that appearance-based optimization could be used in semi-automated modeling tools. When visualizing the optimization process, 
it is often evident where problems occur due to, e.g., under-tessellated regions or insufficient mesh genus, and an artist could likely clean up such areas 
given rudimentary interactive meshing tools.

\bibliographystyle{ACM-Reference-Format}
\bibliography{paper}


\begin{thebibliography}{66}


\ifx \showCODEN    \undefined \def \showCODEN     #1{\unskip}     \fi
\ifx \showDOI      \undefined \def \showDOI       #1{#1}\fi
\ifx \showISBNx    \undefined \def \showISBNx     #1{\unskip}     \fi
\ifx \showISBNxiii \undefined \def \showISBNxiii  #1{\unskip}     \fi
\ifx \showISSN     \undefined \def \showISSN      #1{\unskip}     \fi
\ifx \showLCCN     \undefined \def \showLCCN      #1{\unskip}     \fi
\ifx \shownote     \undefined \def \shownote      #1{#1}          \fi
\ifx \showarticletitle \undefined \def \showarticletitle #1{#1}   \fi
\ifx \showURL      \undefined \def \showURL       {\relax}        \fi
\providecommand\bibfield[2]{#2}
\providecommand\bibinfo[2]{#2}
\providecommand\natexlab[1]{#1}
\providecommand\showeprint[2][]{arXiv:#2}

\bibitem[\protect\citeauthoryear{Bruneton and Neyret}{Bruneton and
  Neyret}{2012}]%
        {Bruneton2012}
\bibfield{author}{\bibinfo{person}{Eric Bruneton} {and}
  \bibinfo{person}{Fabrice Neyret}.} \bibinfo{year}{2012}\natexlab{}.
\newblock \showarticletitle{{A Survey of Non-linear Pre-filtering Methods for
  Efficient and Accurate Surface Shading}}.
\newblock \bibinfo{journal}{\emph{IEEE Transactions on Visualization and
  Computer Graphics}} \bibinfo{volume}{18}, \bibinfo{number}{2}
  (\bibinfo{year}{2012}), \bibinfo{pages}{242--260}.
\newblock


\bibitem[\protect\citeauthoryear{Burley}{Burley}{2012}]%
        {Burley12}
\bibfield{author}{\bibinfo{person}{Brent Burley}.}
  \bibinfo{year}{2012}\natexlab{}.
\newblock \showarticletitle{{Physically Based Shading at Disney}}. In
  \bibinfo{booktitle}{\emph{SIGGRAPH Courses: Practical Physically Based
  Shading in Film and Game Production}}.
\newblock


\bibitem[\protect\citeauthoryear{Chen, Ling, Gao, Smith, Lehtinen, Jacobson,
  and Fidler}{Chen et~al\mbox{.}}{2019}]%
        {Chen2019}
\bibfield{author}{\bibinfo{person}{Wenzheng Chen}, \bibinfo{person}{Huan Ling},
  \bibinfo{person}{Jun Gao}, \bibinfo{person}{Edward Smith},
  \bibinfo{person}{Jaakko Lehtinen}, \bibinfo{person}{Alec Jacobson}, {and}
  \bibinfo{person}{Sanja Fidler}.} \bibinfo{year}{2019}\natexlab{}.
\newblock \showarticletitle{Learning to Predict 3D Objects with an
  Interpolation-based Differentiable Renderer}.
\newblock In \bibinfo{booktitle}{\emph{Advances in Neural Information
  Processing Systems 32}}. \bibinfo{pages}{9609--9619}.
\newblock


\bibitem[\protect\citeauthoryear{Chen, Tagliasacchi, and Zhang}{Chen
  et~al\mbox{.}}{2020}]%
        {Chen2020}
\bibfield{author}{\bibinfo{person}{Zhiqin Chen}, \bibinfo{person}{Andrea
  Tagliasacchi}, {and} \bibinfo{person}{Hao Zhang}.}
  \bibinfo{year}{2020}\natexlab{}.
\newblock \showarticletitle{{BSP-Net: Generating Compact Meshes via Binary
  Space Partitioning}}.
\newblock \bibinfo{journal}{\emph{Proceedings of IEEE Conference on Computer
  Vision and Pattern Recognition (CVPR)}} (\bibinfo{year}{2020}).
\newblock


\bibitem[\protect\citeauthoryear{Cohen, Varshney, Manocha, Turk, Weber,
  Agarwal, Brooks, and Wright}{Cohen et~al\mbox{.}}{1996}]%
        {Cohen1996}
\bibfield{author}{\bibinfo{person}{Jonathan Cohen}, \bibinfo{person}{Amitabh
  Varshney}, \bibinfo{person}{Dinesh Manocha}, \bibinfo{person}{Greg Turk},
  \bibinfo{person}{Hans Weber}, \bibinfo{person}{Pankaj Agarwal},
  \bibinfo{person}{Frederick Brooks}, {and} \bibinfo{person}{William Wright}.}
  \bibinfo{year}{1996}\natexlab{}.
\newblock \showarticletitle{Simplification Envelopes}. In
  \bibinfo{booktitle}{\emph{Proceedings of the 23rd Annual Conference on
  Computer Graphics and Interactive Techniques}}
  \emph{(\bibinfo{series}{SIGGRAPH '96})}. \bibinfo{pages}{119--128}.
\newblock


\bibitem[\protect\citeauthoryear{Cook, Halstead, Planck, and Ryu}{Cook
  et~al\mbox{.}}{2007}]%
        {Cook2007}
\bibfield{author}{\bibinfo{person}{Robert~L. Cook}, \bibinfo{person}{John
  Halstead}, \bibinfo{person}{Maxwell Planck}, {and} \bibinfo{person}{David
  Ryu}.} \bibinfo{year}{2007}\natexlab{}.
\newblock \showarticletitle{{Stochastic Simplification of Aggregate Detail}}.
\newblock \bibinfo{journal}{\emph{ACM Trans. Graph.}} \bibinfo{volume}{26},
  \bibinfo{number}{3} (\bibinfo{year}{2007}).
\newblock


\bibitem[\protect\citeauthoryear{Corsini, Larabi, Lavou{\'e}, Petr{\'i}k,
  V{\'a}sa, and Wang}{Corsini et~al\mbox{.}}{2013}]%
        {Corsini2013}
\bibfield{author}{\bibinfo{person}{Massimiliano Corsini},
  \bibinfo{person}{Mohamed-Chaker Larabi}, \bibinfo{person}{Guillaume
  Lavou{\'e}}, \bibinfo{person}{Oldrich Petr{\'i}k}, \bibinfo{person}{Libor
  V{\'a}sa}, {and} \bibinfo{person}{Kai Wang}.}
  \bibinfo{year}{2013}\natexlab{}.
\newblock \showarticletitle{{Perceptual metrics for static and dynamic triangle
  meshes}}.
\newblock \bibinfo{journal}{\emph{{Computer Graphics Forum}}}
  \bibinfo{volume}{32}, \bibinfo{number}{1} (\bibinfo{year}{2013}),
  \bibinfo{pages}{101--125}.
\newblock


\bibitem[\protect\citeauthoryear{D\'{e}coret, Durand, Sillion, and
  Dorsey}{D\'{e}coret et~al\mbox{.}}{2003}]%
        {Decoret2003}
\bibfield{author}{\bibinfo{person}{Xavier D\'{e}coret},
  \bibinfo{person}{Fr\'{e}do Durand}, \bibinfo{person}{Fran\c{c}ois Sillion},
  {and} \bibinfo{person}{Julie Dorsey}.} \bibinfo{year}{2003}\natexlab{}.
\newblock \showarticletitle{Billboard clouds for extreme model simplification}.
\newblock \bibinfo{journal}{\emph{ACM Trans. Graph.}} \bibinfo{volume}{22},
  \bibinfo{number}{3} (\bibinfo{year}{2003}).
\newblock


\bibitem[\protect\citeauthoryear{Dong}{Dong}{2019}]%
        {Dong2019}
\bibfield{author}{\bibinfo{person}{Yue Dong}.} \bibinfo{year}{2019}\natexlab{}.
\newblock \showarticletitle{{Deep appearance modeling: A survey}}.
\newblock \bibinfo{journal}{\emph{Visual Informatics}} \bibinfo{volume}{3},
  \bibinfo{number}{2} (\bibinfo{year}{2019}), \bibinfo{pages}{59--68}.
\newblock


\bibitem[\protect\citeauthoryear{Dupuy, Heitz, Iehl, Poulin, Neyret, and
  Ostromoukhov}{Dupuy et~al\mbox{.}}{2013}]%
        {Dupuy2013}
\bibfield{author}{\bibinfo{person}{Jonathan Dupuy}, \bibinfo{person}{Eric
  Heitz}, \bibinfo{person}{Jean-Claude Iehl}, \bibinfo{person}{Pierre Poulin},
  \bibinfo{person}{Fabrice Neyret}, {and} \bibinfo{person}{Victor
  Ostromoukhov}.} \bibinfo{year}{2013}\natexlab{}.
\newblock \showarticletitle{{Linear Efficient Antialiased Displacement and
  Reflectance Mapping}}.
\newblock \bibinfo{journal}{\emph{ACM Trans. Graph.}} \bibinfo{volume}{32},
  \bibinfo{number}{6}, Article \bibinfo{articleno}{211} (\bibinfo{year}{2013}).
\newblock


\bibitem[\protect\citeauthoryear{{Epic Games}}{{Epic Games}}{2018}]%
        {Paragon2018}
\bibfield{author}{\bibinfo{person}{{Epic Games}}.}
  \bibinfo{year}{2018}\natexlab{}.
\newblock \bibinfo{title}{Epic Games Paragon Assets}.
\newblock
\newblock
\newblock
\shownote{\url{https://www.unrealengine.com/en-US/paragon}.}


\bibitem[\protect\citeauthoryear{{Epic Games}}{{Epic Games}}{2020}]%
        {Nanite2020}
\bibfield{author}{\bibinfo{person}{{Epic Games}}.}
  \bibinfo{year}{2020}\natexlab{}.
\newblock \bibinfo{title}{Unreal Engine 5: Nanite}.
\newblock
\newblock
\newblock
\shownote{\url{https://www.unrealengine.com/en-US/blog/a-first-look-at-unreal-engine-5}.}


\bibitem[\protect\citeauthoryear{Everitt}{Everitt}{2001}]%
        {Everitt2001}
\bibfield{author}{\bibinfo{person}{Cass Everitt}.}
  \bibinfo{year}{2001}\natexlab{}.
\newblock \bibinfo{title}{{Interactive Order-Independent Transparency}}.
\newblock
\newblock


\bibitem[\protect\citeauthoryear{Fournier}{Fournier}{1992}]%
        {Fournier1992}
\bibfield{author}{\bibinfo{person}{Alain Fournier}.}
  \bibinfo{year}{1992}\natexlab{}.
\newblock \bibinfo{booktitle}{\emph{{Filtering Normal Maps and Creating
  Multiple Surfaces}}}.
\newblock \bibinfo{type}{{T}echnical {R}eport}.
\newblock


\bibitem[\protect\citeauthoryear{Gao, Li, Dong, Peers, Xu, and Tong}{Gao
  et~al\mbox{.}}{2019}]%
        {Gao2019}
\bibfield{author}{\bibinfo{person}{Duan Gao}, \bibinfo{person}{Xiao Li},
  \bibinfo{person}{Yue Dong}, \bibinfo{person}{Pieter Peers},
  \bibinfo{person}{Kun Xu}, {and} \bibinfo{person}{Xin Tong}.}
  \bibinfo{year}{2019}\natexlab{}.
\newblock \showarticletitle{{Deep Inverse Rendering for High-Resolution SVBRDF
  Estimation from an Arbitrary Number of Images}}.
\newblock \bibinfo{journal}{\emph{ACM Trans. Graph.}} \bibinfo{volume}{38},
  \bibinfo{number}{4}, Article \bibinfo{articleno}{134} (\bibinfo{year}{2019}).
\newblock


\bibitem[\protect\citeauthoryear{Gardner, Tchou, Hawkins, and Debevec}{Gardner
  et~al\mbox{.}}{2003}]%
        {Gardner2003}
\bibfield{author}{\bibinfo{person}{Andrew Gardner}, \bibinfo{person}{Chris
  Tchou}, \bibinfo{person}{Tim Hawkins}, {and} \bibinfo{person}{Paul Debevec}.}
  \bibinfo{year}{2003}\natexlab{}.
\newblock \showarticletitle{{Linear Light Source Reflectometry}}.
\newblock \bibinfo{journal}{\emph{ACM Trans. Graph.}} \bibinfo{volume}{22},
  \bibinfo{number}{3} (\bibinfo{year}{2003}), \bibinfo{pages}{749--758}.
\newblock


\bibitem[\protect\citeauthoryear{Garland and Heckbert}{Garland and
  Heckbert}{1997}]%
        {Garland97}
\bibfield{author}{\bibinfo{person}{Michael Garland} {and}
  \bibinfo{person}{Paul~S. Heckbert}.} \bibinfo{year}{1997}\natexlab{}.
\newblock \showarticletitle{{Surface Simplification Using Quadric Error
  Metrics}}. In \bibinfo{booktitle}{\emph{SIGGRAPH '97: Proceedings of the 24th
  Annual Conference on Computer Graphics and Interactive Techniques}}.
  \bibinfo{pages}{209--216}.
\newblock


\bibitem[\protect\citeauthoryear{Ghosh, Chen, Peers, Wilson, and Debevec}{Ghosh
  et~al\mbox{.}}{2009}]%
        {Ghosh2009}
\bibfield{author}{\bibinfo{person}{Abhijeet Ghosh}, \bibinfo{person}{Tongbo
  Chen}, \bibinfo{person}{Pieter Peers}, \bibinfo{person}{Cyrus~A. Wilson},
  {and} \bibinfo{person}{Paul Debevec}.} \bibinfo{year}{2009}\natexlab{}.
\newblock \showarticletitle{{Estimating Specular Roughness and Anisotropy from
  Second Order Spherical Gradient Illumination}}.
\newblock \bibinfo{journal}{\emph{Computer Graphics Forum}}
  \bibinfo{volume}{28}, \bibinfo{number}{4} (\bibinfo{year}{2009}),
  \bibinfo{pages}{1161--1170}.
\newblock


\bibitem[\protect\citeauthoryear{Guarnera, Guarnera, Ghosh, Denk, and
  Glencross}{Guarnera et~al\mbox{.}}{2016}]%
        {Guarnera2016}
\bibfield{author}{\bibinfo{person}{D. Guarnera}, \bibinfo{person}{G.~C.
  Guarnera}, \bibinfo{person}{A. Ghosh}, \bibinfo{person}{C. Denk}, {and}
  \bibinfo{person}{M. Glencross}.} \bibinfo{year}{2016}\natexlab{}.
\newblock \showarticletitle{{BRDF Representation and Acquisition}}. In
  \bibinfo{booktitle}{\emph{Proceedings of the 37th Annual Conference of the
  European Association for Computer Graphics: State of the Art Reports}}.
  \bibinfo{pages}{625--650}.
\newblock


\bibitem[\protect\citeauthoryear{Guo, Smith, Ha\v{s}an, Sunkavalli, and
  Zhao}{Guo et~al\mbox{.}}{2020}]%
        {Guo2020}
\bibfield{author}{\bibinfo{person}{Yu Guo}, \bibinfo{person}{Cameron Smith},
  \bibinfo{person}{Milo\v{s} Ha\v{s}an}, \bibinfo{person}{Kalyan Sunkavalli},
  {and} \bibinfo{person}{Shuang Zhao}.} \bibinfo{year}{2020}\natexlab{}.
\newblock \showarticletitle{{MaterialGAN: Reflectance Capture Using a
  Generative SVBRDF Model}}.
\newblock \bibinfo{journal}{\emph{ACM Trans. Graph.}} \bibinfo{volume}{39},
  \bibinfo{number}{6}, Article \bibinfo{articleno}{254} (\bibinfo{year}{2020}).
\newblock


\bibitem[\protect\citeauthoryear{Han, Sun, Ramamoorthi, and Grinspun}{Han
  et~al\mbox{.}}{2007}]%
        {Han2007}
\bibfield{author}{\bibinfo{person}{Charles Han}, \bibinfo{person}{Bo Sun},
  \bibinfo{person}{Ravi Ramamoorthi}, {and} \bibinfo{person}{Eitan Grinspun}.}
  \bibinfo{year}{2007}\natexlab{}.
\newblock \showarticletitle{{Frequency Domain Normal Map Filtering}}. In
  \bibinfo{booktitle}{\emph{ACM SIGGRAPH 2007 Papers}}.
  \bibinfo{pages}{28--40}.
\newblock


\bibitem[\protect\citeauthoryear{James and Twigg}{James and Twigg}{2005}]%
        {James2005}
\bibfield{author}{\bibinfo{person}{Doug James} {and}
  \bibinfo{person}{Christopher Twigg}.} \bibinfo{year}{2005}\natexlab{}.
\newblock \showarticletitle{Skinning mesh animations}.
\newblock \bibinfo{journal}{\emph{ACM Trans. Graph.}} \bibinfo{volume}{24},
  \bibinfo{number}{3} (\bibinfo{year}{2005}).
\newblock


\bibitem[\protect\citeauthoryear{Jatavallabhula, Smith, Lafleche, {Fuji Tsang},
  Rozantsev, Chen, Xiang, Lebaredian, and Fidler}{Jatavallabhula
  et~al\mbox{.}}{2019}]%
        {Murthy2019}
\bibfield{author}{\bibinfo{person}{{Krishna Murthy} Jatavallabhula},
  \bibinfo{person}{Edward Smith}, \bibinfo{person}{Jean-Francois Lafleche},
  \bibinfo{person}{Clement {Fuji Tsang}}, \bibinfo{person}{Artem Rozantsev},
  \bibinfo{person}{Wenzheng Chen}, \bibinfo{person}{Tommy Xiang},
  \bibinfo{person}{Rev Lebaredian}, {and} \bibinfo{person}{Sanja Fidler}.}
  \bibinfo{year}{2019}\natexlab{}.
\newblock \showarticletitle{{Kaolin: A PyTorch Library for Accelerating 3D Deep
  Learning Research}}.
\newblock \bibinfo{journal}{\emph{arXiv:1911.05063}} (\bibinfo{year}{2019}).
\newblock


\bibitem[\protect\citeauthoryear{Jere\-mias and Quilez}{Jere\-mias and
  Quilez}{2014}]%
        {Jeremias2014}
\bibfield{author}{\bibinfo{person}{Pol Jere\-mias} {and} \bibinfo{person}{Inigo
  Quilez}.} \bibinfo{year}{2014}\natexlab{}.
\newblock \showarticletitle{{Shadertoy: Learn to Create Everything in a
  Fragment Shader}}. In \bibinfo{booktitle}{\emph{SIGGRAPH Asia 2014 Courses}}.
  Article \bibinfo{articleno}{18}.
\newblock


\bibitem[\protect\citeauthoryear{Karis}{Karis}{2013}]%
        {Karis2013}
\bibfield{author}{\bibinfo{person}{Brian Karis}.}
  \bibinfo{year}{2013}\natexlab{}.
\newblock \showarticletitle{Real Shading in Unreal Engine 4}.
\newblock \bibinfo{journal}{\emph{SIGGRAPH 2013 Course: Physically Based
  Shading in Theory and Practice}} (\bibinfo{year}{2013}).
\newblock


\bibitem[\protect\citeauthoryear{Kingma and Ba}{Kingma and Ba}{2015}]%
        {Kingma2014}
\bibfield{author}{\bibinfo{person}{Diederik~P. Kingma} {and}
  \bibinfo{person}{Jimmy Ba}.} \bibinfo{year}{2015}\natexlab{}.
\newblock \showarticletitle{{Adam: A Method for Stochastic Optimization}}. In
  \bibinfo{booktitle}{\emph{Proceedings of the 3rd International Conference for
  Learning Representations}}.
\newblock


\bibitem[\protect\citeauthoryear{Lab}{Lab}{2020}]%
        {CMU2020}
\bibfield{author}{\bibinfo{person}{CMU~Graphics Lab}.}
  \bibinfo{year}{2020}\natexlab{}.
\newblock \bibinfo{title}{{CMU Graphics Lab Motion Capture Database}}.
\newblock
\newblock
\newblock
\shownote{http://mocap.cs.cmu.edu/.}


\bibitem[\protect\citeauthoryear{Lagarde and de~Rousiers}{Lagarde and
  de~Rousiers}{2014}]%
        {Lagarde2014}
\bibfield{author}{\bibinfo{person}{Sebastien Lagarde} {and}
  \bibinfo{person}{Charles de Rousiers}.} \bibinfo{year}{2014}\natexlab{}.
\newblock \showarticletitle{Moving Frostbite to Physically Based Rendering
  3.0}.
\newblock \bibinfo{journal}{\emph{SIGGRAPH 2014 Course: Physically Based
  Shading in Theory and Practice}} (\bibinfo{year}{2014}).
\newblock


\bibitem[\protect\citeauthoryear{Laine, Hellsten, Karras, Seol, Lehtinen, and
  Aila}{Laine et~al\mbox{.}}{2020}]%
        {Laine2020}
\bibfield{author}{\bibinfo{person}{Samuli Laine}, \bibinfo{person}{Janne
  Hellsten}, \bibinfo{person}{Tero Karras}, \bibinfo{person}{Yeongho Seol},
  \bibinfo{person}{Jaakko Lehtinen}, {and} \bibinfo{person}{Timo Aila}.}
  \bibinfo{year}{2020}\natexlab{}.
\newblock \showarticletitle{Modular Primitives for High-Performance
  Differentiable Rendering}.
\newblock \bibinfo{journal}{\emph{ACM Transactions on Graphics}}
  \bibinfo{volume}{39}, \bibinfo{number}{6}, Article \bibinfo{articleno}{194}
  (\bibinfo{year}{2020}).
\newblock


\bibitem[\protect\citeauthoryear{Ledig, Theis, Huszar, Caballero, Cunningham,
  Acosta, Aitken, Tejani, Totz, Wang, and Shi}{Ledig et~al\mbox{.}}{2017}]%
        {Ledig2016}
\bibfield{author}{\bibinfo{person}{Christian Ledig}, \bibinfo{person}{Lucas
  Theis}, \bibinfo{person}{Ferenc Huszar}, \bibinfo{person}{Jose Caballero},
  \bibinfo{person}{Andrew Cunningham}, \bibinfo{person}{Alejandro Acosta},
  \bibinfo{person}{Andrew Aitken}, \bibinfo{person}{Alykhan Tejani},
  \bibinfo{person}{Johannes Totz}, \bibinfo{person}{Zehan Wang}, {and}
  \bibinfo{person}{Wenzhe Shi}.} \bibinfo{year}{2017}\natexlab{}.
\newblock \showarticletitle{Photo-Realistic Single Image Super-Resolution Using
  a Generative Adversarial Network}. In \bibinfo{booktitle}{\emph{Proceedings
  of the IEEE Conference on Computer Vision and Pattern Recognition (CVPR)}}.
\newblock


\bibitem[\protect\citeauthoryear{Lensch, Kautz, Goesele, Heidrich, and
  Seidel}{Lensch et~al\mbox{.}}{2003}]%
        {Lensch2003}
\bibfield{author}{\bibinfo{person}{Hendrik P.~A. Lensch}, \bibinfo{person}{Jan
  Kautz}, \bibinfo{person}{Michael Goesele}, \bibinfo{person}{Wolfgang
  Heidrich}, {and} \bibinfo{person}{Hans-Peter Seidel}.}
  \bibinfo{year}{2003}\natexlab{}.
\newblock \showarticletitle{{Image-Based Reconstruction of Spatial Appearance
  and Geometric Detail}}.
\newblock \bibinfo{journal}{\emph{ACM Trans. Graph.}} \bibinfo{volume}{22},
  \bibinfo{number}{2} (\bibinfo{year}{2003}), \bibinfo{pages}{234--257}.
\newblock


\bibitem[\protect\citeauthoryear{Li, Aittala, Durand, and Lehtinen}{Li
  et~al\mbox{.}}{2018}]%
        {Li2018}
\bibfield{author}{\bibinfo{person}{Tzu-Mao Li}, \bibinfo{person}{Miika
  Aittala}, \bibinfo{person}{Fr{\'e}do Durand}, {and} \bibinfo{person}{Jaakko
  Lehtinen}.} \bibinfo{year}{2018}\natexlab{}.
\newblock \showarticletitle{{Differentiable Monte Carlo Ray Tracing through
  Edge Sampling}}.
\newblock \bibinfo{journal}{\emph{ACM Trans. Graph. (Proc. SIGGRAPH Asia)}}
  \bibinfo{volume}{37}, \bibinfo{number}{6} (\bibinfo{year}{2018}),
  \bibinfo{pages}{222:1--222:11}.
\newblock


\bibitem[\protect\citeauthoryear{Lindstrom and Turk}{Lindstrom and
  Turk}{2000}]%
        {Lindstrom2000}
\bibfield{author}{\bibinfo{person}{Peter Lindstrom} {and} \bibinfo{person}{Greg
  Turk}.} \bibinfo{year}{2000}\natexlab{}.
\newblock \showarticletitle{Image-Driven Simplification}.
\newblock \bibinfo{journal}{\emph{ACM Transactions on Graphics}}
  \bibinfo{volume}{19}, \bibinfo{number}{3} (\bibinfo{year}{2000}),
  \bibinfo{pages}{204--241}.
\newblock


\bibitem[\protect\citeauthoryear{Loubet and Neyret}{Loubet and Neyret}{2017}]%
        {Loubet2017}
\bibfield{author}{\bibinfo{person}{Guillaume Loubet} {and}
  \bibinfo{person}{Fabrice Neyret}.} \bibinfo{year}{2017}\natexlab{}.
\newblock \showarticletitle{{Hybrid mesh-volume LoDs for all-scale
  pre-filtering of complex 3D assets}}.
\newblock \bibinfo{journal}{\emph{Computer Graphics Forum}}
  \bibinfo{volume}{36}, \bibinfo{number}{2} (\bibinfo{year}{2017}),
  \bibinfo{pages}{431--442}.
\newblock


\bibitem[\protect\citeauthoryear{Low and Tan}{Low and Tan}{1997}]%
        {Low1997}
\bibfield{author}{\bibinfo{person}{Kok-Lim Low} {and}
  \bibinfo{person}{Tiow-Seng Tan}.} \bibinfo{year}{1997}\natexlab{}.
\newblock \showarticletitle{{Model Simplification Using Vertex-Clustering}}. In
  \bibinfo{booktitle}{\emph{Proceedings of the 1997 Symposium on Interactive 3D
  Graphics}}. \bibinfo{pages}{75--82}.
\newblock


\bibitem[\protect\citeauthoryear{Luebke and Erikson}{Luebke and
  Erikson}{1997}]%
        {Luebke1997}
\bibfield{author}{\bibinfo{person}{David Luebke} {and} \bibinfo{person}{Carl
  Erikson}.} \bibinfo{year}{1997}\natexlab{}.
\newblock \showarticletitle{{View-Dependent Simplification of Arbitrary
  Polygonal Environments}}. In \bibinfo{booktitle}{\emph{SIGGRAPH '97:
  Proceedings of the 24th Annual Conference on Computer Graphics and
  Interactive Techniques}}. \bibinfo{pages}{199--208}.
\newblock


\bibitem[\protect\citeauthoryear{Luebke, Watson, Cohen, Reddy, and
  Varshney}{Luebke et~al\mbox{.}}{2002}]%
        {Luebke2002}
\bibfield{author}{\bibinfo{person}{David Luebke}, \bibinfo{person}{Benjamin
  Watson}, \bibinfo{person}{Jonathan~D. Cohen}, \bibinfo{person}{Martin Reddy},
  {and} \bibinfo{person}{Amitabh Varshney}.} \bibinfo{year}{2002}\natexlab{}.
\newblock \bibinfo{booktitle}{\emph{{Level of Detail for 3D Graphics}}}.
\newblock \bibinfo{publisher}{Elsevier Science Inc.}, \bibinfo{address}{USA}.
\newblock
\showISBNx{1558608389}


\bibitem[\protect\citeauthoryear{Mikkelsen}{Mikkelsen}{2008}]%
        {Mikkelsen08}
\bibfield{author}{\bibinfo{person}{Morten Mikkelsen}.}
  \bibinfo{year}{2008}\natexlab{}.
\newblock \bibinfo{title}{Simulation of Wrinkled Surfaces Revisited}.
\newblock
\newblock


\bibitem[\protect\citeauthoryear{Mildenhall, Srinivasan, Tancik, Barron,
  Ramamoorthi, and Ng}{Mildenhall et~al\mbox{.}}{2020}]%
        {Mildenhall2020}
\bibfield{author}{\bibinfo{person}{Ben Mildenhall}, \bibinfo{person}{Pratul~P.
  Srinivasan}, \bibinfo{person}{Matthew Tancik}, \bibinfo{person}{Jonathan~T.
  Barron}, \bibinfo{person}{Ravi Ramamoorthi}, {and} \bibinfo{person}{Ren Ng}.}
  \bibinfo{year}{2020}\natexlab{}.
\newblock \showarticletitle{{NeRF: Representing Scenes as Neural Radiance
  Fields for View Synthesis}}. In \bibinfo{booktitle}{\emph{ECCV}}.
\newblock


\bibitem[\protect\citeauthoryear{Morrical, Tremblay, Birchfield, and
  Wald}{Morrical et~al\mbox{.}}{2020}]%
        {Morrical2020}
\bibfield{author}{\bibinfo{person}{Nathan Morrical}, \bibinfo{person}{Jonathan
  Tremblay}, \bibinfo{person}{Stan Birchfield}, {and} \bibinfo{person}{Ingo
  Wald}.} \bibinfo{year}{2020}\natexlab{}.
\newblock \bibinfo{title}{{ViSII}: VIrtual Scene Imaging Interface}.
\newblock
\newblock
\newblock
\shownote{\url{ https://github.com/owl-project/ViSII/}.}


\bibitem[\protect\citeauthoryear{M\"{u}ller, Mcwilliams, Rousselle, Gross, and
  Nov\'{a}k}{M\"{u}ller et~al\mbox{.}}{2019}]%
        {Muller2019}
\bibfield{author}{\bibinfo{person}{Thomas M\"{u}ller}, \bibinfo{person}{Brian
  Mcwilliams}, \bibinfo{person}{Fabrice Rousselle}, \bibinfo{person}{Markus
  Gross}, {and} \bibinfo{person}{Jan Nov\'{a}k}.}
  \bibinfo{year}{2019}\natexlab{}.
\newblock \showarticletitle{Neural Importance Sampling}.
\newblock \bibinfo{journal}{\emph{ACM Trans. Graph.}} \bibinfo{volume}{38},
  \bibinfo{number}{5}, Article \bibinfo{articleno}{145} (\bibinfo{year}{2019}).
\newblock


\bibitem[\protect\citeauthoryear{Munkberg and Hasselgren}{Munkberg and
  Hasselgren}{2020}]%
        {Munkberg2020}
\bibfield{author}{\bibinfo{person}{Jacob Munkberg} {and} \bibinfo{person}{Jon
  Hasselgren}.} \bibinfo{year}{2020}\natexlab{}.
\newblock \showarticletitle{Neural Denoising with Layer Embeddings}.
\newblock \bibinfo{journal}{\emph{Computer Graphics Forum}}
  \bibinfo{volume}{39} (\bibinfo{year}{2020}), \bibinfo{pages}{1--12}.
\newblock


\bibitem[\protect\citeauthoryear{Nam, Lee, Gutierrez, and Kim}{Nam
  et~al\mbox{.}}{2018}]%
        {Nam2018}
\bibfield{author}{\bibinfo{person}{Giljoo Nam}, \bibinfo{person}{Joo~Ho Lee},
  \bibinfo{person}{Diego Gutierrez}, {and} \bibinfo{person}{Min~H. Kim}.}
  \bibinfo{year}{2018}\natexlab{}.
\newblock \showarticletitle{{Practical SVBRDF Acquisition of 3D Objects with
  Unstructured Flash Photography}}.
\newblock \bibinfo{journal}{\emph{ACM Trans. Graph.}} \bibinfo{volume}{37},
  \bibinfo{number}{6}, Article \bibinfo{articleno}{267} (\bibinfo{year}{2018}).
\newblock


\bibitem[\protect\citeauthoryear{Nimier-David, Speierer, Ruiz, and
  Jakob}{Nimier-David et~al\mbox{.}}{2020}]%
        {Nimier2020}
\bibfield{author}{\bibinfo{person}{Merlin Nimier-David},
  \bibinfo{person}{S\'{e}bastien Speierer}, \bibinfo{person}{Beno\^{\i}t Ruiz},
  {and} \bibinfo{person}{Wenzel Jakob}.} \bibinfo{year}{2020}\natexlab{}.
\newblock \showarticletitle{{Radiative Backpropagation: An Adjoint Method for
  Lightning-Fast Differentiable Rendering}}.
\newblock \bibinfo{journal}{\emph{ACM Trans. Graph.}} \bibinfo{volume}{39},
  \bibinfo{number}{4}, Article \bibinfo{articleno}{146} (\bibinfo{year}{2020}).
\newblock


\bibitem[\protect\citeauthoryear{Nimier-David, Vicini, Zeltner, and
  Jakob}{Nimier-David et~al\mbox{.}}{2019}]%
        {Nimier2019}
\bibfield{author}{\bibinfo{person}{Merlin Nimier-David}, \bibinfo{person}{Delio
  Vicini}, \bibinfo{person}{Tizian Zeltner}, {and} \bibinfo{person}{Wenzel
  Jakob}.} \bibinfo{year}{2019}\natexlab{}.
\newblock \showarticletitle{{Mitsuba 2: A Retargetable Forward and Inverse
  Renderer}}.
\newblock \bibinfo{journal}{\emph{ACM Trans. Graph.}} \bibinfo{volume}{38},
  \bibinfo{number}{6}, Article \bibinfo{articleno}{203} (\bibinfo{year}{2019}).
\newblock


\bibitem[\protect\citeauthoryear{Olano and Baker}{Olano and Baker}{2010}]%
        {Olano2010}
\bibfield{author}{\bibinfo{person}{Marc Olano} {and} \bibinfo{person}{Dan
  Baker}.} \bibinfo{year}{2010}\natexlab{}.
\newblock \showarticletitle{{LEAN Mapping}}. In
  \bibinfo{booktitle}{\emph{Proceedings of the 2010 ACM SIGGRAPH Symposium on
  Interactive 3D Graphics and Games}}. \bibinfo{pages}{181--188}.
\newblock


\bibitem[\protect\citeauthoryear{Paszke, Gross, Chintala, Chanan, Yang, DeVito,
  Lin, Desmaison, Antiga, and Lerer}{Paszke et~al\mbox{.}}{2017}]%
        {Paszke2017}
\bibfield{author}{\bibinfo{person}{Adam Paszke}, \bibinfo{person}{Sam Gross},
  \bibinfo{person}{Soumith Chintala}, \bibinfo{person}{Gregory Chanan},
  \bibinfo{person}{Edward Yang}, \bibinfo{person}{Zachary DeVito},
  \bibinfo{person}{Zeming Lin}, \bibinfo{person}{Alban Desmaison},
  \bibinfo{person}{Luca Antiga}, {and} \bibinfo{person}{Adam Lerer}.}
  \bibinfo{year}{2017}\natexlab{}.
\newblock \showarticletitle{{Automatic differentiation in PyTorch}}. In
  \bibinfo{booktitle}{\emph{NIPS-W}}.
\newblock


\bibitem[\protect\citeauthoryear{{Pixar Animation Studios}}{{Pixar Animation
  Studios}}{2016}]%
        {USD2016}
\bibfield{author}{\bibinfo{person}{{Pixar Animation Studios}}.}
  \bibinfo{year}{2016}\natexlab{}.
\newblock \bibinfo{title}{{Universal Scene Description Website}}.
\newblock
\newblock
\newblock
\shownote{http://www.openusd.org.}


\bibitem[\protect\citeauthoryear{Radford, Metz, and Chintala}{Radford
  et~al\mbox{.}}{2015}]%
        {Radford2015}
\bibfield{author}{\bibinfo{person}{Alec Radford}, \bibinfo{person}{Luke Metz},
  {and} \bibinfo{person}{Soumith Chintala}.} \bibinfo{year}{2015}\natexlab{}.
\newblock \showarticletitle{{Unsupervised representation learning with deep
  convolutional generative adversarial networks}}.
\newblock \bibinfo{journal}{\emph{arXiv:1511.06434}} (\bibinfo{year}{2015}).
\newblock


\bibitem[\protect\citeauthoryear{Ravi, Reizenstein, Novotny, Gordon, Lo,
  Johnson, and Gkioxari}{Ravi et~al\mbox{.}}{2020}]%
        {Ravi2020}
\bibfield{author}{\bibinfo{person}{Nikhila Ravi}, \bibinfo{person}{Jeremy
  Reizenstein}, \bibinfo{person}{David Novotny}, \bibinfo{person}{Taylor
  Gordon}, \bibinfo{person}{Wan-Yen Lo}, \bibinfo{person}{Justin Johnson},
  {and} \bibinfo{person}{Georgia Gkioxari}.} \bibinfo{year}{2020}\natexlab{}.
\newblock \showarticletitle{{Accelerating 3D Deep Learning with PyTorch3D}}.
\newblock \bibinfo{journal}{\emph{arXiv:2007.08501}} (\bibinfo{year}{2020}).
\newblock


\bibitem[\protect\citeauthoryear{{RenderPeople}}{{RenderPeople}}{2020}]%
        {RenderPeople2020}
\bibfield{author}{\bibinfo{person}{{RenderPeople}}.}
  \bibinfo{year}{2020}\natexlab{}.
\newblock \bibinfo{title}{RenderPeople}.
\newblock
\newblock
\newblock
\shownote{\url{https://renderpeople.com/3d-people/}.}


\bibitem[\protect\citeauthoryear{Schroeder}{Schroeder}{1997}]%
        {Schroeder1997}
\bibfield{author}{\bibinfo{person}{William Schroeder}.}
  \bibinfo{year}{1997}\natexlab{}.
\newblock \showarticletitle{{A topology modifying progressive decimation
  algorithm}}. In \bibinfo{booktitle}{\emph{In VIS '97: 8th conference on
  Visualization}}. \bibinfo{pages}{205--212}.
\newblock


\bibitem[\protect\citeauthoryear{Smithsonian}{Smithsonian}{2018}]%
        {Smithsonian2020}
\bibfield{author}{\bibinfo{person}{Smithsonian}.}
  \bibinfo{year}{2018}\natexlab{}.
\newblock \bibinfo{title}{{Smithsonian 3D Digitization}}.
\newblock
\newblock
\newblock
\shownote{https://3d.si.edu/.}


\bibitem[\protect\citeauthoryear{Sorkine}{Sorkine}{2005}]%
        {Sorkine2005}
\bibfield{author}{\bibinfo{person}{Olga Sorkine}.}
  \bibinfo{year}{2005}\natexlab{}.
\newblock \showarticletitle{Laplacian Mesh Processing}. In
  \bibinfo{booktitle}{\emph{Eurographics 2005 - State of the Art Reports}}.
\newblock


\bibitem[\protect\citeauthoryear{Stokes, Anderson, Chandrasekar, and
  Motta}{Stokes et~al\mbox{.}}{1996}]%
        {srgb96}
\bibfield{author}{\bibinfo{person}{Michael Stokes}, \bibinfo{person}{Matthew
  Anderson}, \bibinfo{person}{Srinivasan Chandrasekar}, {and}
  \bibinfo{person}{Ricardo Motta}.} \bibinfo{year}{1996}\natexlab{}.
\newblock \bibinfo{title}{{A Standard Default Color Space for the Internet -
  sRGB}}.
\newblock
\newblock
\urldef\tempurl%
\url{https://www.w3.org/Graphics/Color/sRGB.html}
\showURL{%
\tempurl}


\bibitem[\protect\citeauthoryear{Sztrajman, K\v{r}iv\'anek, Wilkie, and
  Weyrich}{Sztrajman et~al\mbox{.}}{2017}]%
        {Sztrajman2017}
\bibfield{author}{\bibinfo{person}{Alejandro Sztrajman},
  \bibinfo{person}{Jaroslav K\v{r}iv\'anek}, \bibinfo{person}{Alexander
  Wilkie}, {and} \bibinfo{person}{Tim Weyrich}.}
  \bibinfo{year}{2017}\natexlab{}.
\newblock \showarticletitle{{Image-based Remapping of Material Appearance}}. In
  \bibinfo{booktitle}{\emph{Proc. 5th Workshop on Material Appearance
  Modeling}}. \bibinfo{pages}{5--8}.
\newblock


\bibitem[\protect\citeauthoryear{Thies, Zollh\"{o}fer, and Nie\ss{}ner}{Thies
  et~al\mbox{.}}{2019}]%
        {Thies2019}
\bibfield{author}{\bibinfo{person}{Justus Thies}, \bibinfo{person}{Michael
  Zollh\"{o}fer}, {and} \bibinfo{person}{Matthias Nie\ss{}ner}.}
  \bibinfo{year}{2019}\natexlab{}.
\newblock \showarticletitle{{Deferred Neural Rendering: Image Synthesis Using
  Neural Textures}}.
\newblock \bibinfo{journal}{\emph{ACM Trans. Graph.}} \bibinfo{volume}{38},
  \bibinfo{number}{4}, Article \bibinfo{articleno}{66} (\bibinfo{year}{2019}).
\newblock


\bibitem[\protect\citeauthoryear{Toksvig}{Toksvig}{2005}]%
        {Toksvig2005}
\bibfield{author}{\bibinfo{person}{Michael Toksvig}.}
  \bibinfo{year}{2005}\natexlab{}.
\newblock \showarticletitle{{Mipmapping normal maps}}.
\newblock \bibinfo{journal}{\emph{Journal of Graphics Tools}}
  \bibinfo{volume}{10}, \bibinfo{number}{3} (\bibinfo{year}{2005}),
  \bibinfo{pages}{65--71}.
\newblock


\bibitem[\protect\citeauthoryear{Valentin, Keskin, Pidlypenskyi, Makadia, Sud,
  and Bouaziz}{Valentin et~al\mbox{.}}{2019}]%
        {Valentin2019}
\bibfield{author}{\bibinfo{person}{Julien Valentin}, \bibinfo{person}{Cem
  Keskin}, \bibinfo{person}{Pavel Pidlypenskyi}, \bibinfo{person}{Ameesh
  Makadia}, \bibinfo{person}{Avneesh Sud}, {and} \bibinfo{person}{Sofien
  Bouaziz}.} \bibinfo{year}{2019}\natexlab{}.
\newblock \showarticletitle{{TensorFlow Graphics: Computer Graphics Meets Deep
  Learning}}.
\newblock


\bibitem[\protect\citeauthoryear{{Walt Disney Animation Studios}}{{Walt Disney
  Animation Studios}}{2018}]%
        {Moana2018}
\bibfield{author}{\bibinfo{person}{{Walt Disney Animation Studios}}.}
  \bibinfo{year}{2018}\natexlab{}.
\newblock \bibinfo{title}{Moana Island Scene (v1.1)}.
\newblock
\newblock
\newblock
\shownote{\url{http://technology.disneyanimation.com/islandscene/}.}


\bibitem[\protect\citeauthoryear{Walter, Marschner, Li, and Torrance}{Walter
  et~al\mbox{.}}{2007}]%
        {Walter2007}
\bibfield{author}{\bibinfo{person}{Bruce Walter}, \bibinfo{person}{Stephen~R.
  Marschner}, \bibinfo{person}{Hongsong Li}, {and} \bibinfo{person}{Kenneth~E.
  Torrance}.} \bibinfo{year}{2007}\natexlab{}.
\newblock \showarticletitle{{Microfacet Models for Refraction through Rough
  Surfaces}}. In \bibinfo{booktitle}{\emph{Proceedings of the 18th Eurographics
  Conference on Rendering Techniques}}. \bibinfo{pages}{195–206}.
\newblock


\bibitem[\protect\citeauthoryear{Wang, Zhang, Li, Fu, Liu, and Jiang}{Wang
  et~al\mbox{.}}{2018}]%
        {Wang2018b}
\bibfield{author}{\bibinfo{person}{Nanyang Wang}, \bibinfo{person}{Yinda
  Zhang}, \bibinfo{person}{Zhuwen Li}, \bibinfo{person}{Yanwei Fu},
  \bibinfo{person}{Wei Liu}, {and} \bibinfo{person}{Yu-Gang Jiang}.}
  \bibinfo{year}{2018}\natexlab{}.
\newblock \showarticletitle{{Pixel2Mesh: Generating 3D Mesh Models from Single
  RGB Images}}. In \bibinfo{booktitle}{\emph{ECCV}}.
\newblock


\bibitem[\protect\citeauthoryear{Weinmann and Klein}{Weinmann and
  Klein}{2015}]%
        {Weinmann2015}
\bibfield{author}{\bibinfo{person}{Michael Weinmann} {and}
  \bibinfo{person}{Reinhard Klein}.} \bibinfo{year}{2015}\natexlab{}.
\newblock \showarticletitle{{Advances in Geometry and Reflectance Acquisition
  (Course Notes)}}. In \bibinfo{booktitle}{\emph{SIGGRAPH Asia 2015 Courses}}.
  Article \bibinfo{articleno}{1}.
\newblock


\bibitem[\protect\citeauthoryear{Wu, Zhao, Yan, and Ramamoorthi}{Wu
  et~al\mbox{.}}{2019}]%
        {Wu2019}
\bibfield{author}{\bibinfo{person}{Lifan Wu}, \bibinfo{person}{Shuang Zhao},
  \bibinfo{person}{Ling-Qi Yan}, {and} \bibinfo{person}{Ravi Ramamoorthi}.}
  \bibinfo{year}{2019}\natexlab{}.
\newblock \showarticletitle{{Accurate Appearance Preserving Prefiltering for
  Rendering Displacement-Mapped Surfaces}}.
\newblock \bibinfo{journal}{\emph{ACM Trans. Graph.}} \bibinfo{volume}{38},
  \bibinfo{number}{4}, Article \bibinfo{articleno}{137} (\bibinfo{year}{2019}).
\newblock


\bibitem[\protect\citeauthoryear{Zhang, Chen, Ling, Gao, Zhang, Torralba, and
  Fidler}{Zhang et~al\mbox{.}}{2020}]%
        {Zhang2020image}
\bibfield{author}{\bibinfo{person}{Yuxuan Zhang}, \bibinfo{person}{Wenzheng
  Chen}, \bibinfo{person}{Huan Ling}, \bibinfo{person}{Jun Gao},
  \bibinfo{person}{Yinan Zhang}, \bibinfo{person}{Antonio Torralba}, {and}
  \bibinfo{person}{Sanja Fidler}.} \bibinfo{year}{2020}\natexlab{}.
\newblock \showarticletitle{{Image GANs meet Differentiable Rendering for
  Inverse Graphics and Interpretable 3D Neural Rendering}}.
\newblock \bibinfo{journal}{\emph{arXiv:2010.09125}} (\bibinfo{year}{2020}).
\newblock


\bibitem[\protect\citeauthoryear{Zhao, Wu, Durand, and Ramamoorthi}{Zhao
  et~al\mbox{.}}{2016}]%
        {Zhao2016}
\bibfield{author}{\bibinfo{person}{Shuang Zhao}, \bibinfo{person}{Lifan Wu},
  \bibinfo{person}{Fr\'{e}do Durand}, {and} \bibinfo{person}{Ravi
  Ramamoorthi}.} \bibinfo{year}{2016}\natexlab{}.
\newblock \showarticletitle{{Downsampling Scattering Parameters for Rendering
  Anisotropic Media}}.
\newblock \bibinfo{journal}{\emph{ACM Trans. Graph.}} \bibinfo{volume}{35},
  \bibinfo{number}{6}, Article \bibinfo{articleno}{166} (\bibinfo{year}{2016}).
\newblock


\end{thebibliography}

\appendix


\newcommand{\figBaking}{
\begin{figure*}
	\setlength{\tabcolsep}{1pt}
	\begin{tabular}{ccc}
		\includegraphics[width=0.33\textwidth]{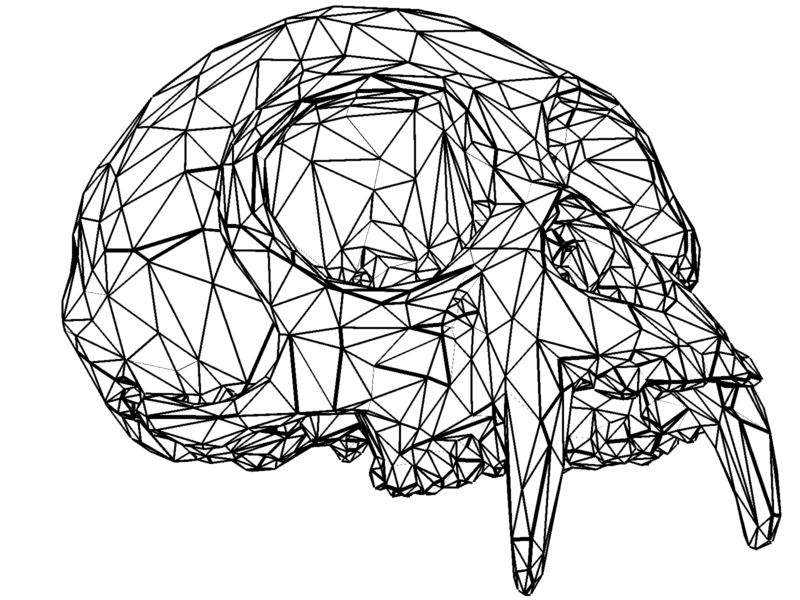} &
		\includegraphics[width=0.33\textwidth]{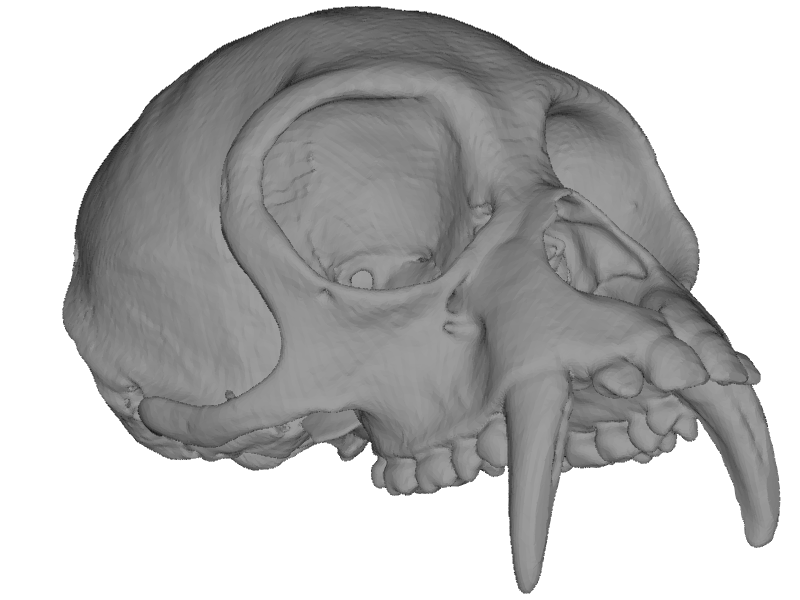} &
		\includegraphics[width=0.33\textwidth]{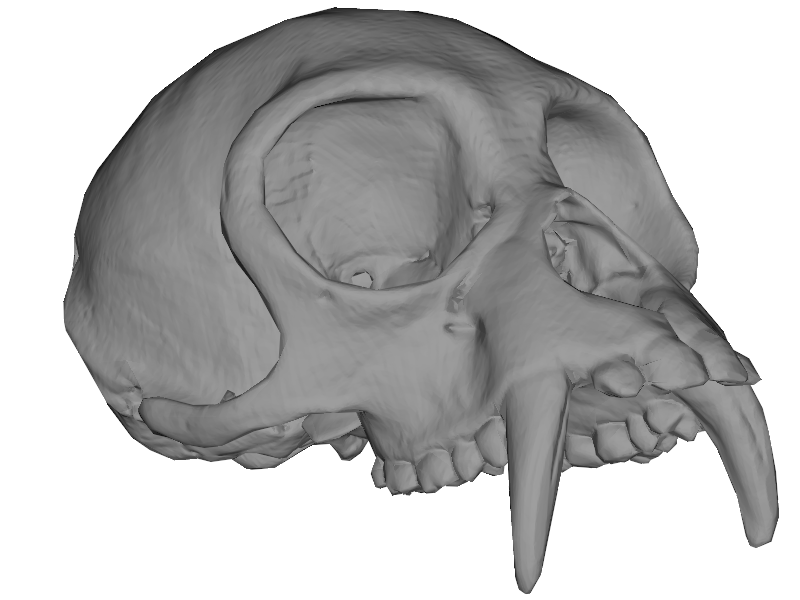} \\
		Input (9k tris) & 
		Reference (735k tris) &
		Simplygon (9k tris) \\
		& & PSNR: 27.07~dB \\
		\includegraphics[width=0.33\textwidth]{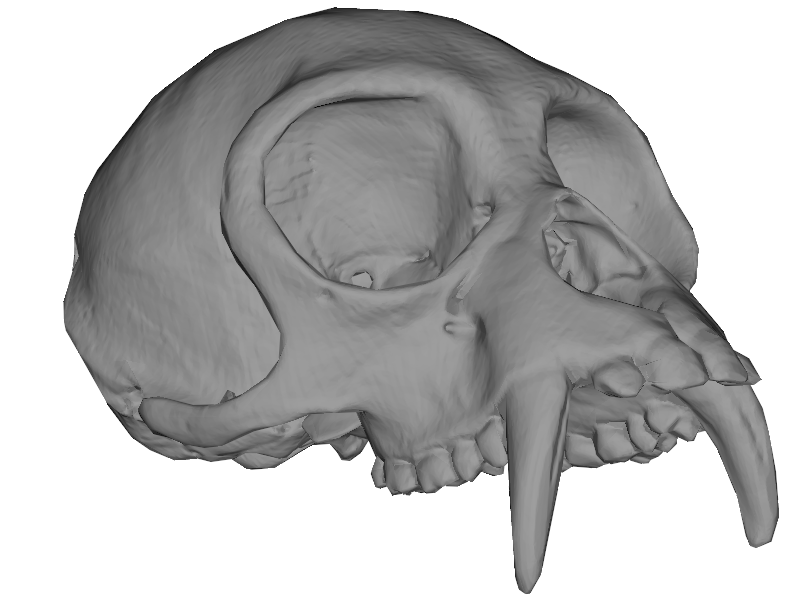} &
		\includegraphics[width=0.33\textwidth]{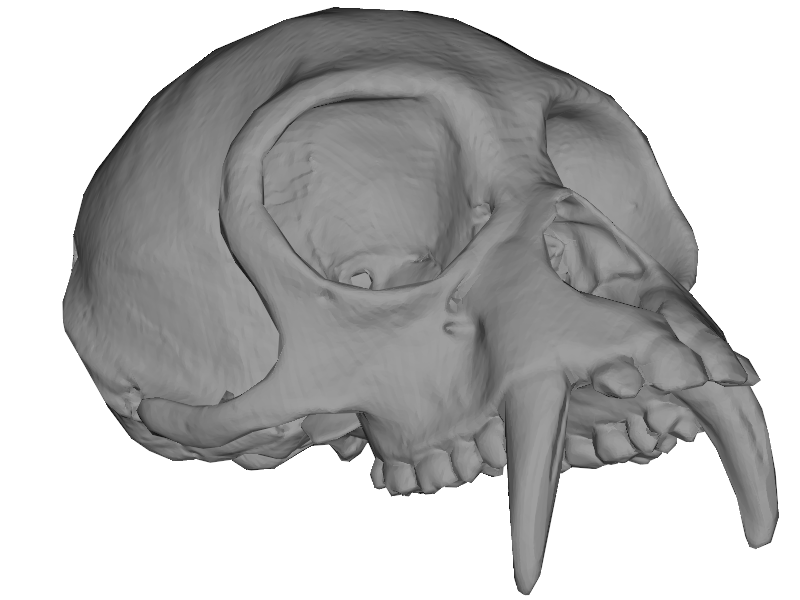} &
		\includegraphics[width=0.33\textwidth]{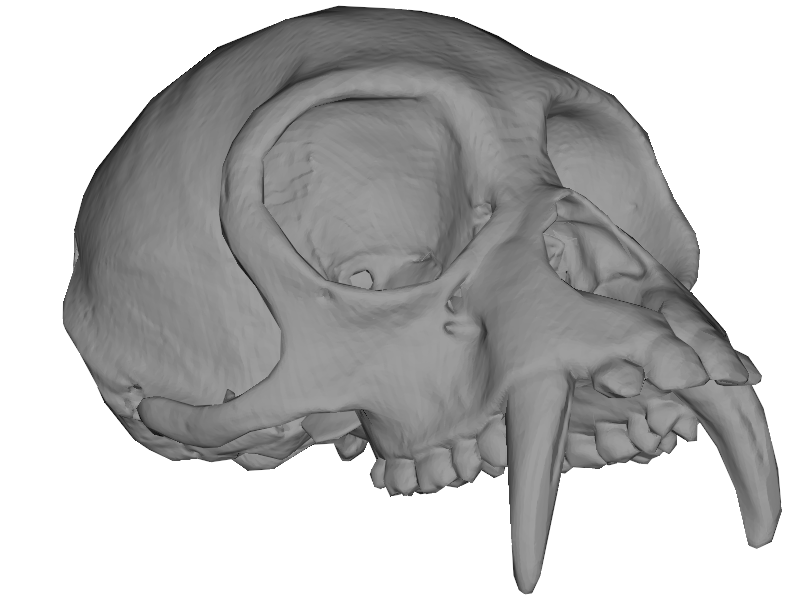} \\
		Substance (9k tris) &
		Our, normals only (9k tris) &
		Our, normals \& positions (9k tris) \\
		PSNR: 27.11~dB &
		PSNR: 26.22~dB &
		\textbf{PSNR: 28.30~dB}
	\end{tabular}
	\caption{A comparison with normal map baking. The starting point is a reduced base mesh with 9k triangles (reduced from the 735k triangle reference in Simplygon 8.3) and we bake a normal map from a 735k triangle reference using the normal bakers of Simplygon 8.3 and Substance Painter v2020.2.2. To our knowledge, these bakers only optimize the normal map, and leave the base mesh geometry unmodified. In our version, we jointly optimize the base geometry and normals based on rendered image observations. 
	}
	\label{fig:baking}
\end{figure*}

}


\newcommand{\figConvergence}{
	\begin{figure}
		\centering
		\includegraphics[width=0.99\columnwidth]{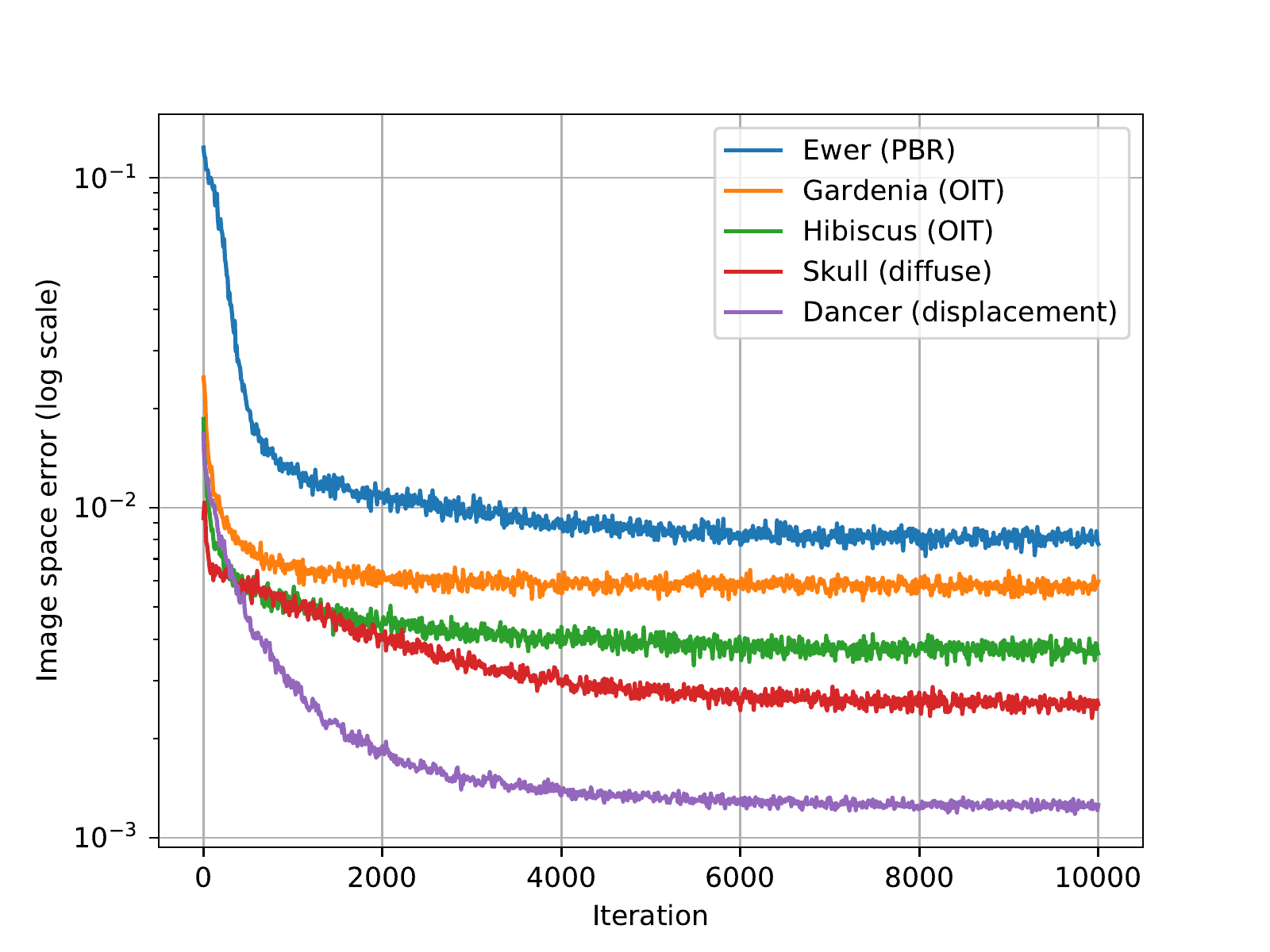}
		\caption{Training convergence plots for five optimization examples from the paper.}
		\label{fig:convergence}
	\end{figure}	
}


\newcommand{\figLaplace}{
	\begin{figure*}
		\setlength{\tabcolsep}{1pt}
		\begin{tabular}{ccccc}
			\includegraphics[width=0.19\textwidth]{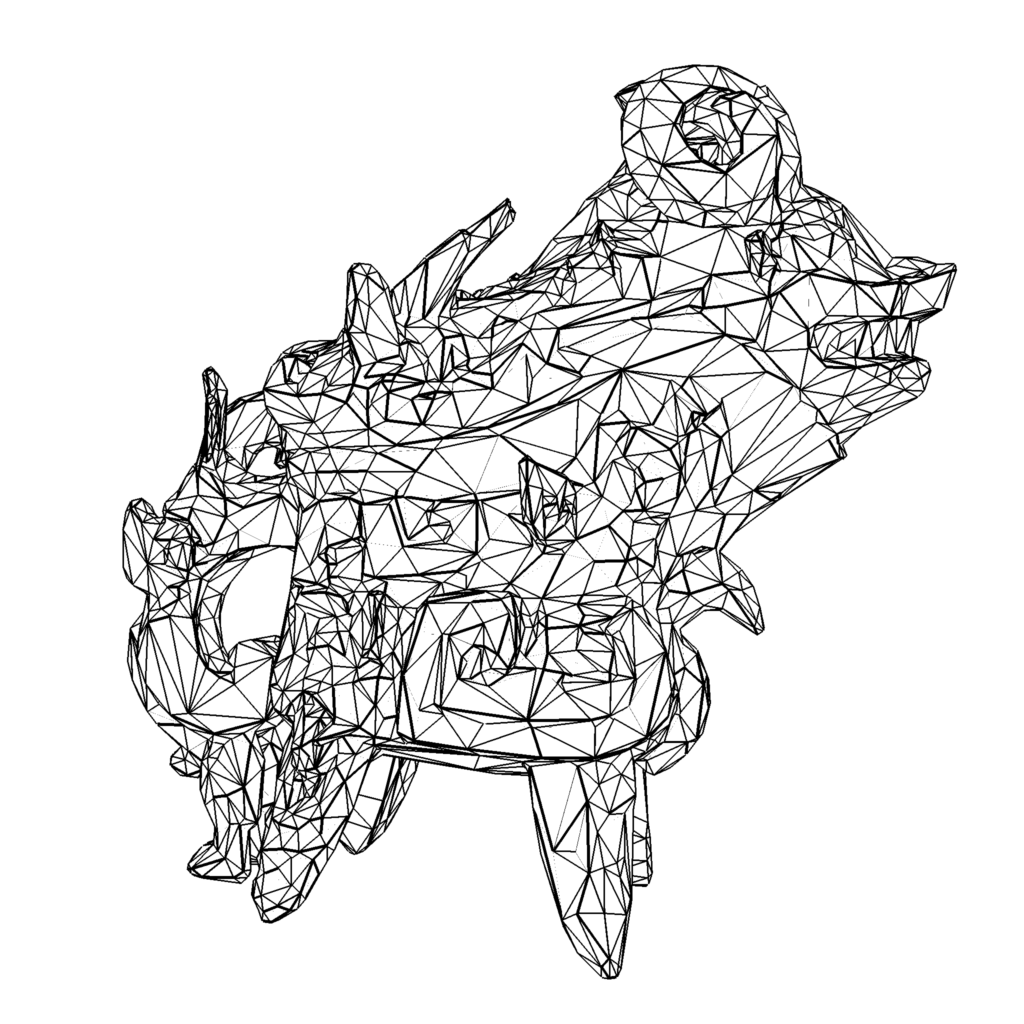} &
			\includegraphics[width=0.19\textwidth]{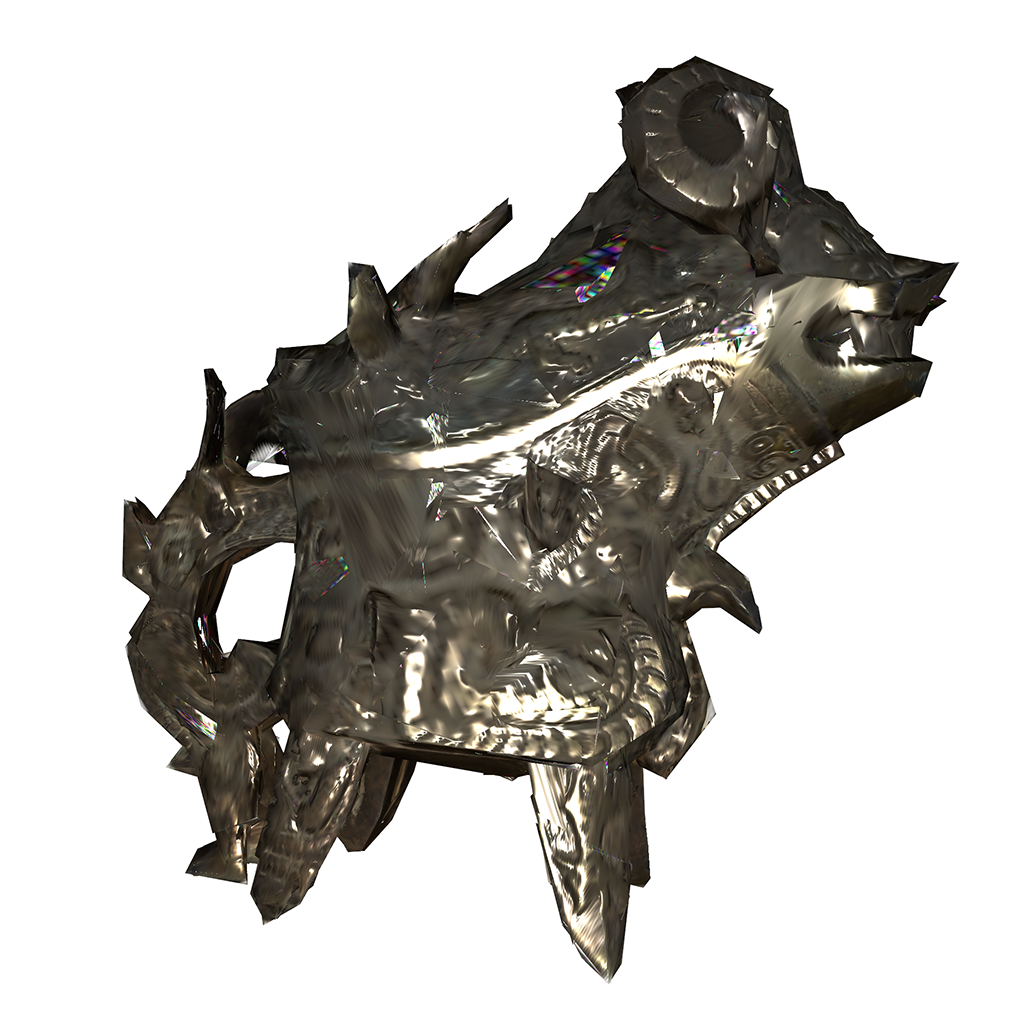} &
			\includegraphics[width=0.19\textwidth]{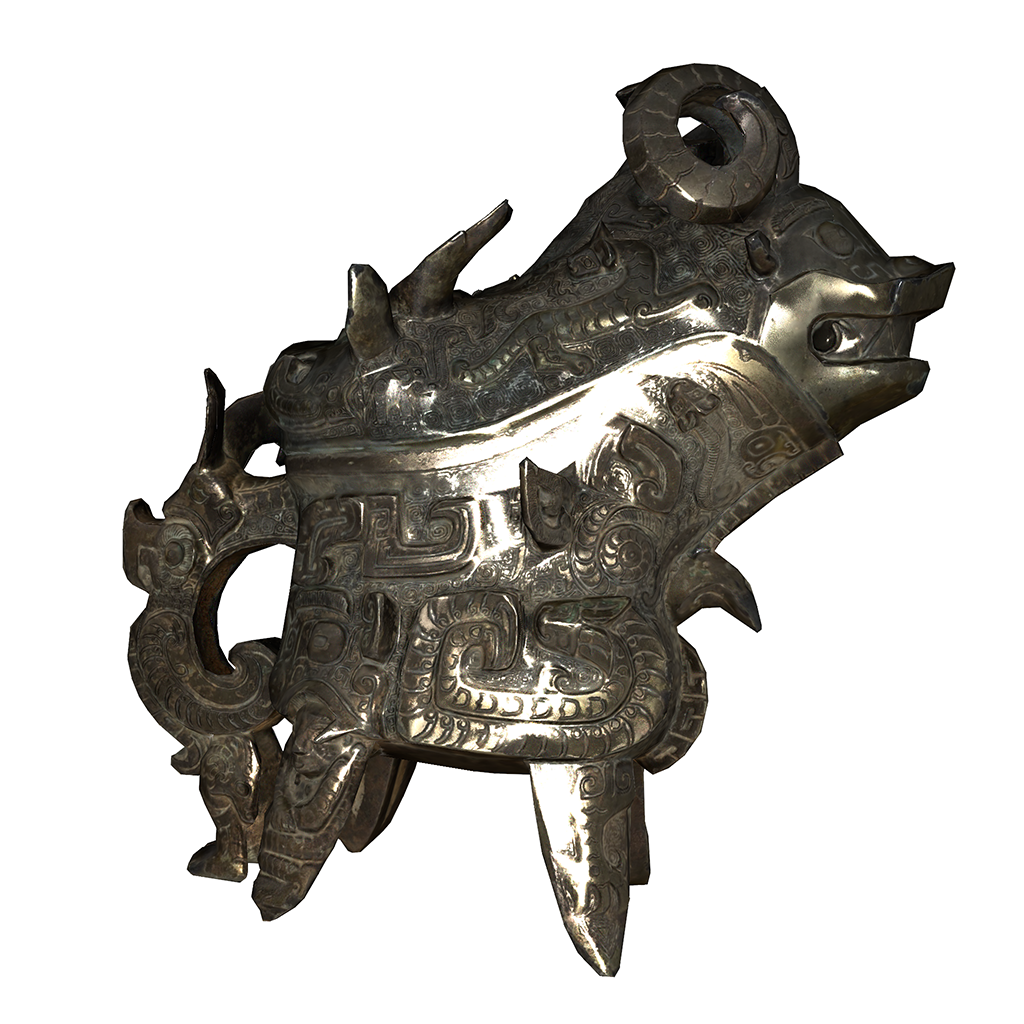} &
			\includegraphics[width=0.19\textwidth]{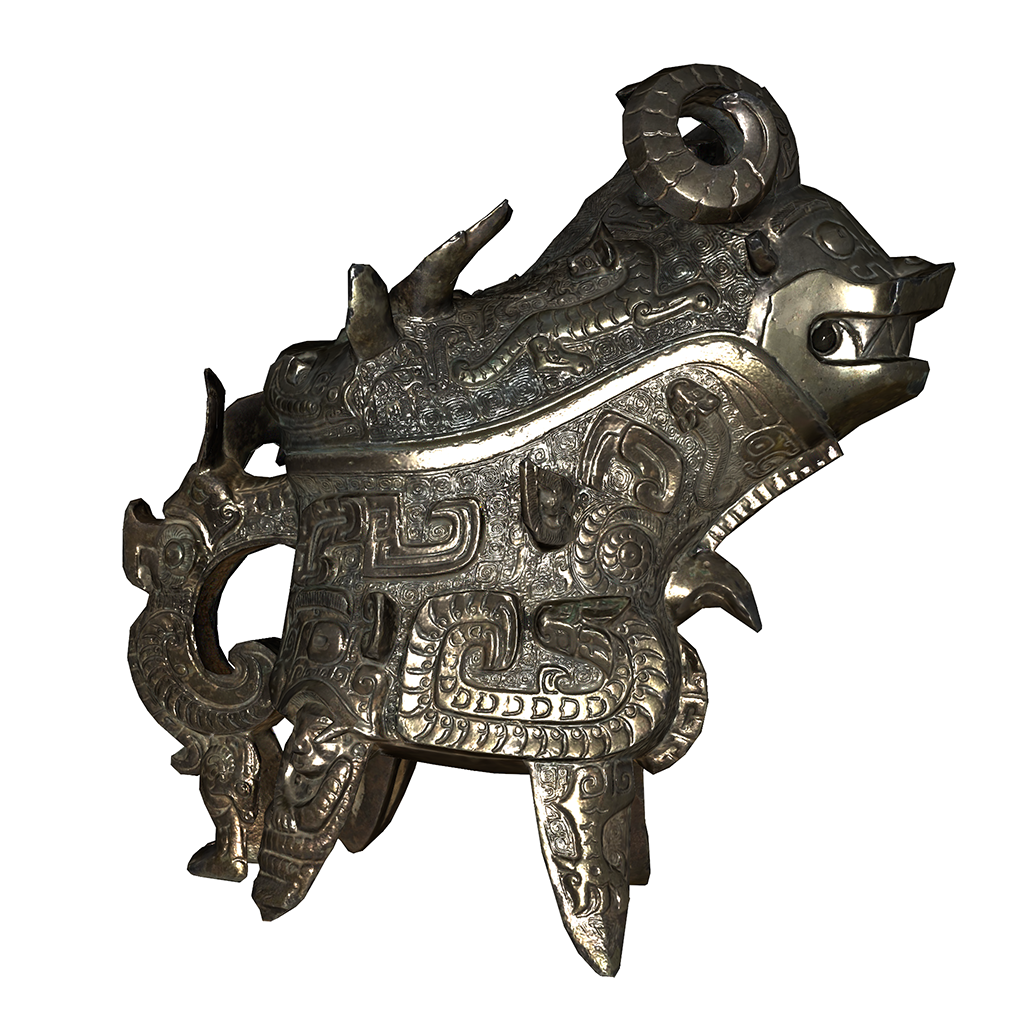} &
			\includegraphics[width=0.19\textwidth]{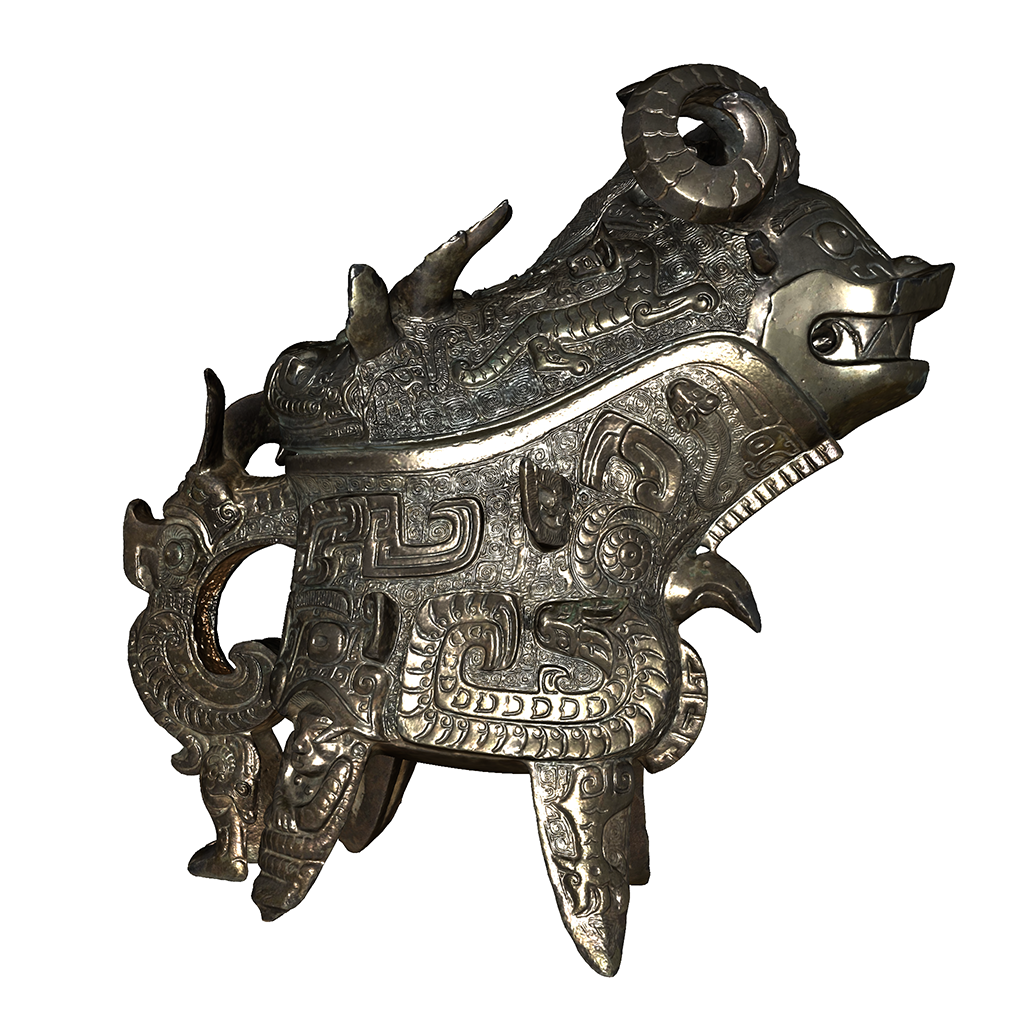} \\
			\includegraphics[width=0.19\textwidth]{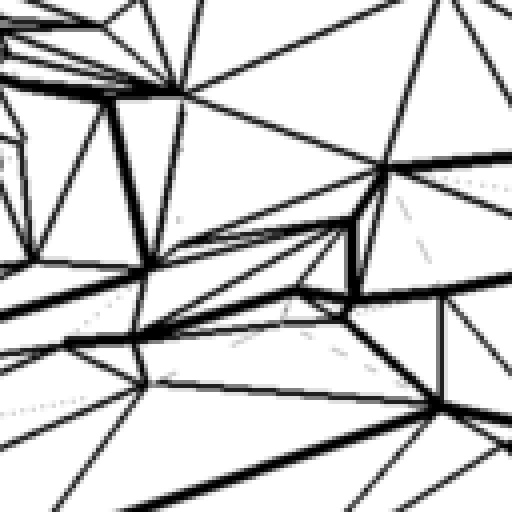} &
			\includegraphics[width=0.19\textwidth]{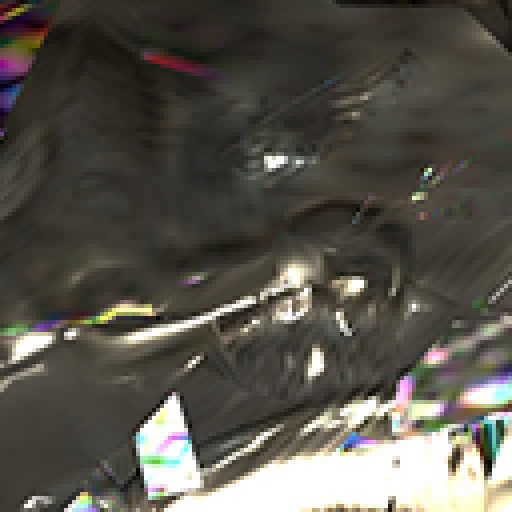} &
			\includegraphics[width=0.19\textwidth]{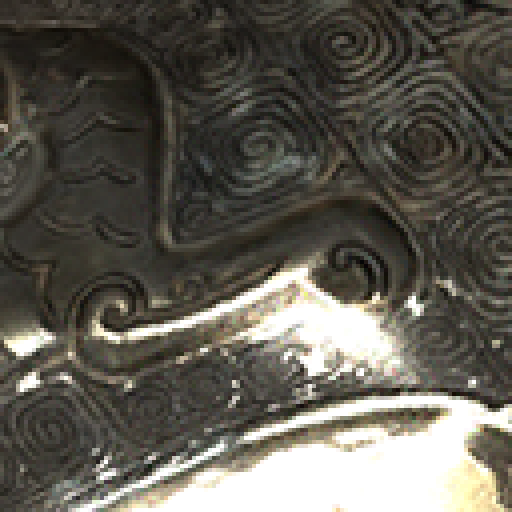} &
			\includegraphics[width=0.19\textwidth]{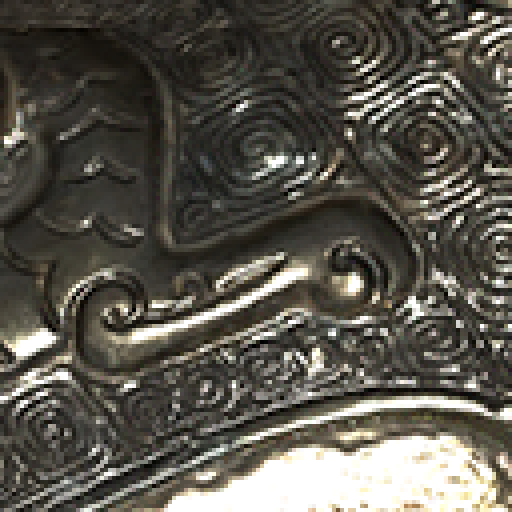} &
			\includegraphics[width=0.19\textwidth]{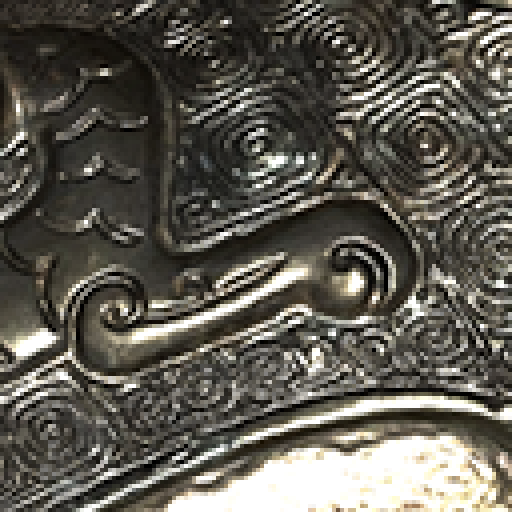} \\
			\small{Initial guess (7k tris)} & 
			\small{w/o Laplacian} &
			\small{w/o normal map} & 
			\small{Our} & 
			\small{Reference (300k tris)} \\ \hline
			\small{PSNR (dB)} & 18.13 & 18.15 & 24.49 & \\
			\small{FLIP} & 0.124 & 0.127 & 0.0588 & \\
		\end{tabular}
		\vspace{-5mm}
		\caption{The Ewer bronze sculpture, courtesy of the Smithsonian 3D Digitization project~\shortcite{Smithsonian2020}. 
			The reference mesh consists of 300k triangles, normal maps, textured base color and a bronze metal material.
			We start from a reduced version of the reference mesh with 7k triangles, with randomized material parameters.	In this case, we obtain a high quality result, closely approximating the reference.
			The insets highlight the results obtained when we disable either the normal map or the Laplacian regularizer during optimization. We note that both components are critical to obtain high quality results.
			Without the normal map, we lose high-frequency detail. Without the Laplacian regularizer, the mesh is malformed, 
			which subsequently also blurs the material parameters.}
		\label{fig:ewer_normal}
	\end{figure*}
}


\newcommand{\figInitialGuessA}{
	\begin{figure*}
		\setlength{\tabcolsep}{1pt}
		\begin{tabular}{ccccccc}    
			\includegraphics[width=0.14\textwidth]{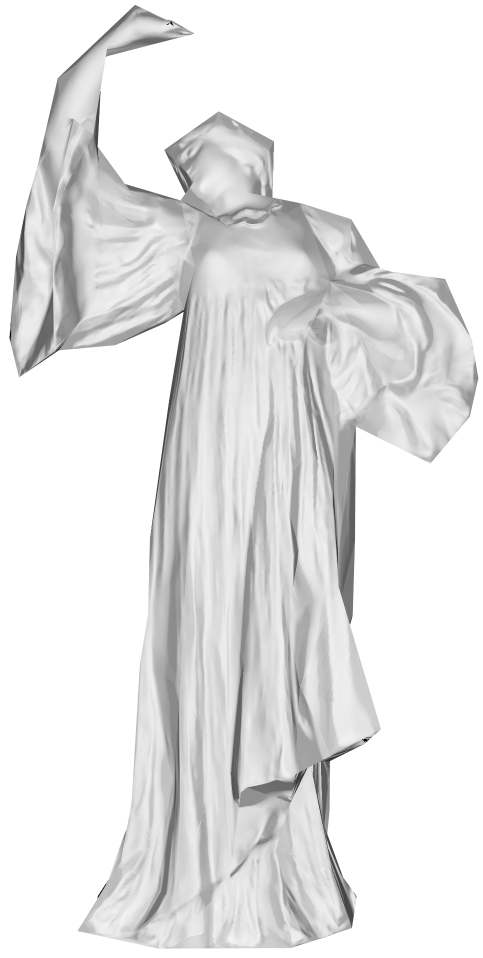} &
			\includegraphics[width=0.14\textwidth]{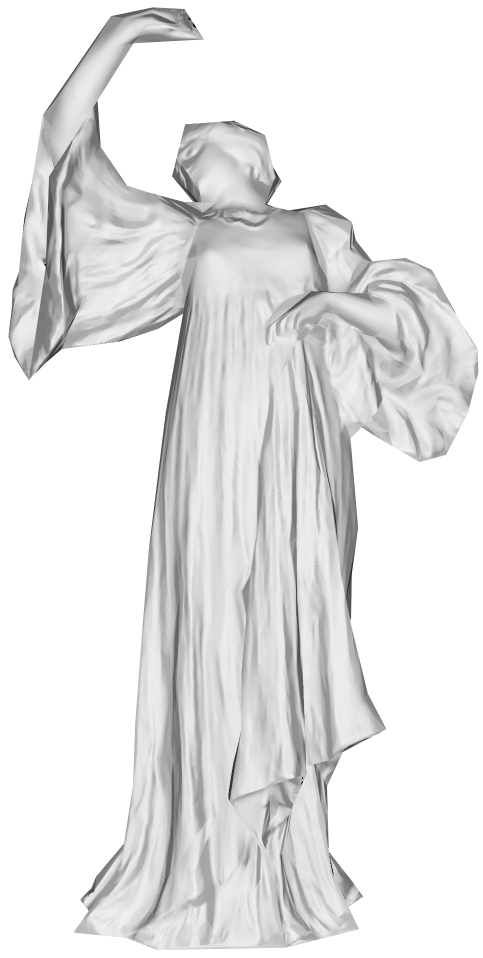} &
			\includegraphics[width=0.14\textwidth]{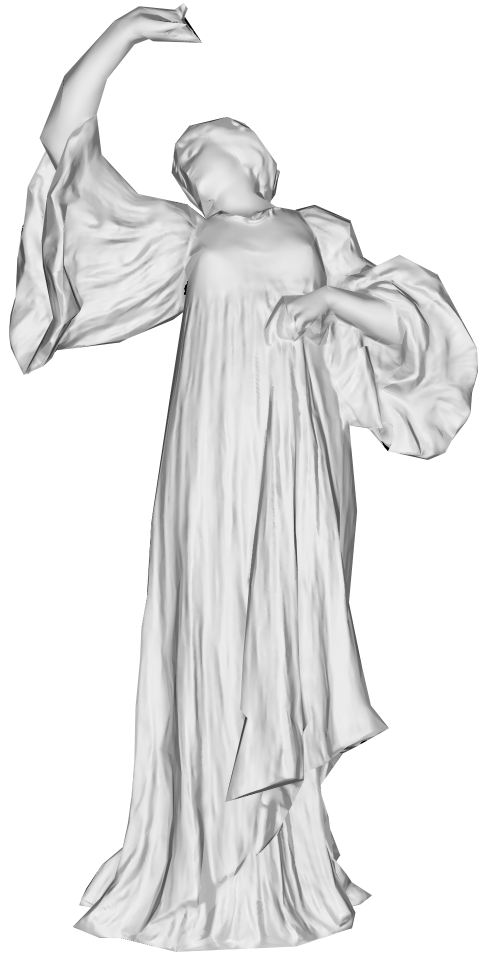} &
			\includegraphics[width=0.14\textwidth]{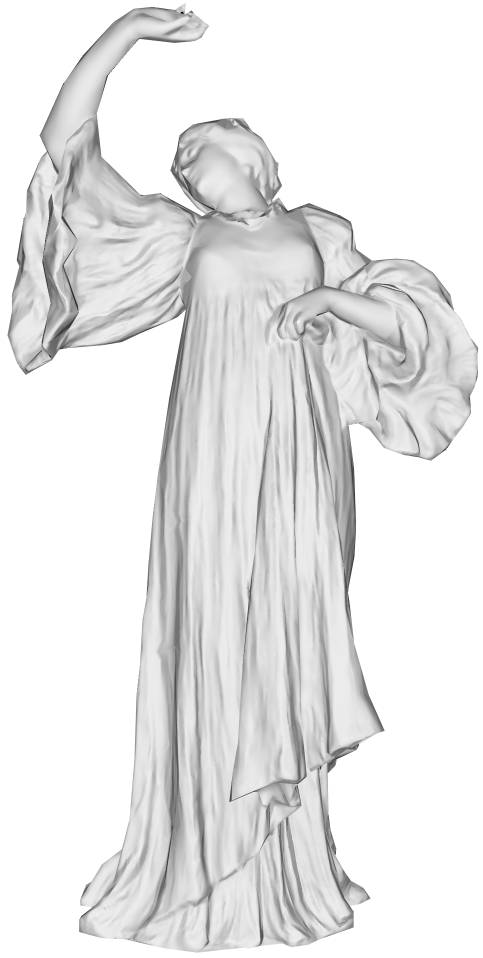} &
			\includegraphics[width=0.14\textwidth]{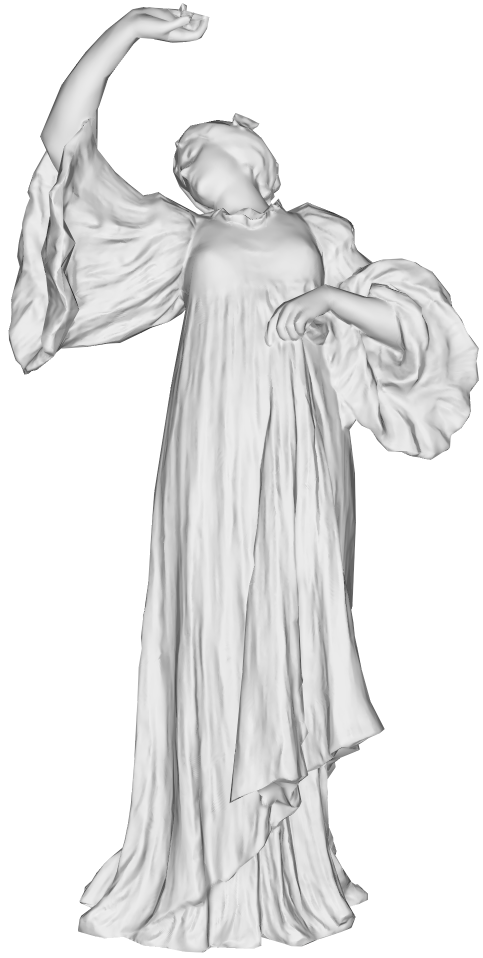} &
			\includegraphics[width=0.14\textwidth]{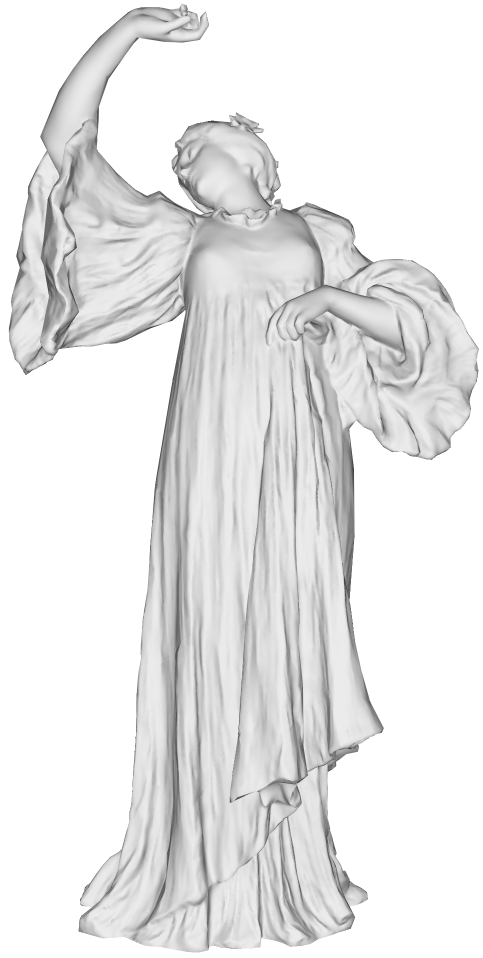} &
			\includegraphics[width=0.14\textwidth]{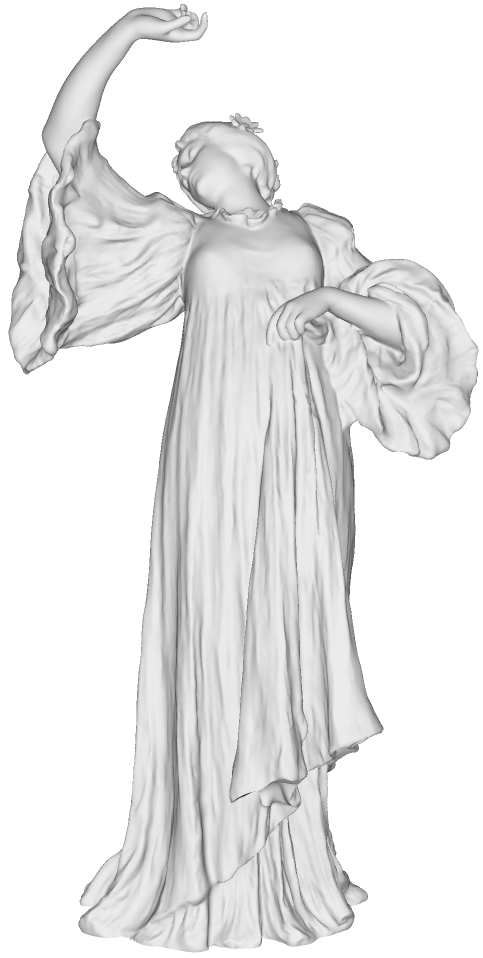} \\
			\includegraphics[width=0.14\textwidth]{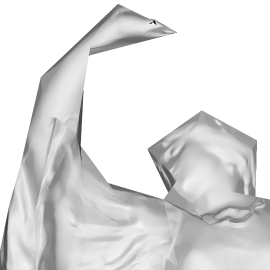} &
			\includegraphics[width=0.14\textwidth]{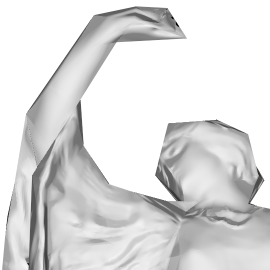} &
			\includegraphics[width=0.14\textwidth]{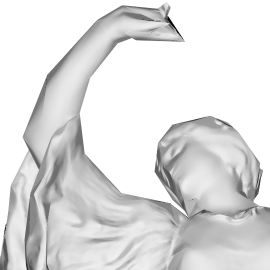} &
			\includegraphics[width=0.14\textwidth]{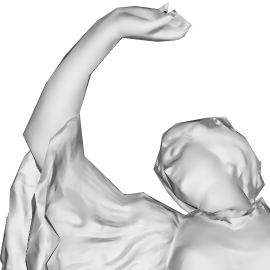} &
			\includegraphics[width=0.14\textwidth]{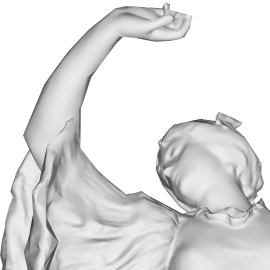} &
			\includegraphics[width=0.14\textwidth]{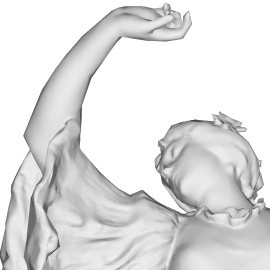} &
			\includegraphics[width=0.14\textwidth]{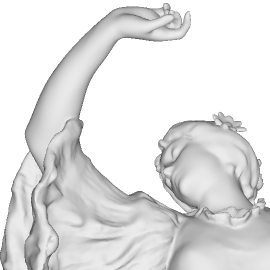} \\
			\small{500 tris} & \small{1k tris} & \small{2k tris} & \small{4k tris} & \small{8k tris} & \small{16k tris} & \small{Reference (370k tris)} \\
			22.90 & 24.19 & 25.78 & 28.25 & 29.39 & 31.61 & PSNR \\
			0.1411 & 0.0874 & 0.0458 & 0.0284 & 0.0159 & 0.0112 & Chamfer-$L^1$   
		\end{tabular}
		\caption{Starting from an initial guess with varying triangle count, we optimize shape and normal map to match the appearance 
			of the reference. As can be seen in the insets, 
			small details and silhouettes benefit greatly from increased triangle count. The importance of normal 
			mapping is evident. Details not part of the silhouette, e.g., the folds in cloth, are captured 
			even at low triangle counts. The Chamfer-$L^1$ scores are computed from 1M point samples over the meshes, and are multiplied with a factor $10^3$. The dancer model is courtesy of the Smithsonian 3D Digitization project~\shortcite{Smithsonian2020}
		}
		\label{fig:tri_sweep}
	\end{figure*}
}

\newcommand{\figInitialGuessB}{
	\begin{figure*}
		\vspace*{5mm}
		\setlength{\tabcolsep}{1pt}
		\begin{tabular}{ccccccc}    
			\includegraphics[width=0.14\textwidth]{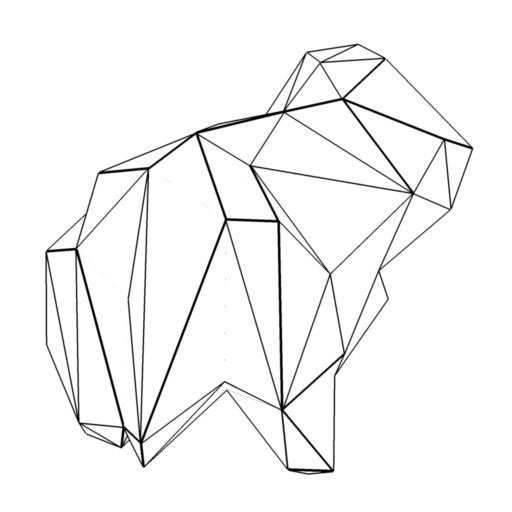} &
			\includegraphics[width=0.14\textwidth]{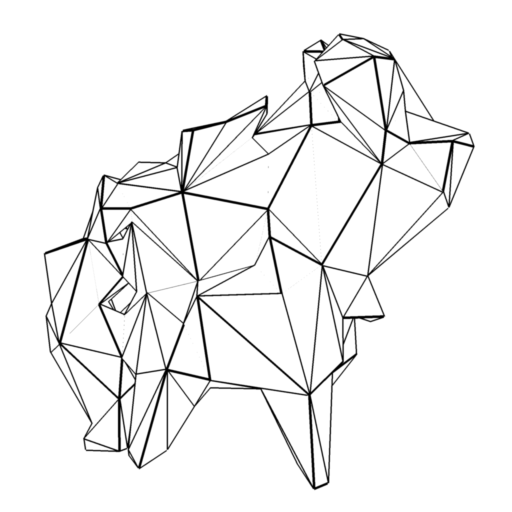} &
			\includegraphics[width=0.14\textwidth]{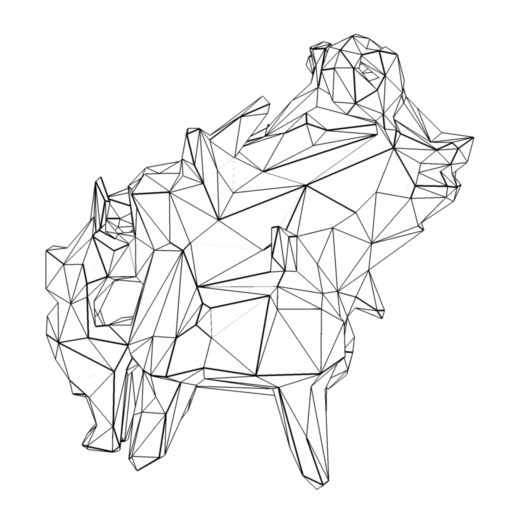} &
			\includegraphics[width=0.14\textwidth]{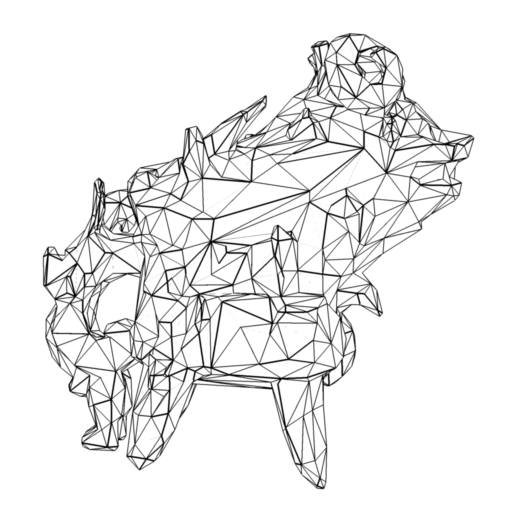} &
			\includegraphics[width=0.14\textwidth]{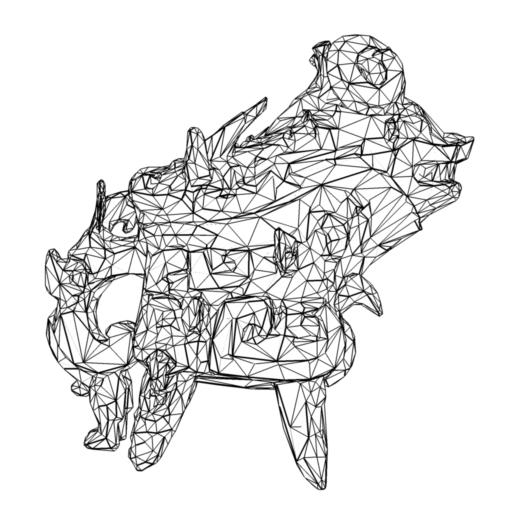} &
			\includegraphics[width=0.14\textwidth]{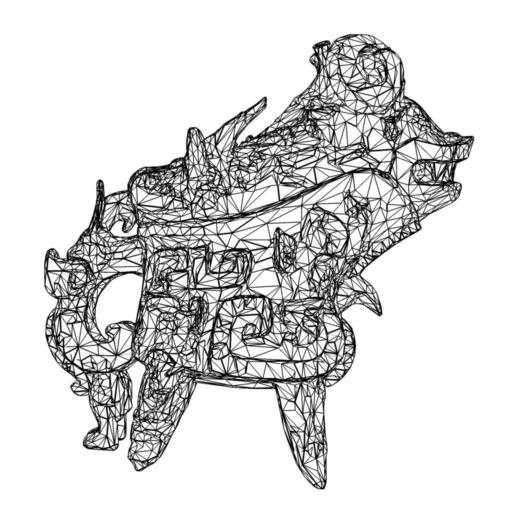} &
			\includegraphics[width=0.14\textwidth]{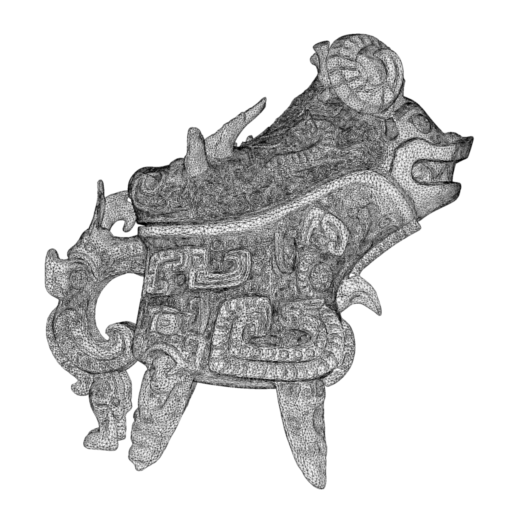} \\
			\includegraphics[width=0.14\textwidth]{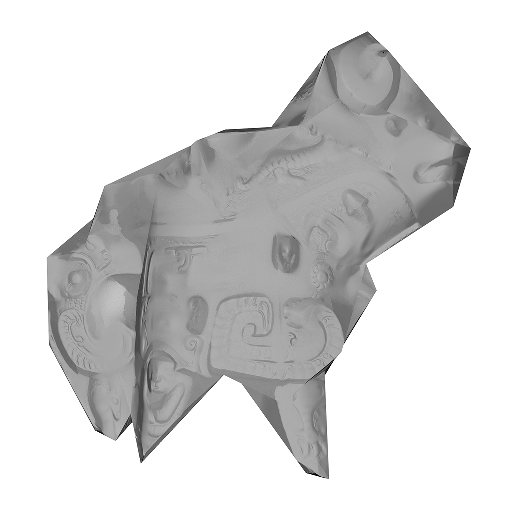} &
			\includegraphics[width=0.14\textwidth]{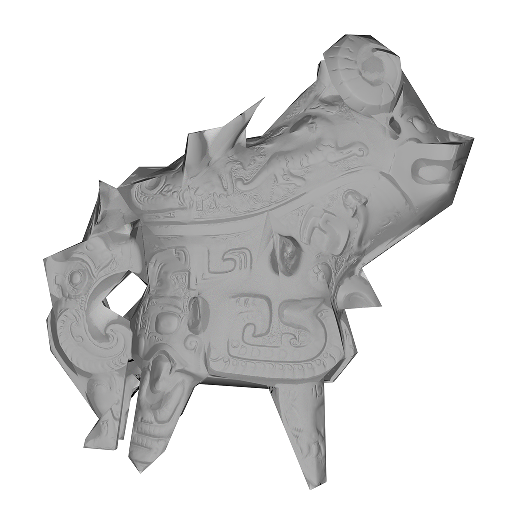} &
			\includegraphics[width=0.14\textwidth]{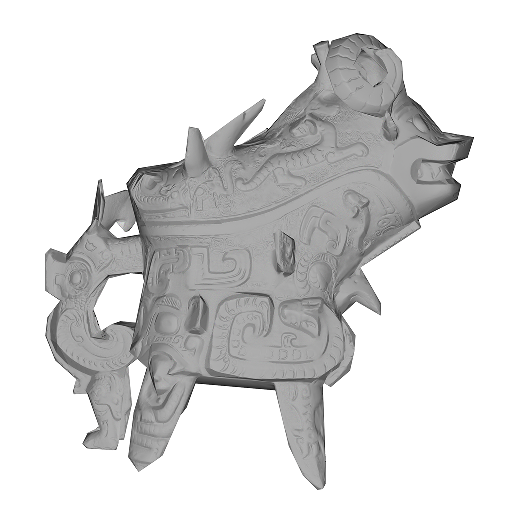} &
			\includegraphics[width=0.14\textwidth]{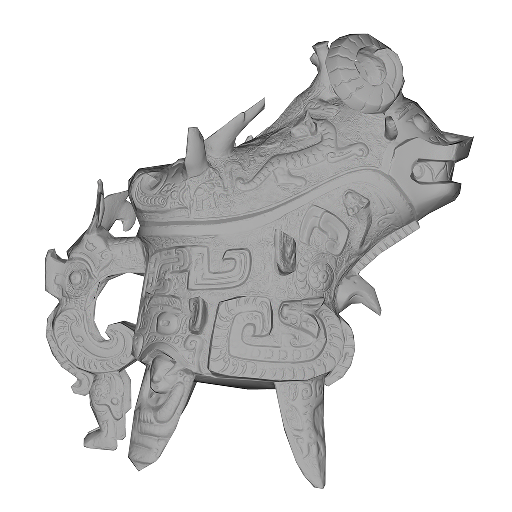} &
			\includegraphics[width=0.14\textwidth]{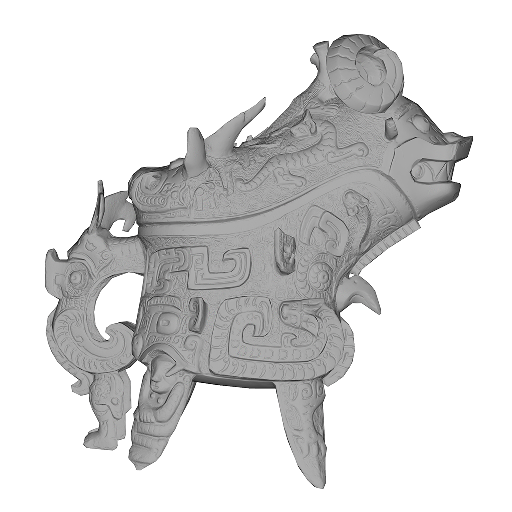} &
			\includegraphics[width=0.14\textwidth]{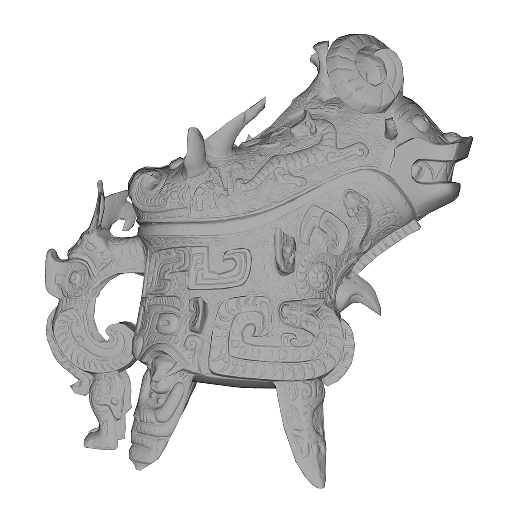} &
			\includegraphics[width=0.14\textwidth]{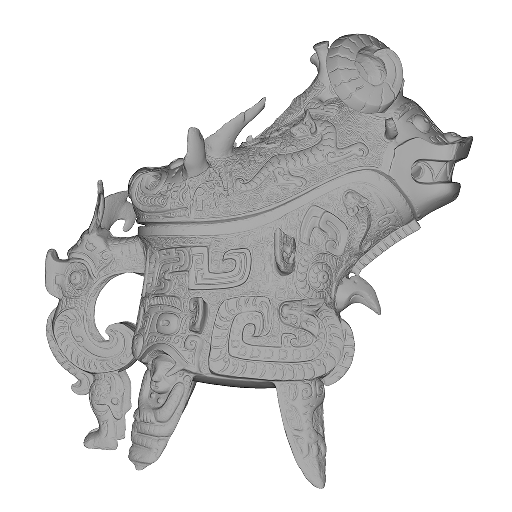} \\        
			\small{132 tris} & \small{500 tris} & \small{1.2k tris} & \small{2.9k tris} & \small{7k tris} & \small{18k tris} & \small{Reference} \\
			18.42 & 20.78 & 23.77 & 26.36 & 29.44 & 30.14 & PSNR \\
			0.3825 & 0.1086 & 0.0404 & 0.0209 & 0.0149 & 0.0137 & Chamfer-$L^1$   
		\end{tabular}
		\vspace{-5mm}
		\caption{Influence of initial guess. We show six different input meshes, ranging from 130 to 18k triangles, 
			all trying to approximate a reference mesh with 300k triangles (Ewer model courtesy of the Smithsonian 3D Digitization project~\shortcite{Smithsonian2020}). The Chamfer-$L^1$ scores are multiplied with a factor $10^3$.
		}
		\label{fig:ewer_sweep}
	\end{figure*}
}


\newcommand{\figTessellationLOD}{
	\begin{figure*}
		\setlength{\tabcolsep}{1pt}
		\begin{tabular}{ccccc}
			\includegraphics[width=0.2\textwidth]{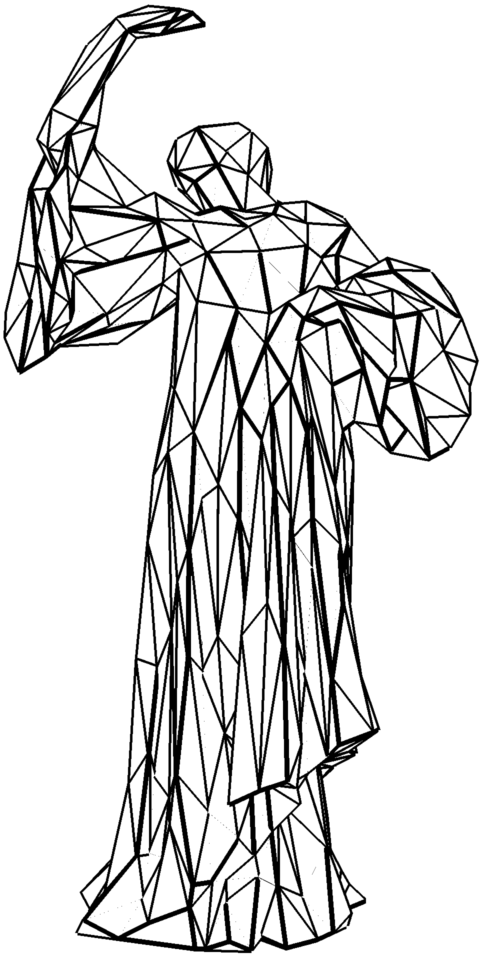} &
			\includegraphics[width=0.2\textwidth]{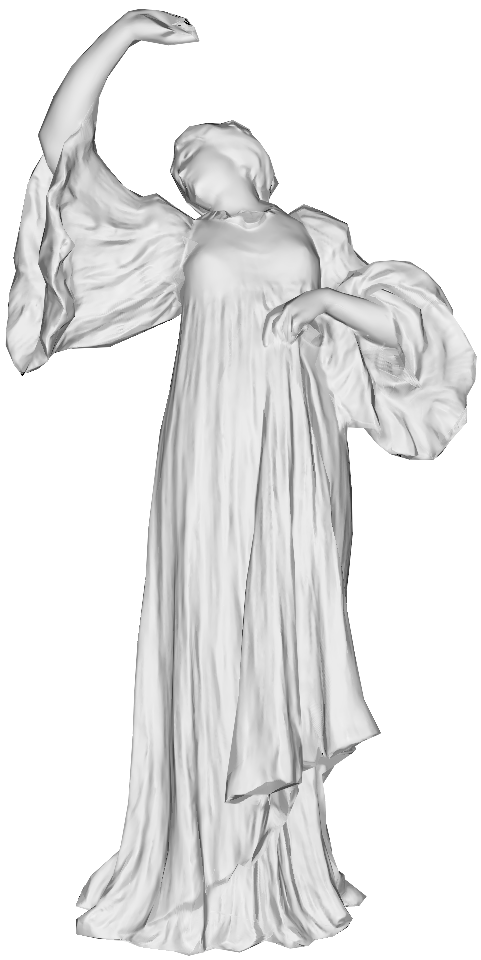} &
			\includegraphics[width=0.2\textwidth]{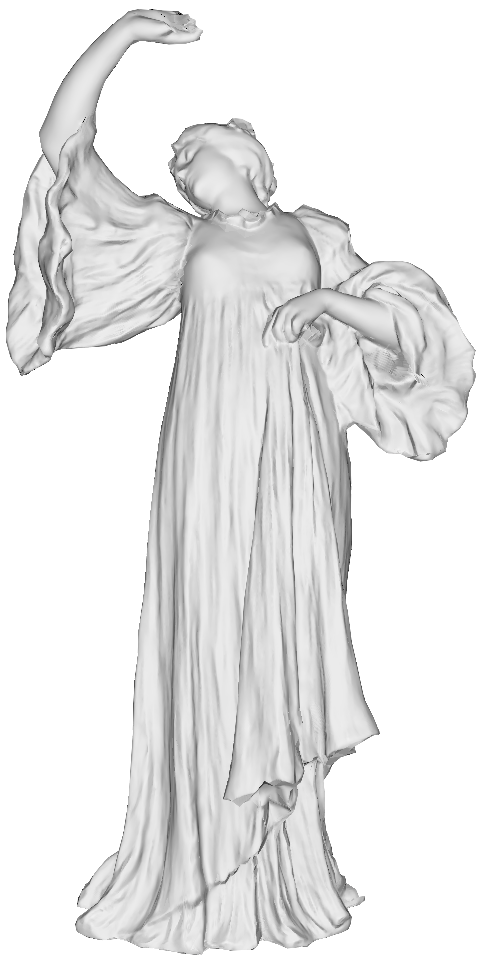} &
			\includegraphics[width=0.2\textwidth]{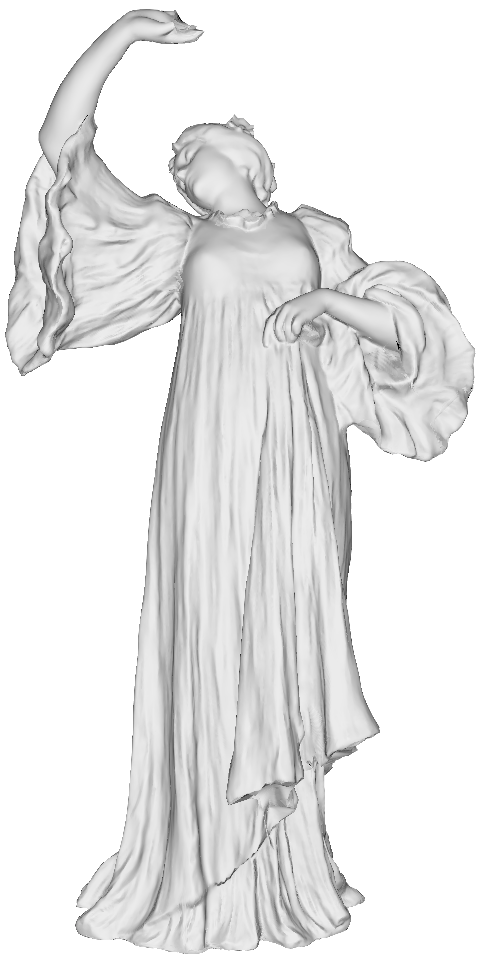} &
			\includegraphics[width=0.2\textwidth]{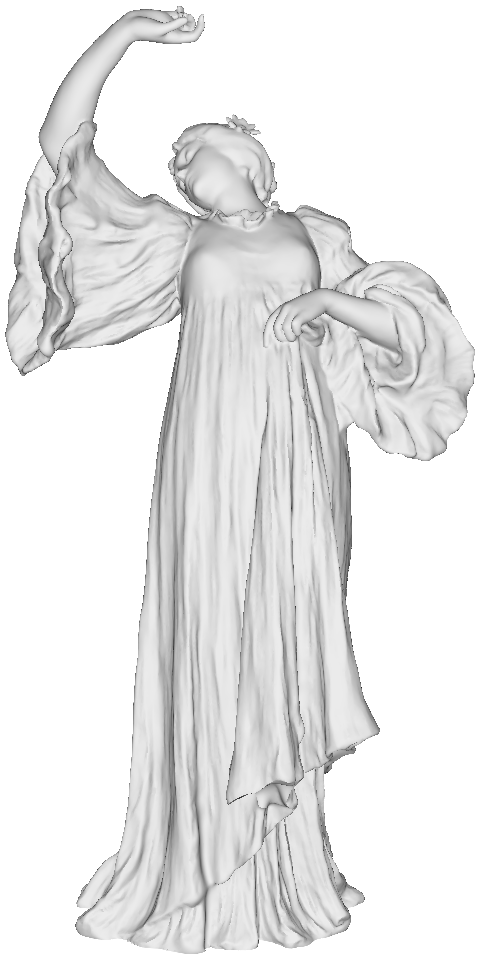} \\
			\includegraphics[width=0.2\textwidth]{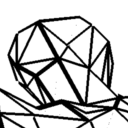} &
			\includegraphics[width=0.2\textwidth]{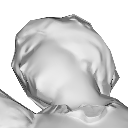} &
			\includegraphics[width=0.2\textwidth]{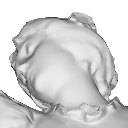} &
			\includegraphics[width=0.2\textwidth]{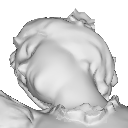} &
			\includegraphics[width=0.2\textwidth]{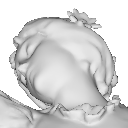} \\
			\small{Base mesh} & \small{One subdivision} & \small{Two subdivisions} & \small{Three subdivisions} & \small{Reference} \\
			\small{1k tris} & \small{4k tris} & \small{16k tris} & \small{64k tris} & \small{370k tris} \\
		\end{tabular}
		\caption{A level of detail example with different levels of tessellation rendered from a single base mesh. 
			All levels of tessellation use the same base mesh, displacement map and normal map. As can be seen in the insets, silhouette and details are improved with increased tessellation.}
		\label{fig:tess_sweep}
	\end{figure*}
}


\newcommand{\figAggregateCmp}{
	\begin{figure*}
		\centering
		\setlength{\tabcolsep}{1pt}
		\begin{tabular}{ccc}
			\includegraphics[width=0.33\textwidth]{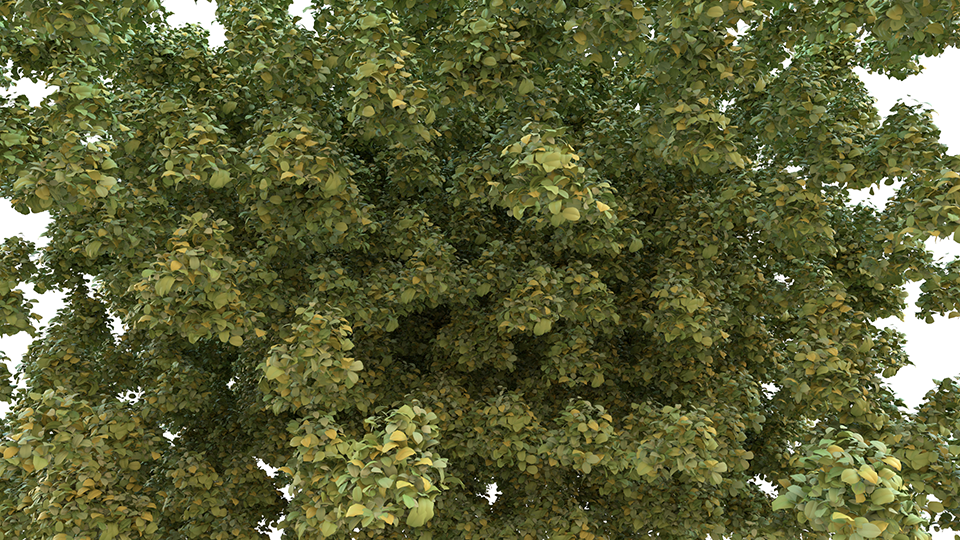} &
			\includegraphics[width=0.33\textwidth]{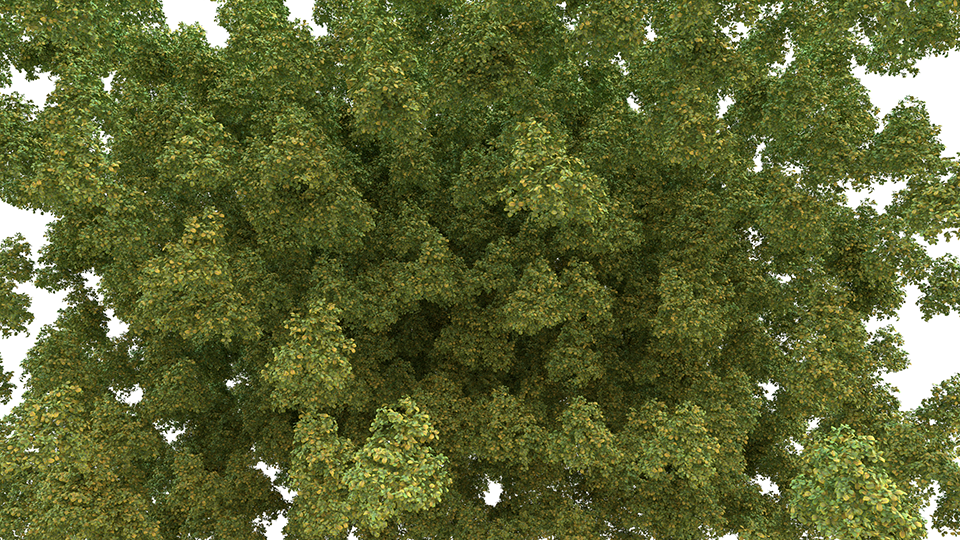} &    
			\includegraphics[width=0.33\textwidth]{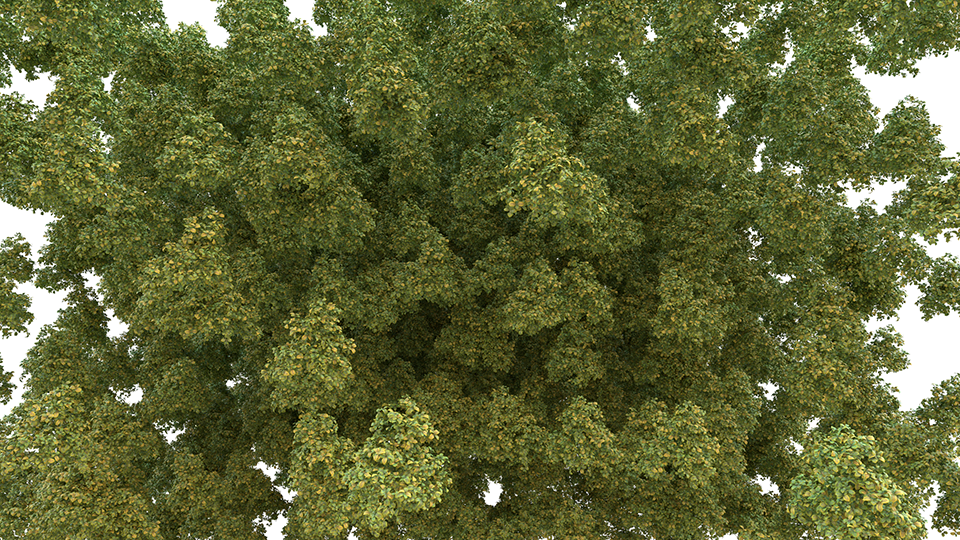} \\
			\small{Stochastic simplification: 510M tris, PSNR: 14.49~dB} & \small{Our: 20M tris, PSNR: 23.89~dB} & \small{Reference: 5.1B tris} \\
			\includegraphics[width=0.30\columnwidth]{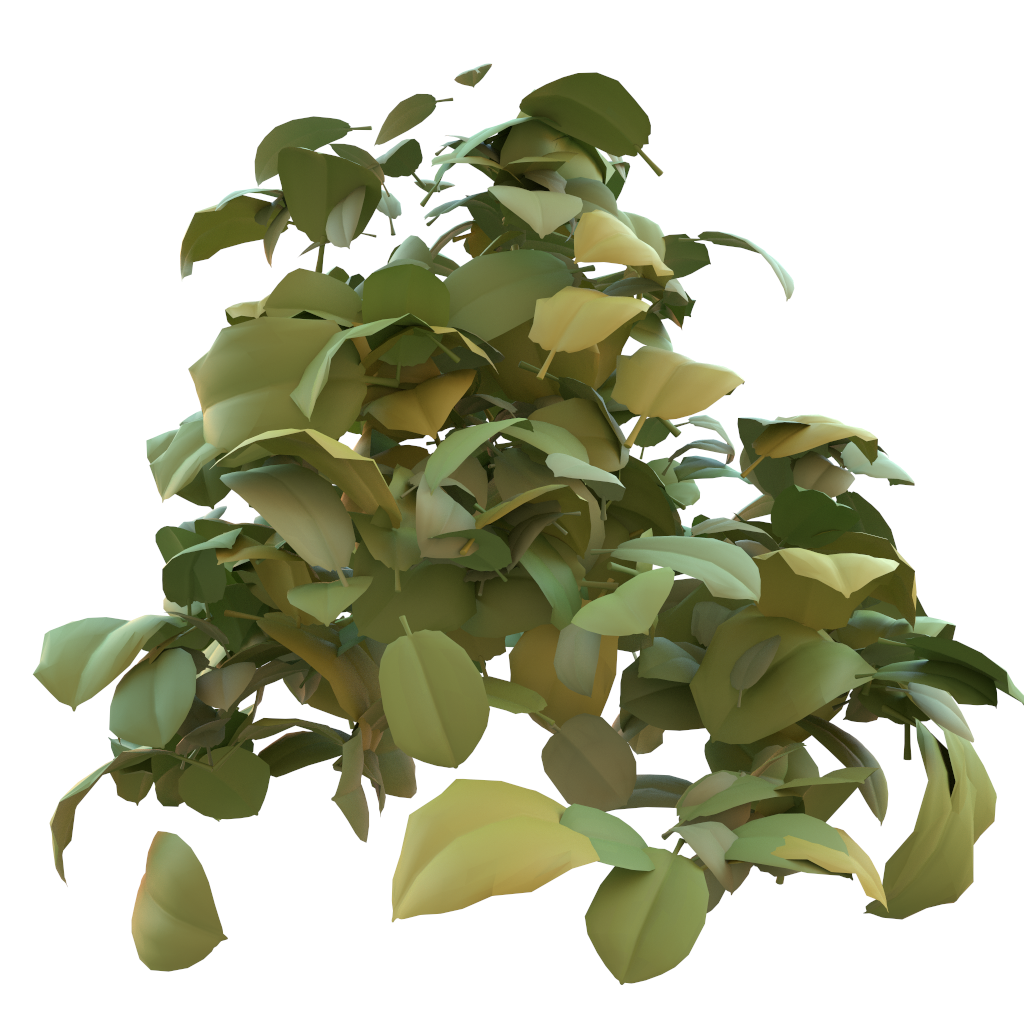} &
			\includegraphics[width=0.30\columnwidth]{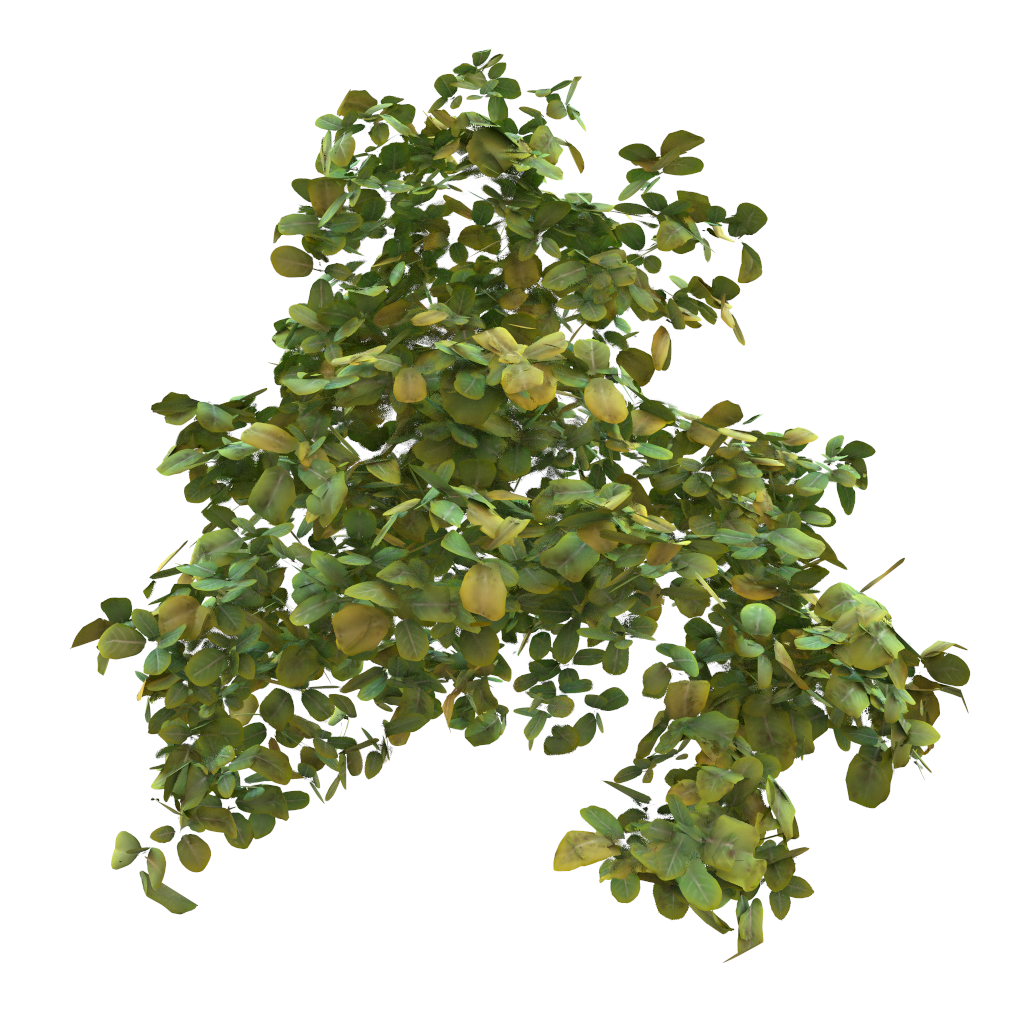} &    
			\includegraphics[width=0.30\columnwidth]{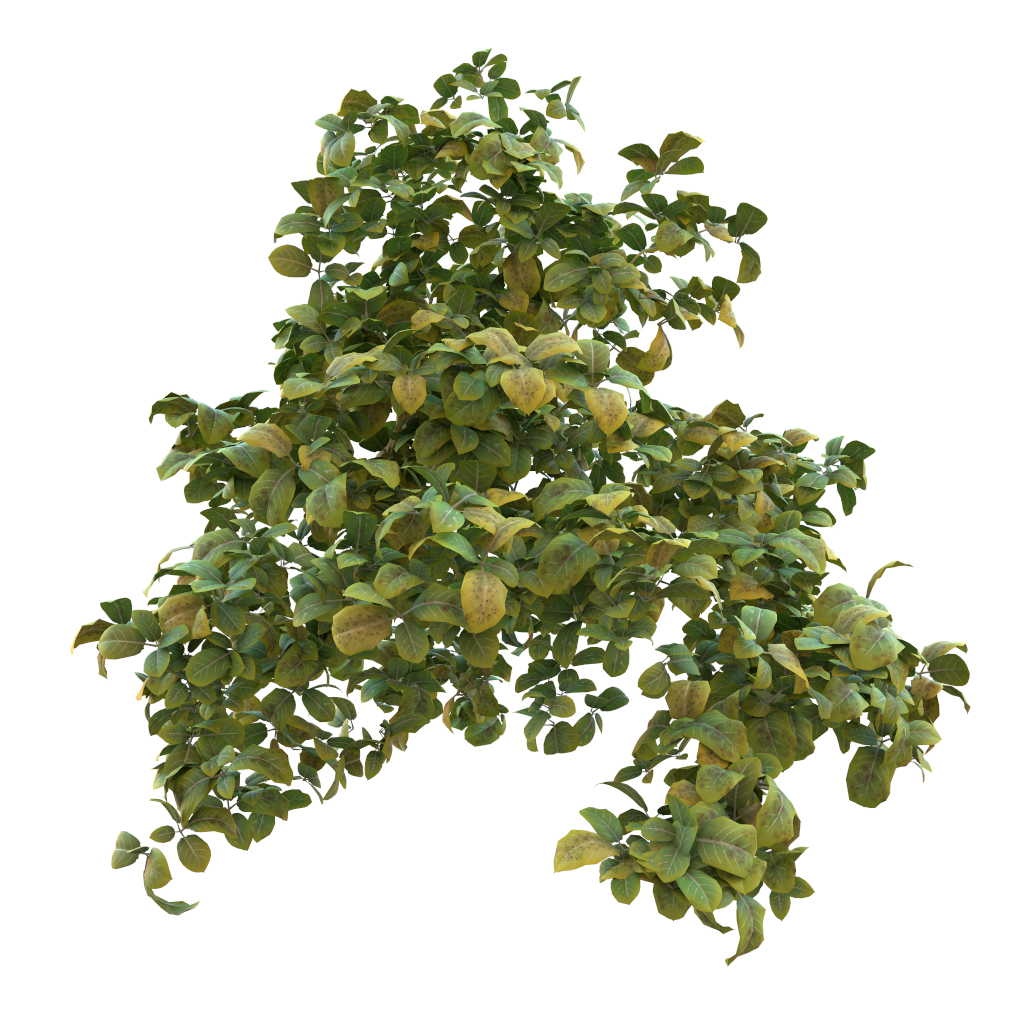} \\
			\small{170k tris} & \small{6.5k tris} & \small{1.7M tris} \\
			\small{Stochastic simplification} & \small{Our model} & \small{Reference}
		\end{tabular}
		\caption{Our approach compared to stochastic simplification of aggregate detail~\cite{Cook2007}.}
		\label{fig:stochsimpl}
	\end{figure*}
}


\newcommand{\figPrefilterLOD}{
	\begin{figure}
		\centering
		\setlength{\tabcolsep}{1pt}
		\begin{tabular}{ccccc}
			\rotatebox[origin=c]{90}{\small{Ref: 1~spp}} &     
			\raisebox{-0.5\height}{\includegraphics[width=0.22\columnwidth]{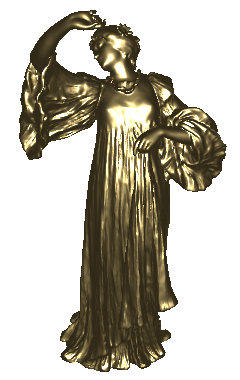}} &
			\raisebox{-0.5\height}{\includegraphics[width=0.22\columnwidth]{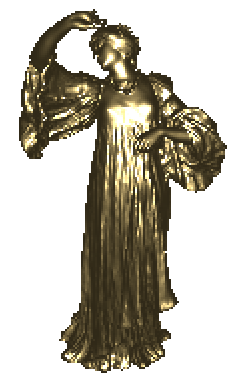}} &
			\raisebox{-0.5\height}{\includegraphics[width=0.22\columnwidth]{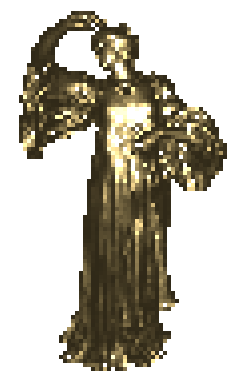}} &
			\raisebox{-0.5\height}{\includegraphics[width=0.22\columnwidth]{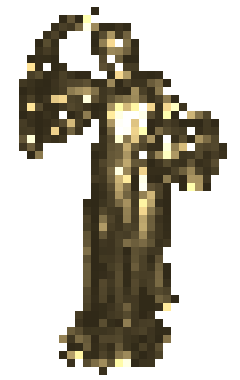}} \\
			\rotatebox[origin=c]{90}{\small{Our: 1~spp}} &     
			\raisebox{-0.5\height}{\includegraphics[width=0.22\columnwidth]{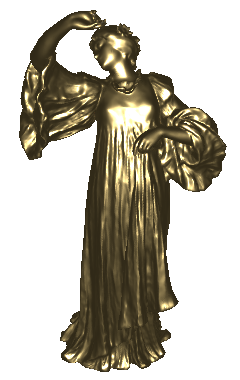}} &
			\raisebox{-0.5\height}{\includegraphics[width=0.22\columnwidth]{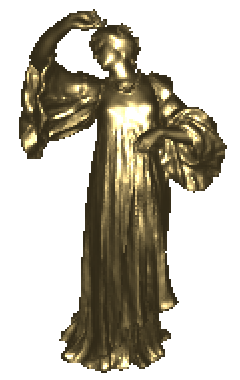}} &
			\raisebox{-0.5\height}{\includegraphics[width=0.22\columnwidth]{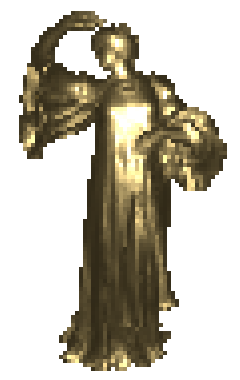}} &
			\raisebox{-0.5\height}{\includegraphics[width=0.22\columnwidth]{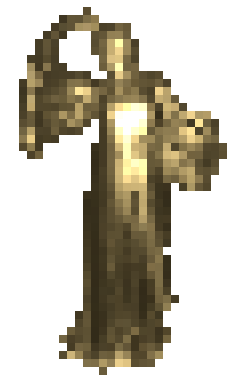}} \\
			\rotatebox[origin=c]{90}{\small{Ref: 256~spp}} &     
			\raisebox{-0.5\height}{\includegraphics[width=0.22\columnwidth]{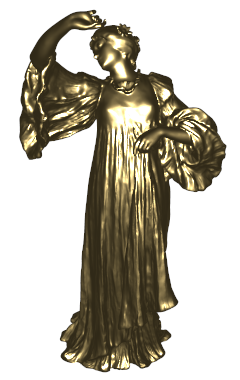}} &
			\raisebox{-0.5\height}{\includegraphics[width=0.22\columnwidth]{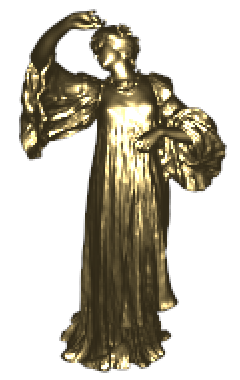}} &
			\raisebox{-0.5\height}{\includegraphics[width=0.22\columnwidth]{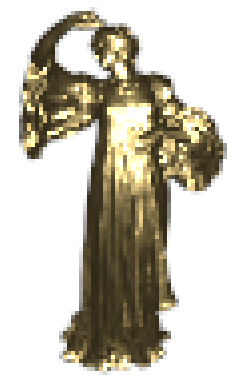}} &
			\raisebox{-0.5\height}{\includegraphics[width=0.22\columnwidth]{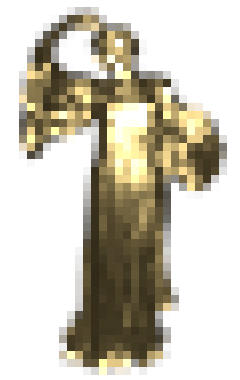}} \\
			& \small{512$\times$512} & \small{256$\times$256} & \small{128$\times$128} & \small{64$\times$64} 
		\end{tabular}
		\caption{We perform shape and appearance prefiltering by optimizing for a particular rendering resolution. Here, we 
			show our results for four different resolutions. As expected, geometric details and shading are smoothed as 
			rendering resolution decrease. 
			\textbf{Top:} The reference dancer model rendered at 1~spp, note the aliasing.
			\textbf{Middle:} Our optimized model with prefiltered shape and appearance rendered at 1~spp.
			\textbf{Bottom:} The reference dancer model rendered, with antialiasing, at 256~spp.
		}
		\label{fig:lodzoom}
	\end{figure}
}


\newcommand{\figSDFA}{
	\begin{figure}
		\setlength{\tabcolsep}{1pt}
		\begin{tabular}{ccc}
			\includegraphics[width=0.32\columnwidth]{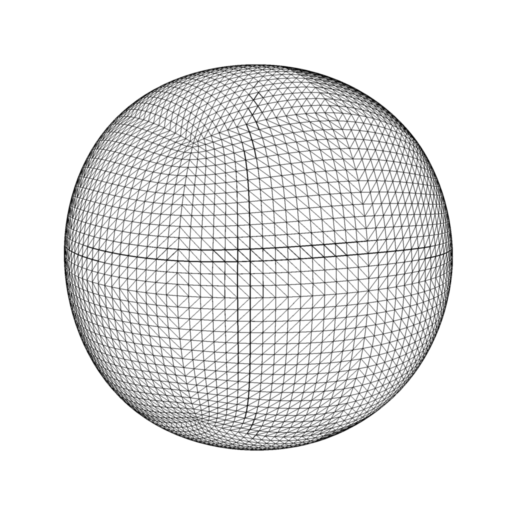} &
			\includegraphics[width=0.32\columnwidth]{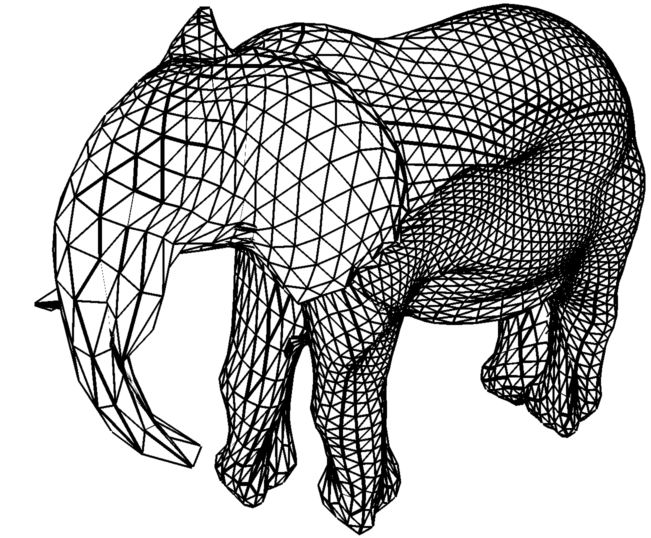} &    
			\includegraphics[width=0.32\columnwidth]{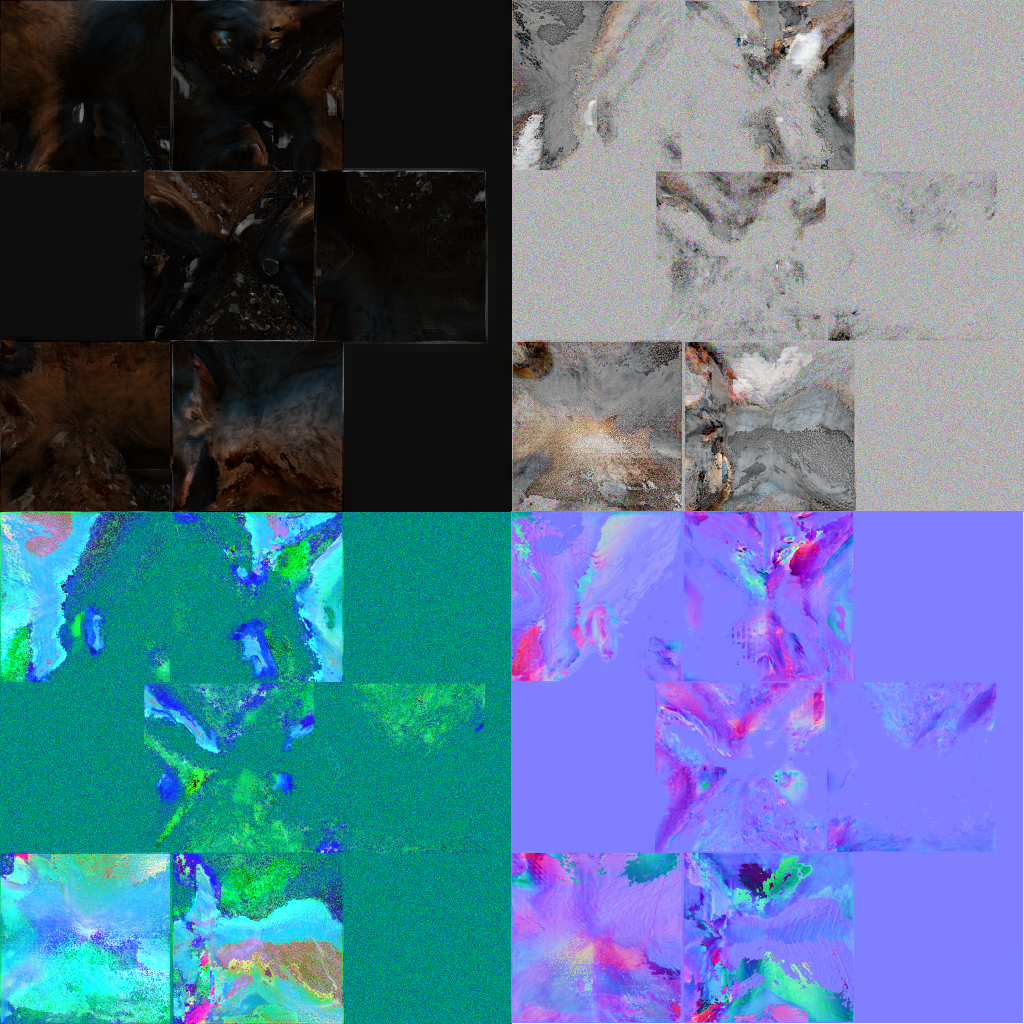} \\
			\small{Initial guess} & \small{Optimized mesh} & \small{Optimized materials}
		\end{tabular}
		\begin{tabular}{cc}
			\includegraphics[width=0.49\columnwidth]{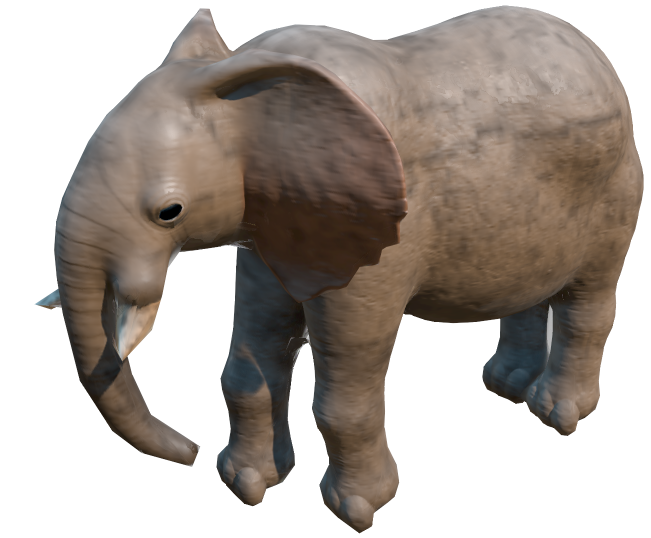} &
			\includegraphics[width=0.49\columnwidth]{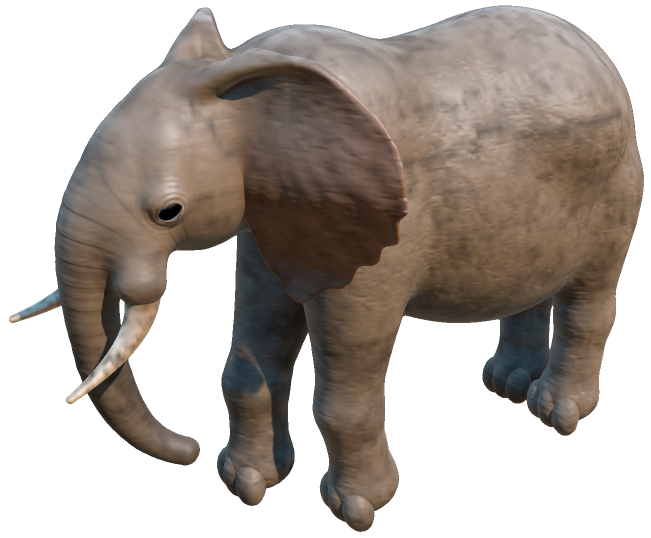} \\
			\small{Our rasterized model} &  \small{Ray marched reference} 
		\end{tabular}
		\captionof{figure}{We extract a mesh and materials from a ray marched distance field: the ShaderToy ``Elephant,'' from 
			Inigo Quilez. We initialize the optimization process by a sphere with random material parameters and learn
			shape and material parameters such that our rasterized model resembles the ray marched reference.}
		\label{fig:elephant}
	\end{figure}
}

\newcommand{\figSDFB}{
	\begin{figure}
		\setlength{\tabcolsep}{1pt}
		\begin{tabular}{ccccc}
			\includegraphics[width=0.32\columnwidth]{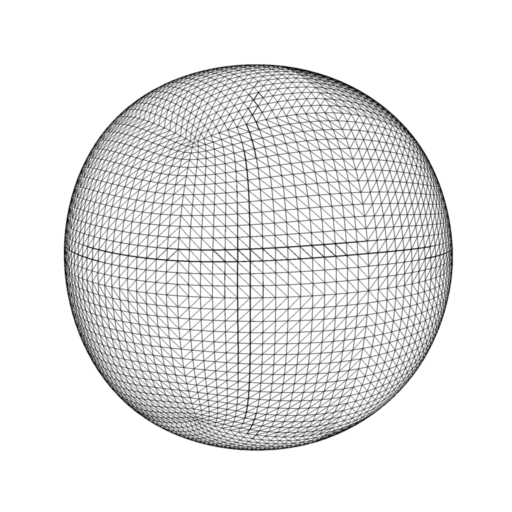} &
			\includegraphics[width=0.32\columnwidth]{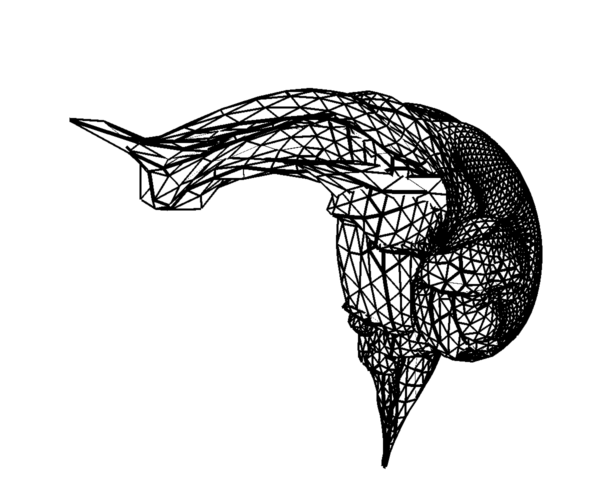} &    
			\includegraphics[width=0.32\columnwidth]{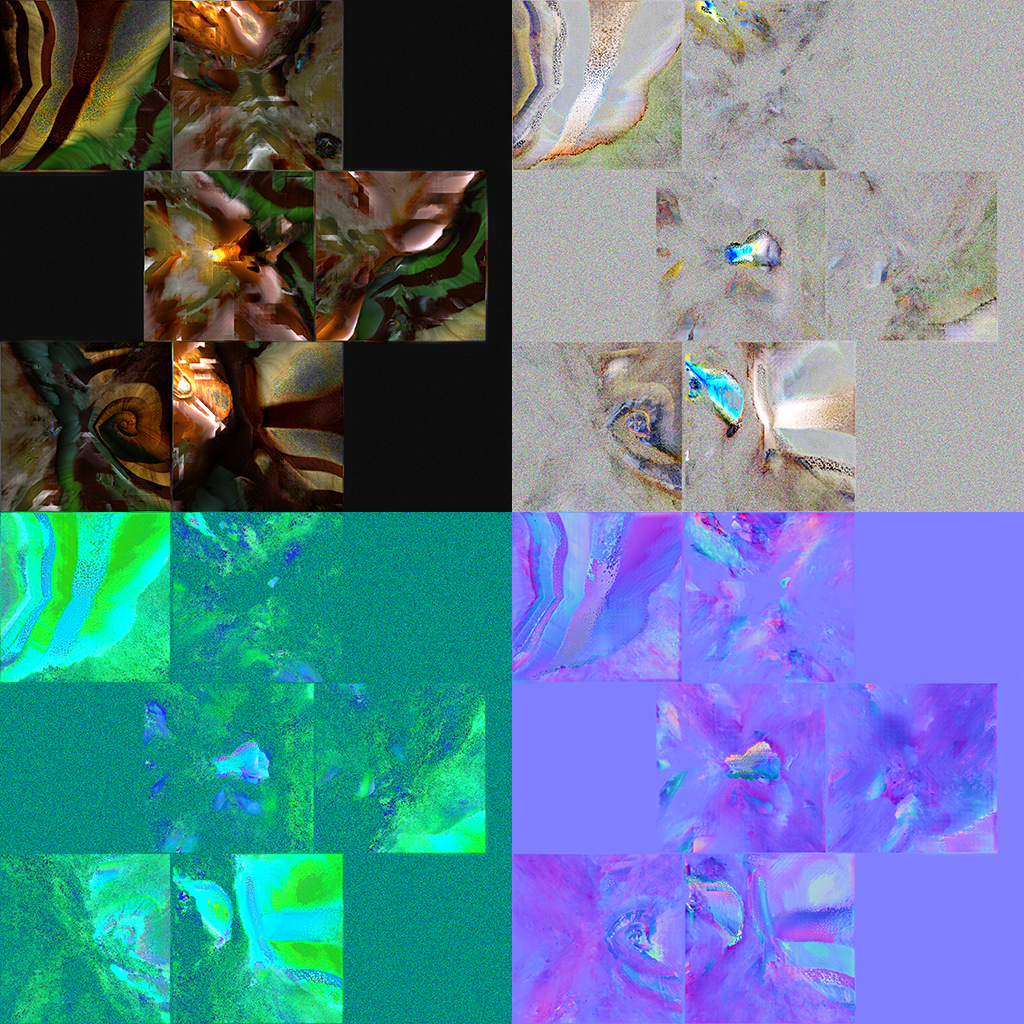} \\
			\small{Initial guess} & \small{Optimized mesh} & \small{Optimized materials}
		\end{tabular}
		\begin{tabular}{cc}
			\includegraphics[width=0.49\columnwidth]{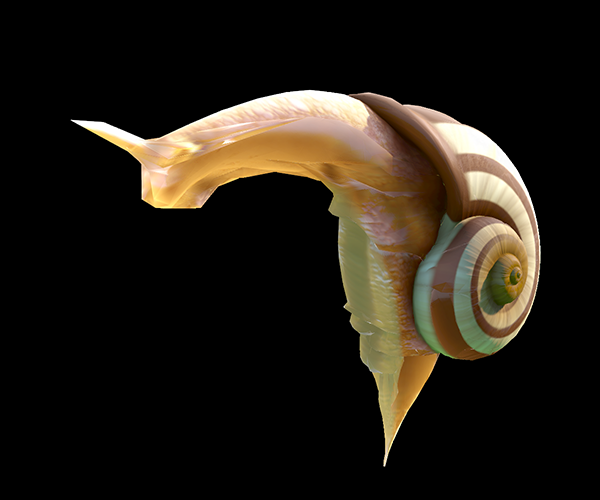} &
			\includegraphics[width=0.49\columnwidth]{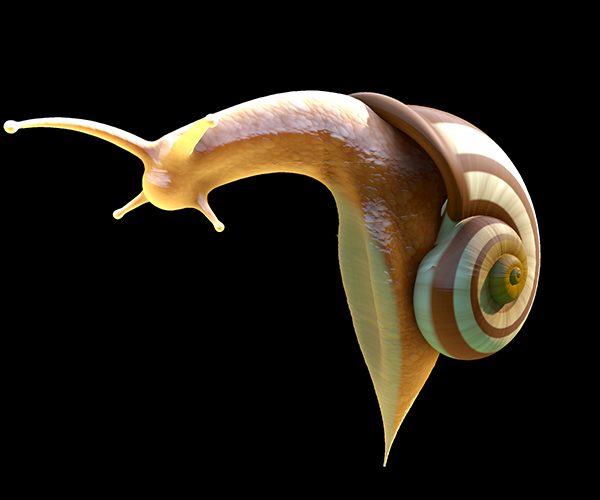} \\
			\small{Our rasterized model} &  \small{Ray marched reference} 
		\end{tabular}
		\captionof{figure}{Similar to Figure~\ref{fig:elephant}, we extract a mesh and materials from a 
			ray marched distance field: the ShaderToy ``Snail,'' from Inigo Quilez.}
		\label{fig:snail}
	\end{figure}
}

\figBaking

\pagebreak

\section{Comparison with Normal Map Bakers}
\label{sup:bakers}

In Figure~\ref{fig:baking} we compare our approach with two production quality normal map bakers (Simplygon 8.3 and
Substance Painter v2020.2.2) on a scanned skull, Hylobates sp.: Cranium\footnote{\url{https://3d.si.edu/object/3d/hylobates-klossii-cranium:4b963018-7642-431e-99ba-603b51bd158f}}, courtesy of the Smithsonian~\shortcite{Smithsonian2020}. 
The normal map resolution is $2048\times 2048$ texels for all versions and all versions use the same reduced input mesh with 9k tris 
(generated in Simplygon). To our knowledge, both Simplygon and Substance generate the normal map by ray tracing 
from the reduced mesh to the high resolution reference. 
In contrast, our version is optimized from image observations using a resolution of $2048\times 2048$ pixels for 10k steps
(randomized viewing conditions and lighting) in our differentiable rasterizer. 
As can be seen, given that we optimize both vertex positions of the base mesh and normal map texels from image observations, 
we obtain a slightly higher-quality result on the same triangle budget. If we restrict the optimization to only 
normal map texels, the quality is slightly lower (cf. ``Our, normals only'' versus ``Ours, normals \& positions'').

\figConvergence

\figLaplace

\section{Convergence}
\label{sup:convergence}

In Figure~\ref{fig:convergence} we show training convergence plots for five examples from the paper. 
The normal map baking example (Skull) is the easiest task, with diffuse shading, low dynamic range and 
joint optimization of the vertex positions and the tangent space normal map texels. 
The displacement map example (Dancer) performs joint optimization of shape, normal- and displacement maps
and uses a lower initial learning rate. The aggregate geometry 
examples, Gardenia and Hibiscus, are harder, as they include order-independent transparency and PBR shading. 
Additionally, in those examples, we used an MSE loss function to minimize the differences against the reference
when transferring the optimized assets to a path tracer, and the MSE loss is more sensitive to outliers than the
tone mapped $L_1$ variant (see main paper) used in the other examples. Finally, we show the Ewer sculpture, 
which includes high-frequency specular materials and high dynamic range lighting. Note that the convergence
plot show the image space \emph{training} error, and as we randomize the light and camera position each training iteration, 
some remaining noise is expected throughout the training. We note that all examples converge nicely.

\figInitialGuessB

\section{Influence of Normal Map and Laplacian}
\label{sup:ablation}

In Figure~\ref{fig:ewer_normal} we show how the use of normal map
and Laplacian regularizer during the optimization influences the quality of our results. 
We run 10k steps of optimization at a resolution of $2048\times 2048$ pixels.
All material textures are initialized to random values, except for the normal map,
which is initialized to $(0,0,1)$. 
Normal maps help capture micro-detail, which is clearly visible in the insets. 
The Laplacian regularizer helps stabilize optimization and improves mesh quality. Without it, 
large initial optimization steps may cause mesh corruption or self-intersections which are hard to 
recover from.

\figInitialGuessA

\section{Mesh Decimation: Varying Triangle Count in the Reduced Mesh}
\label{sup:decimation}

In Figure~\ref{fig:ewer_sweep} we study quality as function of the triangle count in our initial guess for the Ewer model. 
All reduced versions are genererated in Simplygon 8.3. 

In Figure~\ref{fig:tri_sweep} we study quality as function of the triangle count in our initial guess for the reduced mesh. 
All versions of the dancer are optimized for 5k iterations using a resolution of $2048\times 2048$ pixels.
As can be seen, small details and silhouettes benefit greatly from increased triangle count.
This result also shows the importance of optimizing the normal map. Details that are not
part of the silhouette, e.g., folds in cloth, are captured well even at low triangle counts. 
Even the statue with 500 triangles does a good job estimating the overall appearance of the reference mesh\footnote{\url{https://3d.si.edu/object/3d/figure-dancer:88de08dd-b8ab-470a-b987-ed6fe35def04}} and could be used as a distant level of detail.

\section{Tessellation as Level of Detail}
\label{sup:tessellation}

\figTessellationLOD
\figAggregateCmp

Tessellation is often used as a level of detail scheme. Here, we optimize for a single \emph{common} base mesh, displacement map and normal map, with the objective that their renderings reproduce the reference at all levels of tessellation.
Figure~\ref{fig:tess_sweep} shows increasing levels of tessellation on the dancer model from a base mesh with 1k triangles, optimized with $2048\times 2048$ pixels resolution and 5k iterations.
We use edge-midpoint-tessellation, wherein each subdivision step quadruples the triangle count.
From the insets we note that, as expected, silhouette edges and details are improved as tessellation is increased. 

Note that the fingers on the right hand of the figure are never accurately captured even at the highest level of tessellation. Here, the
limitation is in the base mesh. Displacement mapping cannot introduce concavities, and since the hand is not sufficiently modeled in the base mesh it cannot be recreated through displacement mapping. This would require increasing the polygon count of the base mesh, or 
possibly through an artist improving the initial guess to have higher tessellation in this region.

\figPrefilterLOD
\figSDFA
\figSDFB

\section{Approximating Aggregate Detail: Comparison with Stochastic Simplification}
\label{sup:stochsimplification}

In Figure~\ref{fig:stochsimpl}, we compare against stochastic simplification of aggregate detail.
Following Cook et~al.~\shortcite{Cook2007}, we stochastically remove 90\% ($\lambda = 0.1$) of the leaves from one instance of the Disney Moana Island 
Gardenia asset~\shortcite{Moana2018}, and adjust the element area of the reduced model by scaling each leaf uniformly with a factor
$\sqrt{1/\lambda}$. 
Finally, the contrast of each leaf texture is adjusted by modifying the color of its texels as $c'_i = \bar{c} + \sqrt{\lambda} (c_i - \bar{c})$,
where $\bar{c}$ is the average color of the texture.

Our version is optimized from image observations using a resolution of $2048\times 2048$ pixels for 10k steps
(randomized viewing conditions and lighting) in our differentiable rasterizer. We
start from an initial guess with 6.5k triangles where each leaf geometry is replaced with a quad 
and material parameters are randomized. Please refer to the paper for a visual example of the input mesh.
We visualize both a single model, and a larger scene with 3000 instances,
rendered in a path tracer. Our version, with more aggressive reduction (99.6\% of the triangles removed), 
still produces a high quality approximation by automatically moving geometry details into transparency- and normal maps. 
The shape and material parameters are optimized from image observations, so no heuristic for scaling or 
contrast adjustments is needed.

\section{Shape and Appearance Prefiltering}
\label{sup:material}

Figure~\ref{fig:lodzoom} is an extension of Figure 2 in the main paper, and shows the dancer model specifically optimized 
for four different rendering resolutions. Our results match the appearance of the antialiased reference well, considering the 
difference in sample count. As expected, geometric detail and shading are gradually smoothed as the rendering resolutions 
decrease, even though all results are generated from the same initial guess.

\section{Learning Mesh and Materials from Implicit Surfaces}
\label{sup:implicit}

In Figure~\ref{fig:elephant} we show an example of learning shape and materials to approximate
a signed distance field rendered using ray marching. We adapt a version of the ShaderToy 
``Elephant`` from Inigo Quilez\footnote{\url{https://www.shadertoy.com/view/MsXGWr}}, 
modified to isolate the main object.
Figure~\ref{fig:snail} shows a harder example with subsurface scattering, based on the ShaderToy 
``Snail`` from Inigo Quilez\footnote{\url{https://www.shadertoy.com/view/ld3Gz2}}, 
modified to isolate the main object.

We use a sphere with 12k triangles as an initial guess for the rasterizer, and optimize at a resolution of 
$2048\times 2048$ pixels for 10k steps. The appearance of the shaded result matches the reference well, 
and the sphere deforms to a reasonable mesh (please refer to the wireframe inset). However, we note that this 
example is limited by the quality of the initial guess, and further efforts would be required to generalize 
to more complex assets.

\end{document}